\documentclass[acmsmall]{acmart}
\usepackage{color}
\usepackage{xcolor}
\usepackage{threeparttable}
\usepackage{multirow}
\usepackage{multicol}
\usepackage{graphicx}
\usepackage{lscape}
\usepackage{soul}
\usepackage{ulem}
\usepackage{booktabs}
\usepackage{tcolorbox}
\usepackage{longtable} 
\usepackage{pdflscape} 
\usepackage{colortbl}
\usepackage{tabularx} 
\usepackage{booktabs} 
\usepackage{caption} 
\usepackage{algorithmic}
\usepackage{algorithm}
\AtBeginDocument{%
  \providecommand\BibTeX{{%
    \normalfont B\kern-0.5em{\scshape i\kern-0.25em b}\kern-0.8em\TeX}}}

\setcopyright{acmcopyright}
\copyrightyear{XXXX}
\acmYear{XXXX}
\acmDOI{XXXXXXX.XXXXXXX}

\begin{document}


\markboth{X. Wan et al.}{Data Complexity: A New Perspective for Analyzing the Difficulty of Defect Prediction Tasks}

\title{Data Complexity: A New Perspective for Analyzing the Difficulty of Defect Prediction Tasks}

\author{Xiaohui Wan}
\email{xhwan@buaa.edu.cn}
\affiliation{
    \department{School of Automation Science and Electrical Engineering}
    \institution{Beihang University}
    \city{Beijing}
    \postcode{100191}
    \country{China}
}
\author{Zheng Zheng}
\authornotemark[1]
\email{zhengz@buaa.edu.cn}
\affiliation{
    \department{School of Automation Science and Electrical Engineering}
    \institution{Beihang University}
    \city{Beijing}
    \postcode{100191}
    \country{China}
}
\author{Fangyun Qin}
\email{fyqin@cnu.edu.cn}
\affiliation{
    \department{College of Information Engineering}
    \institution{Capital Normal University}
    \city{Beijing}
    \postcode{100048}
    \country{China}
}
\author{Xuhui Lu}
\email{luxuhui@buaa.edu.cn}
\affiliation{
    \department{School of Automation Science and Electrical Engineering}
    \institution{Beihang University}
    \city{Beijing}
    \postcode{100191}
    \country{China}
}
\authorsaddresses{This work was funded by the National Natural Science Foundation of China, under Grant 61772055 and Grant 61872169. \\ Authors'addresses: Xiaohui Wan, Zheng Zheng (corresponding author), and Xuhui Lu, School of Automation Science and Electrical Engineering, Beihang University, Beijing, China; emails: \{xhwan, zhengz, luxuhui\}@buaa.edu.cn; Fangyun Qin, College of Information Engineering, Capital Normal University, Beijing, China; email: fyqin@cnu.edu.cn.}


%
%
%
\renewcommand{\shortauthors}{X. Wan et al.}

\begin{abstract}
Defect prediction is crucial for software quality assurance and has been extensively researched over recent decades. However, prior studies rarely focus on data complexity in defect prediction tasks, and even less on understanding the difficulties of these tasks from the perspective of data complexity. In this paper, we conduct an empirical study to estimate the hardness of over 33,000 instances, employing a set of measures to characterize the inherent difficulty of instances and the characteristics of defect datasets. Our findings indicate that: (1) instance hardness in both classes displays a right-skewed distribution, with the defective class exhibiting a more scattered distribution; (2) class overlap is the primary factor influencing instance hardness and can be characterized through feature, structural, instance, and multiresolution overlap; (3) no universal preprocessing technique is applicable to all datasets, and it may not consistently reduce data complexity, fortunately, dataset complexity measures can help identify suitable techniques for specific datasets; (4) integrating data complexity information into the learning process can enhance an algorithm's learning capacity. In summary, this empirical study highlights the crucial role of data complexity in defect prediction tasks, and provides a novel perspective for advancing research in defect prediction techniques.
\end{abstract}


%


\maketitle

\section{Introduction}
\label{sect:intro}

As software systems continue to grow in size and complexity, software quality assurance has become increasingly crucial in the software development life cycle. In such a context, defect prediction techniques have emerged as a crucial means of ensuring software quality \cite{catal2009systematic, hall2012systematic, menzies2010defect}. These techniques typically employ machine learning (ML) algorithms to train a prediction model from the collected historical defect data. This model can then be used to predict whether or not a given instance of code regions could exist defects. An accurate defect prediction model can greatly assist software testers in optimizing the allocation of limited testing resources by focusing more effort on the defect-prone code regions \cite{2016Defect, li2018progress}.  Over the past few decades, numerous approaches have been proposed, with the majority involving the application of ML algorithms \cite{menzies2006data, nagappan2006mining, zimmermann2008predicting, kim2008classifying, hassan2009predicting}. Hence, defect prediction has emerged as an important application domain for ML techniques.

Despite the breakthroughs of ML techniques and their applications in defect prediction, previous studies pursue optimal performance by modifying ML algorithms without understanding the data being modeled \cite{li2018progress}. As such, algorithmic development for defect prediction has largely been measured by evaluation measures (i.e., Matthews correlation coefficient) on defect datasets \cite{agrawal2018better, li2018progress}. However, these evaluation measures hold both advantages and disadvantages, and which one is more appropriate has been debated without reaching a consensus \cite{menzies2006data, moussa2022use}. Moreover, they provide only aggregated information about the ML algorithm and the task it performs. They do not offer detailed insights into which instances are misclassified, nor do they explain why this occurs \cite{smith2014instance}. The current understanding in the research field regarding the effects of data preprocessing techniques (e.g., data normalization, feature selection, and data re-sampling) on the complexity of defect prediction tasks is limited. In addition, it remains unclear whether there exist instances that all ML models misclassify, or whether some instances are inherently more difficult to classify, and what factors contribute to this difficulty. All of these unresolved problems strongly motivate us to conduct a thorough investigation into data complexity in defect prediction tasks.



Data complexity is a crucial concept in ML research that encompasses both instance hardness and dataset complexity. Instance hardness refers to the challenge of correctly classifying individual instances within a dataset, whereas dataset complexity concerns the overall difficulty of learning a concept from a dataset \cite{ho2002complexity, smith2014instance}. Ho and Basu identified three primary factors contributing to the complexity of a classification task \cite{ho2002complexity}: (1) class ambiguity due to ill-defined concepts or non-discriminative features; (2) high dimensionality and data sparsity with irrelevant or redundant features; and (3) complex decision boundaries lacking concise descriptions. In defect prediction, data complexity primarily manifests as class overlap \cite{chen2018tackling, gong2019empirical}, small disjuncts \cite{weiss2009impact}, noisy data \cite{shepperd2013data, tantithamthavorn2015impact}, irrelevant or redundant features \cite{lu2012software, shivaji2012reducing}, and class imbalance \cite{tantithamthavorn2018impact1}. Unfortunately, they are intertwined rather than separate, complicating defect prediction \cite{chen2018tackling}. Dataset characteristics significantly influence the process \cite{lopez2013insight}, necessitating consideration of multiple factors when tackling class imbalance challenges in defect prediction \cite{bennin2017mahakil, chen2018tackling, song2018comprehensive, wan2022spe}. Recognizing a task's intricacies is the first step toward resolving it, prompting numerous researchers to propose dataset complexity and instance hardness measures that capture the data complexity from both dataset and instance perspectives \cite{smith2014instance, lorena2019complex}. These measures help identify the domain of competence for each classification algorithm, enabling automated algorithm selection. Incorporating these measures into the learning process has led to considerable enhancements in ML model performance \cite{britto2014dynamic, luengo2015automatic, brun2018framework, walmsley2018ensemble, souza2019online}. Such practices highlight the importance of data complexity analysis in defect prediction, yielding valuable insights that advance defect prediction techniques.

To the best of our knowledge, no comprehensive empirical study on data complexity in defect prediction has been conducted thus far. To fill this gap, we perform an empirical study involving 36 open-source defect datasets and 25 commonly-used ML algorithms from previous defect prediction studies \cite{ghotra2015revisiting, tantithamthavorn2018impact1}. Given that estimating instance hardness is largely dependent on the selection of representative learning algorithms, 11 diverse ML algorithms are ultimately selected by using hierarchical agglomerative clustering. A systematic examination of instance hardness across all defect datasets is then conducted. Moreover, 15 instance-level measures \cite{smith2014instance} and 23 dataset-level measures \cite{lorena2019complex} are employed to characterize the data complexities of defect prediction tasks. Utilizing the aforementioned experimental results and correlation analysis, we investigate the sources of classification complexity in defect prediction tasks and the impact of data preprocessing techniques on data complexity. Lastly, we propose some novel ideas for integrating the instance hardness and data complexity measures into the training process and provide preliminary verification of their effectiveness. Hence, the main contributions of this article are as follows: 
\begin{itemize}
\item {We adopt unsupervised meta-learning to select 11 representative algorithms from 25 candidates to estimate the hardness value of each instance within defect datasets;}
\item {We explore the distribution characteristics and differences in hardness values for defective and non-defective instances, as well as for instances from different datasets;}
\item {We employ 15 instance hardness measures to characterize data complexities in defect prediction at the instance level and examine their impacts on defect prediction;}
\item {We apply 23 dataset complexity measures to characterize data complexities in defect prediction at the dataset level, investigating their impacts on defect prediction and the impacts of various data preprocessing techniques on the dataset complexity;}
\item {We provide some novel ideas about integrating data complexity information into the model learning process and experimentally verify their feasibility and effectiveness;}
\item {We provide all experimental data and source code available on GitHub repository\footnote{\url{https://github.com/Wan-xiaohui/DC4SDP}} to ensure the reproducibility of this study and to facilitate other types of future work.}

\end{itemize}

%

The remainder of this article is organized as follows: Section \ref{sect:related} provides an introduction to the background of defect prediction and previous studies on data complexity analysis and its applications. Section \ref{sect:preliminaries} presents the preliminaries of this work, including instance hardness, instance hardness measures, and dataset complexity measures. The experimental methodology and setup of our empirical study are described in Section \ref{sect:experiment}, and then four research questions (RQs) are answered in Section \ref{sect:results}. The potential threats that could affect the validity of this study are discussed in Section \ref{sect:validity}. Our conclusion and suggestions for future work are presented in the last section.

\section{Background and related work}
\label{sect:related}

Numerous defect prediction approaches have emerged in recent decades, primarily employing ML techniques \cite{menzies2006data, nagappan2006mining, zimmermann2008predicting, kim2008classifying, hassan2009predicting, nam2014survey, wang2016automatically, li2018progress}. However, the performance of ML models relies on dataset characteristics and the inherent difficulty of the classification task. In this section, we first review the application of ML techniques in defect prediction and then discuss prior studies related to data complexity analysis and its applications within the ML domain.

\subsection{Background on defect prediction Approaches}
\label{subsect:ml-defect prediction}

\begin{figure}[htbp]
\centering
\includegraphics[width=5.3in,keepaspectratio]{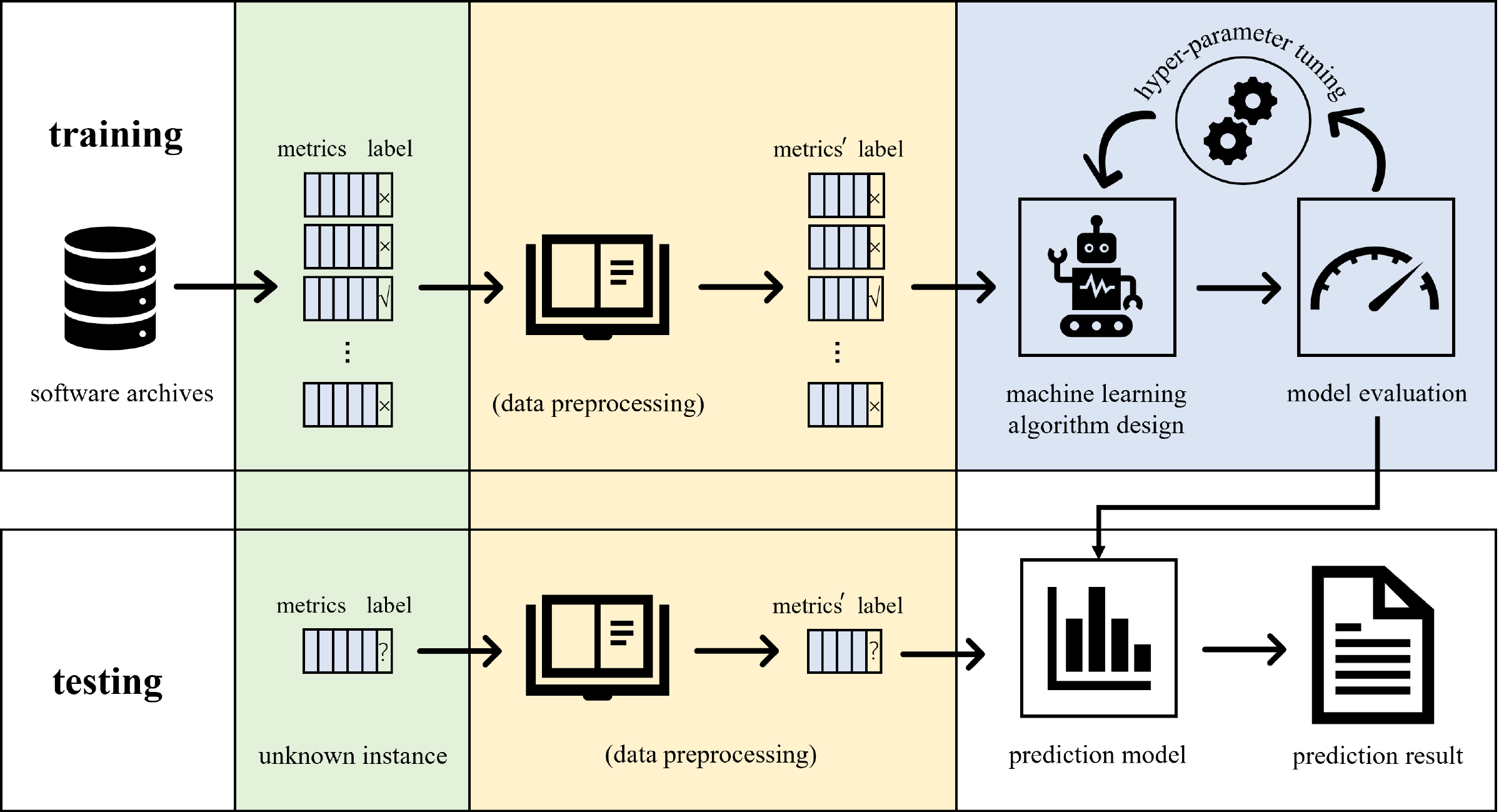}
\caption{A common process of ML-based defect prediction (including the training and testing phases)}
\label{Fig1}
\end{figure}

Fig. \ref{Fig1} presents a typical process for ML-based defect prediction. The first step is to collect instances from software archives, such as version control systems and issue tracking systems. Depending on the prediction granularity, each instance can represent a software component (or package), a file, a class, a function (method), or a code change. An instance has several metrics extracted from software archives and is labeled as either ``defective'' or ``non-defective'' (represented respectively by a tick and a cross). In other words, this paper regards defect prediction as a binary classification task. Once instances with metrics and labels are collected, we can apply data preprocessing techniques, such as data normalization, feature selection, and data re-sampling \cite{menzies2006data, shivaji2012reducing, tantithamthavorn2018impact1}. However, these operations are optional and depend on the characteristics of the defect datasets. Subsequently, plenty of ML algorithms can be utilized or developed to train a prediction model. Furthermore, hyper-parameter optimization techniques can be incorporated into the training process \cite{tantithamthavorn2016automated, fu2016tuning, li2020understanding}. Finally, the trained model can predict whether a new instance is defective or non-defective. Therefore, previous studies on defect prediction can be mainly categorized into three main directions.

The first direction is manually designing more useful metrics to represent defects. Actually, most software metrics can be divided into code metrics (CMs) and process metrics (PMs) \cite{rahman2013and}. CMs are directly collected from source code, and they mainly measure the complexity of source code, such as Halstead metrics \cite{maurice1977elements}, McCabe metrics \cite{mccabe1976complexity}, CK metrics \cite{chidamber1994metrics}, MOOD metrics \cite{harrison1998evaluation}. While PMs are extracted from historical information archived in various software repositories, and they mainly reflect the changes over time and the software development process, such as source code changes and developer information \cite{nam2014survey}. The representative PMs can be roughly divided into five categories: code change-based metrics, developer information-based metrics, dependency analysis-based metrics, project team organization-based metrics, and other metrics such as popularity and anti-pattern \cite{li2018progress}. However, there is a continued debate on if CMs are sufficient for defect prediction and whether PMs are better than CMs. In addition, a more recent trend is to leverage deep learning techniques to automatically learn the semantic representation of programs \cite{wang2016automatically, wang2018deep}.

The second direction is the application of ML techniques in defect prediction. ML techniques are the most popular approaches for defect prediction \cite{catal2009systematic, hall2012systematic, li2018progress}. With the development of ML technology in recent years, more and more emerging algorithms, such as dictionary learning \cite{jing2014dictionary}, collaborative representation learning \cite{jing2014software}, multiple kernel ensemble learning \cite{wang2016multiple}, deep learning \cite{yang2015deep, wang2016automatically} and transfer learning \cite{ma2012transfer, nam2013transfer}, have been applied to build defect prediction models. According to the type of ML algorithms, existing ML-based defect prediction approaches can be roughly divided into supervised, unsupervised, and semi-supervised approaches \cite{li2018progress}. Supervised approaches use labeled training data to train a defect prediction model, are the most extensively studied type of defect prediction approaches \cite{yan2017file}. Unsupervised approaches learn a prediction model by discovering hidden patterns or data groupings without labeled training data \cite{li2020systematic}, while semi-supervised approaches use both labeled and unlabeled data for training \cite{meng2021semi}.

Generally, we build a prediction model by learning from labeled instances within a project and then adopt it to predict the labels of unknown instances within the same project, which is known as within-project defect prediction \cite{turhan2009relative}. According to relevant literature \cite{shepperd2014researcher, herbold2018comparative, tantithamthavorn2018impact2, li2020understanding}, supervised defect prediction approaches can be categorized into seven types: statistical techniques (e.g., naive Bayes, logistic regression), rule-based techniques (e.g., rule lists, rule sets), neural networks (e.g., RBF-net, multi-layer perceptron), nearest neighbors (a.k.a., lazy-learning), support vector machines, decision trees (e.g., C4.5, CART), ensemble learning methods (e.g., bagging, boosting). Moreover, for new projects or projects with limited historical data, we leverage data from other projects (source project) to build a prediction model for the current project (target project), which is also known as cross-project defect prediction. There are two main types of cross-project defect prediction approaches: instance-based and feature-based. Instance-based approaches involve selecting or re-weighting source instances \cite{ma2012transfer, peters2013better, kawata2015improving, ryu2015hybrid, chen2015negative, ryu2016value}, while feature-based approaches make distributions similar by feature selection or feature transformation \cite{he2012investigation, herbold2013training, nam2013transfer, he2015empirical, qin2015cross, qin2018studying, wan2019supervised, qin2020empirical}.

The third direction pertains to data preprocessing techniques for defect prediction. Data preprocessing is one of the primary solutions for addressing data complexity challenges, aiming to simplify machine learning tasks by manipulating the data. At present, the most extensively studied data preprocessing techniques include data normalization \cite{graf2001normalization}, feature selection \cite{li2017feature}, and data re-sampling \cite{chawla2002smote}. First, data normalization is employed to ensure that all the original features in the raw training data are equally weighted, thereby reducing the negative impact of data heterogeneity \cite{graf2001normalization}. Min-max normalization \cite{qin2018studying} and standard normalization \cite{wan2022spe} are commonly used in defect prediction. The former normalizes each raw feature to have a mean of 0 and a variance of 1, while the latter normalizes each raw feature to have values ranging between 0 and 1. Second, feature selection techniques reduce the number of features to address the multicollinearity problem \cite{xu2016impact}. Recent studies have investigated the impact of feature selection techniques on the defect prediction performance \cite{xu2016impact, ghotra2017large}. Third, data re-sampling techniques are frequently employed to address class imbalance in defect prediction studies, involving undersampling, oversampling, and hybrid sampling \cite{he2009learning}. Undersampling removes data from the original dataset, potentially omitting crucial majority class information without appropriate guidelines. Conversely, many studies focus on synthetic oversampling methods \cite{bennin2017mahakil, agrawal2018better}, generating minority instances based on feature space similarities. A recent literature review highlights SMOTE as the most popular method, used by about 85\% of defect prediction approaches to address class imbalance concerns \cite{agrawal2018better}. SMOTE generates synthetic samples by interpolating between existing minority instances, thereby enhancing the balance between two classes while maintaining the distribution \cite{chawla2002smote}. Hybrid sampling combines oversampling and undersampling techniques to achieve a balanced dataset \cite{batista2003balancing, batista2004study}.

\subsection{Prior Research on Data Complexity Analysis}
\label{subsect:dc}

In the field of machine learning, there has been growing interest in characterizing the complexity of both individual instances and entire datasets, known as data complexity analysis. The concept of instance hardness was initially proposed by M. Smith et al. \cite{smith2009empirical}, which estimates the probability that a given instance in a particular dataset will be misclassified. Limited training samples are distributed in a high-dimensional and sparse feature space, which naturally results in certain samples being difficult to classify. For example, overlapping instances and minority instances in small disjuncts tend to have a relatively high instance hardness value, as a ML algorithm has to overfit the limited training instances to learn an accurate decision boundary on the entire dataset. Recently, some researchers have explored incorporating the information of instance hardness into the learning progress to improve the performance of ML models or alleviate data complexity issues \cite{smith2014instance, kabir2018mixed, liu2020self, liu2020mesa, chongomweru2021novel, zhou2022dynamic}. Moreover, Zhou et al. expanded on the notion of instance hardness by proposing dynamic instance hardness and integrated it into curriculum learning, effectively improving the efficiency and performance of curriculum learning approaches \cite{smith2016comparative, zhou2020curriculum}.


Instance hardness values reveal instances that are prone to misclassification, while hardness measures explain why they are difficult to classify. M. Smith et al. \cite{smith2014instance} proposed a set of interpretable hardness measures by examining the learning mechanisms of several simple ML algorithms. These measures provide insights into why particular instances are challenging to classify and how to detect them. Specifically, these measures reveal the difficulty of a classification problem at the instance level, rather than at the aggregate level of the entire dataset. However, these measures can also be averaged to estimate the hardness of the dataset as a whole. After that, Arruda et al. proposed new instance hardness measures as a supplement to the previous work \cite{arruda2020measuring}. An open-source Python library named PyHard\footnote{\url{https://pypi.org/project/pyhard/}} has been developed to implement all of the measures proposed in \cite{smith2014instance} and \cite{arruda2020measuring}. Recent work on dynamic classifier selection has demonstrated the utility of instance hardness measures in identifying classifiers that perform well in confusing or overlapping areas of the dataset, providing indications of local competency. Furthermore, these measures can be valuable for generating pools of classifiers in ensemble learning, where the probability of an instance being selected during the re-sampling process can be determined by its instance hardness measures. Experimental results indicate that incorporating instance hardness measures into the training process can improve the performance of the ensemble model on noisy datasets \cite{walmsley2018ensemble}.


Alongside the aforementioned research, another line of inquiry has proposed numerous dataset complexity measures \cite{ho2002complexity, singh2003multiresolution, li2006data, macia2011data, lorena2019complex}. Ho and Basu \cite{ho2002complexity} first suggested analyzing classification task complexity by examining dataset characteristics, and since then, more measures have been introduced \cite{singh2003multiresolution, li2006data, macia2011data}. Lorena et al. \cite{lorena2019complex} reviewed and expanded upon \cite{ho2002complexity}, assembling 22 dataset complexity measures into the R package ECoL, available on GitHub repository\footnote{\url{https://github.com/lpfgarcia/ECoL}}. Komorniczak et al. \cite{komorniczak2023problexity} developed a Python library, Problexity, for calculating dataset complexity measures, available on GitHub repository\footnote{\url{https://github.com/w4k2/problexity}}. Data complexity measures have found applications in data preprocessing \cite{pranckeviciene2006class, leyva2014set, okimoto2017complexity}, learning algorithms \cite{luengo2015automatic, brun2018framework, sun2019novel}, and meta-learning \cite{leyva2014set, garcia2020boosting}. In data preprocessing, these measures are employed for feature selection \cite{okimoto2017complexity}, instance selection \cite{kim2009using, leyva2014set}, noise identification \cite{saez2013predicting, smith2014instance, garcia2015effect}, and addressing class imbalance \cite{anwar2014measurement, zhang2019instance}. In learning algorithms, these measures help determine the domain of competence \cite{luengo2015automatic, britto2014dynamic} and guide algorithm design \cite{smith2014instance, brun2018framework}. In meta-learning, these measures serve as meta-features to describe the dataset characteristics \cite{leyva2014set, rivolli2022meta}.

Despite the growing attention data complexity has received in the field of ML research, its exploration within the context of defect prediction remains limited. Many researchers have yet to grasp the significance of data complexity in understanding the difficulties of defect prediction tasks and enhancing defect prediction techniques. This gap manifests in the limited knowledge of the distribution of easy and hard instances, the sources of difficulty in defect prediction tasks (including the classification difficulty of individual instances and entire datasets), and the role of data preprocessing techniques (particularly their impact on data complexity). As a result, so far, the field of defect prediction has lacked a comprehensive empirical investigation into data complexity. Moreover, no research has been conducted that analyzes the data complexity of a particular dataset and, selects data preprocessing operations and refines learning algorithms based on data complexity analysis. To address these gaps, this paper will focus on tackling these concerns.

%



\section{Preliminaries on Data Complexity Measures}
\label{sect:preliminaries}
This section reviews the definitions of instance hardness and dataset hardness in classification tasks, as well as data complexity measures from both dataset-level and instance-level perspectives. Each subsection outlines a group of measures and includes references for further details.

\subsection{Definition of Instance Hardness and Dataset Hardness}
\label{sect:ih}

Assume that we have a training set $t\!=\!\left\{\left(x_{i}, y_{i}\right)\!: \!x_{i} \! \in \!X \!\wedge\! y_{i} \!\in \!Y\right\}$, where the pairs in $t$ are drawn i.i.d., a classifier $h$ is induced by a ML algorithm $g$ trained on $t$ with hyper-parameters $\alpha$, i.e., $h=g(t, \alpha)$. For a training instance $\left\langle x_{i}, y_{i}\right\rangle$, the quantity $p\left(y_{i}\! \mid \! x_{i}, h\right)$ estimates the probability that $h$ assigns the label $y_{i}$ to the input $x_{i}$. The larger $p\left(y_{i} \! \mid \! x_{i}, h\right)$ is, the more likely $h$ is to produce the correct label to $x_{i}$. In turn, the smaller $p\left(y_{i} \! \mid \! x_{i}, h\right)$ is, the less likely $h$ is to produce the correct label for $x_{i}$. Hence, the definition of instance hardness with respect to $h$ is as follows \cite{smith2014instance}:
$$ \begin{aligned}
IH_{h}\left(\left\langle x_{i}, y_{i}\right\rangle\right)=1-p\!\left(y_{i} \! \mid \! x_{i}, g(t, a)\right)=1-p\!\left(y_{i} \! \mid \! x_{i}, h\right).
\end{aligned} $$

Apparently, the hardness of an instance is reliant on the instances in the training set and the underlying algorithm used to produce $h$. To eliminate the dependence of instance hardness on a specific hypothesis and understand what causes instances hardness in general, we sum up instance hardness values over the set of hypotheses $\mathcal{H}$ and weighting each $h\in \mathcal{H}$ by $p(h\! \mid \! t)$:
$$ \begin{aligned}
IH\! \left(\left\langle x_{i}, y_{i}\right\rangle\right) &=\sum_{\mathcal{H}}\!\left(1 \! - \! p\!\left(y_{i} \! \mid \! x_{i}, h\right)\right) p(h \! \mid \! t) = 1\!-\!\sum_{\mathcal{H}} p\!\left(y_{i} \! \mid \! x_{i}, h\right) p(h \! \mid \! t) .
\end{aligned} $$

Theoretically, we have to sum over the complete set of hypotheses $\mathcal{H}$, specifically, over the complete set of ML algorithms and hyper-parameters associated with each algorithm. Obviously, it is not feasible in practice. In this paper, we adopted Smith's proposed approach in \cite{smith2014instance} and restricted attention to a carefully chosen set of representative ML algorithms and hyper-parameters. In fact, it is important to estimate $p(h \! \mid \! t)$ because the free lunch theorem indicates that all instances would have the same instance hardness if all hypotheses are equally likely \cite{wolpert1996lack}. A natural way to approximate the unknown distribution $p(h \! \mid \! t)$ is to weight a set of representative ML algorithms and their associated hyper-parameters. Here, assuming that we have such a set $\mathcal{L}$ of ML algorithms, the instance hardness can be estimated with the following equation:
$$ \begin{aligned}
IH_{\mathcal{L}}\!\left(\left\langle x_i, y_i\right\rangle\right)=1-\frac{1}{|\mathcal{L}|} \sum_{j=1}^{|\mathcal{L}|} p\!\left(y_i \! \mid \! x_i, g_j(t, \alpha)\right),
\end{aligned}$$
where $p(h\!\mid\!t)$ is approximated as $\frac{1}{|\mathcal{L}|}$ and the distribution $p\left(y_{i} \! \mid \! x_{i}, g_j(t, \alpha)\right)$ can be estimated using the indicator function and confidence scores, as described in \cite{smith2014instance}. Several previous studies have attempted to incorporate the latter form of instance hardness into the learning progress \cite{kabir2018mixed, liu2020self, liu2020mesa, chongomweru2021novel, zhou2022dynamic}. However, in this paper, we have chosen to use the former manner to estimate the instance hardness, as confidence scores are often poorly calibrated and may lead to over-confident predictions for instances significantly different from the training set \cite{kuleshov2015calibrated, guo2017calibration}. The indicator function establishes the frequency of an instance being misclassified by different models through multiple repetitions. The presumption is that instances systematically misclassified by diverse classification algorithms can be regarded as difficult. For simplicity, we refer to $IH_{\mathcal{L}}$ as simply $IH$ from now on. By this point, the focus of instance hardness estimation has shifted to selecting representative ML algorithms and hyper-parameters, which will be discussed in detail in Section \ref{sect:algorithms}. Furthermore, it is possible to calculate an overall measure of hardness for a complete dataset. DataSet Hardness (DSH) can be defined by averaging instance hardness over the instances in a dataset as follows:
$$ \begin{aligned}
DSH(D)= \frac{\sum_{\left\langle x_i, y_i\right\rangle \in D} IH_{\mathcal{L}}\!\left(\left\langle x_i, y_i\right\rangle\right)}{|D|},
\end{aligned}$$
where $D$ is the dataset. DSH will be used in Section \ref{rq2}. Clearly, DSH does not take into account the class imbalance issue and treats all instances equally. In the context of defect prediction tasks, imbalanced datasets require a new measure to estimate dataset complexity from a dataset-level perspective. A naive approach would be to re-weight the instance hardness of minority and majority classes separately. Here, we propose a novel measure called Imbalanced DataSet Hardness (IDSH) to estimate the dataset complexity of at a dataset-level. The definition of IDSH is as follows:
$$ \begin{aligned}
IDSH(D)=1-\frac{1}{|\mathcal{L}|} \sum_{j=1}^{|\mathcal{L}|} MCC\!\left(D, g_j(t, \alpha)\right),
\end{aligned}$$
where MCC denotes the average Matthews Correlation Coefficient (MCC) of the trained classifier $h=g_j(t, \alpha)$ when tested on the dataset $D$. The MCC is commonly used in imbalanced classification tasks \cite{yao2020assessing, moussa2022use}. The average performance of classifiers on a dataset can provide an indication of dataset hardness: better performance suggests lower dataset hardness, while worse performance indicates greater dataset hardness. Since the MCC values of our classifier consistently exceed 0, we define the IDSH of a dataset as outlined above, which will be employed in Section \ref{rq3}.

\subsection{Instance-Level Analysis and Instance Hardness Measures}
\label{sect:ih}

To provide an instance-level analysis of defect data complexity, we employ 15 instance hardness measures proposed in  \cite{smith2014instance} and \cite{arruda2020measuring}. Moreover, we borrow ideas from \cite{paiva2022relating} which introduce modifications into some measures to limit and standardize their values. Therefore, all measures are constructed so that higher measure values represent harder instances. Each measure can characterize an aspect of why an instance may be misclassified (e.g., class overlap, complex boundary, and class imbalance). For example, kDN measures the local overlap of an instance in the feature space in relation to its nearest neighbors. The kDN of an instance is the percentage of the k neighbors (in terms of Euclidean distance) for an instance that do not share its target class value. kDN is defined as follows, where $\mathrm{kNN}(x)$ is the set of $k$ nearest neighbors of $x$ and $t(x)$ is the target class for $x$:
$$ \begin{aligned}
kDN(x)=\frac{|\{y: \! y \! \in \! \mathrm{kNN}(x) \! \wedge \! t(y) \! \neq \! t(x)\}|}{k}.
\end{aligned}$$

DS measure how tightly a ML algorithm has to divide the task space to correctly classify an instance, hence it is a measure related to the complexity of the decision boundary. Some ML algorithms, such as decision trees or rule-based algorithms, can express the learned concept as a disjunctive description. Therefore, the DS of an instance can be defined as the number of instances in a disjunct divided by the number of instances covered by the largest disjunct in a dataset:
$$ \begin{aligned}
DS(x)=\frac{|disjunct(x)|-1}{\max _{y \in D}|disjunct(y)|-1},
\end{aligned} $$
where $disjunct(x)$ returns the disjunct that covers instance $x$, and $D$ is the data set that contains instance $x$. The disjuncts are formed using a C4.5 decision tree \cite{quinlan2014c4} without pruning. However, the current definition is negatively correlated with instance hardness. Hence, we set 1 minus the value as the DS value in practice. DCP, like DS, measures the overlap of an instance on a subset of the features. More Specific, the TCP of an instance is the number of instances in a disjunct belonging to its class divided by the total number of instances in the disjunct:
$$ \begin{aligned}
DCP(x)=\frac{|\{z: z \in disjunct(x) \wedge t(z)=t(x)\}|}{|disjunct(x)|}.
\end{aligned} $$

CL provides a global measure of overlap and the likelihood of an instance belonging to a class. The CL of an instance belonging to a certain class is defined as follows, where $|x|$ refers to the number of features of instance $x$ and $x_{i}$ represents the value of the $i$-th feature of instance $x$\footnote{Continuous variables are assigned a probability using a kernel density estimation.}:
$$ \begin{aligned}
CL(x)=\mathrm{CL}(x, t(x))=\prod_i^{|x|} P\left(x_i \mid t(x)\right).
\end{aligned} $$

Besides, more measures are defined exactly from various perspectives. For instance, decision trees can estimate the minimum description length required for an instance by using the depth of the leaf node for an instance in an induced C4.5 decision tree (both pruned (TD\_P) and unpruned (TD\_U)) as an estimate. MV (Minority Value) measures the skewness of the class an instance belongs to by taking the ratio of the number of instances sharing its target class value to the number of instances in the majority class. CB (Class Balance) is an alternative measure of class imbalance. A more recent paper presented more instance hardness measures by decomposing the dataset complexity measures discussed in Section \ref{sect:dm}, such as F1, N1, N2, LSC and LSR. Moreover, we adopt two instance hardness measures proposed in \cite{leyva2015three}, namely Usefulness and Harmfulness, denoted as ``U'' and ``H''. Usefulness corresponds to the fraction of instances having a given instance in their local sets, while Harmfulness is defined as the fraction of instances having a given instance as their nearest enemy. Some of these measures can be grouped into class overlap measures, including feature overlap, structural overlap, instance-level overlap, and multiresolution overlap. Feature overlap measures characterise the class overlap of individual features in data,  structural overlap is associated with the internal structure of classes, instance-level overlap is typically associated to the error of the k-nearest neighbor classifier, and multiresolution overlap measures use multiresolution approaches to identify regions of different complexity within the domains \cite{santos2022joint}. It is worth noting that the original definitions of these measures have standardized and limited in value, as described in \cite{paiva2022relating}. Due to space constraints, Table~\ref{table1} provides only a summary of all hardness measures employed in this work. For detailed descriptions and definitions of each measure, please refer to \cite{leyva2015three, arruda2020measuring, paiva2022relating}.


\begin{table*}[htbp]
\caption{A summary of all the instance hardness measures adopted in this paper}
\label{table1}
\centering
\begin{threeparttable}
\footnotesize
\renewcommand\arraystretch{1.0}
\setlength{\tabcolsep}{4.5mm}{
\begin{tabular}{clccl}

\toprule
\textbf{Acronym}               & \textbf{Measure Name}       & \textbf{Min}      & \textbf{Max}         & \textbf{References} \\
\midrule
kDN                            & k-Disagreeing Neighbors     & 0                 & 1                    & \cite{smith2014instance, paiva2022relating} \\
DS                             & Disjunct Size               & 0                 & 1                    & \cite{smith2014instance} \\
DCP                            & Disjunct Class Percentage   & 0                 & 1                    & \cite{smith2014instance, paiva2022relating} \\
TD\_P                          & Tree Depth (Pruned)         & 0                 & 1                    & \cite{smith2014instance, paiva2022relating} \\
TD\_U                          & Tree Depth (Unpruned)       & 0                 & 1                    & \cite{smith2014instance, paiva2022relating} \\
CL                             & Class Likelihood            & 0                 & 1                    & \cite{smith2014instance, paiva2022relating} \\
MV                             & Minority Value              & 0                 & 1                    & \cite{smith2014instance} \\
CB                             & Class Balance               & 0                 & 1                    & \cite{smith2014instance} \\
F1                             & Fraction of features in overlapping areas          & 0                 & 1                    & \cite{arruda2020measuring, paiva2022relating} \\
N1                             & Fraction of nearby instances of different classes  & 0                 & 1                    & \cite{arruda2020measuring, paiva2022relating} \\
N2                             & Ratio of the intra-class and extra-class distances & 0                 & $\approx 1$          & \cite{arruda2020measuring, paiva2022relating} \\
LSC                            & Local Set Cardinality                              & 0                 & 1                    & \cite{arruda2020measuring, paiva2022relating} \\
LSR                            & Local Set Radius                                   & 0                 & 1                    & \cite{arruda2020measuring, paiva2022relating} \\
U                              & Usefulness                                         & $\approx 0 $      & 1                    & \cite{leyva2015three, arruda2020measuring, paiva2022relating} \\
H                              & Harmfulness                                        & 0                 & $\approx 1$          & \cite{leyva2015three, arruda2020measuring, paiva2022relating} \\
\bottomrule
\end{tabular}}
\end{threeparttable}
\end{table*}

\subsection{Dataset-Level Analysis and Dataset Complexity Measures}
\label{sect:dm}


This subsection examines what causes hardness at the dataset level. According to \cite{lorena2019complex}, dataset complexity measures can be grouped into 6 categories as follows:

\begin{itemize}
\item \textbf{Feature-based measures}: They evaluate the discriminative power of the features;
\item \textbf{Linearity measures}: They quantify whether different classes can be linearly separated;
\item \textbf{Neighborhood measures}: They characterize the dataset based on neighboring information;
\item \textbf{Network measures}: They model the dataset as a graph and extract its structural information;
\item \textbf{Dimensionality measures}: They evaluate data sparsity based on the dataset dimensionality;
\item \textbf{Class imbalance measures}: They analyze the ratio of the sample size of different classes.
\end{itemize}

F1 is a feature-based measure, which means it evaluate the complexity of a dataset by analyzing the distribution of each feature. Specifically, F1 uses the maximum Fisher's discriminant ratio to measures the overlap between the values of the features in different classes. It is defined as:
$$ \begin{aligned}
F 1=\frac{1}{1+\max _{i=1}^m r_{f_i}},
\end{aligned} $$
where $r_{f_i}$ is a discriminant ratio for each feature $f_i$, $m$ is the number of features. Here we adopt an alternative for $r_{f_i}$ computation proposed in \cite{arruda2020measuring}. The mathematical definition of  $r_{f_i}$ is as follows:
\setlength\abovedisplayskip{1.5pt}
\setlength\belowdisplayskip{1.5pt}
$$ \begin{aligned}
r_{f_i}=\frac{\sum_{j=1}^{n_c} n_{c_j}\!\left(\mu_{c_j}^{f_i}\!-\mu^{f_i}\right)^2}{\sum_{j=1}^{n_c}\! \sum_{l=1}^{n_{c_j}}\!\left(\!x_{l i}^j-\mu_{c_j}^{f_i}\right)^2},
\end{aligned} $$
where $n_{c_j}$ is the number of instances in class $c_j$, $\mu_{c_j}^{f_i}$ is the mean of feature $f_i$ across instances of class $c_j$, $\mu^{f_i}$ is the mean of the $f_i$ values across all the classes, and $x_{l i}^j$ denotes the individual value of the feature $f_i$ for an instance from class $c_j$. In addition, F1 originally takes the value of the largest discriminant ratio among all the features. If at least one feature discriminates the classes, the dataset can be considered easier than if no such feature exists. Here, this paper uses the inverse of the original F1 value, hence higher values indicate more complex problems.

There are several measures similar to F1, including F1v, F2, F3, and F4. F1v is recommended as a complementary measure to F1, as it searches for a vector that can separate the two classes after projecting instances into it and considers a directional Fisher criterion proposed in \cite{malina2001two}. On the other hand, both F2 and F3 measures account for the overlap of the input feature distributions in each class. Specifically, F2 calculates the overlap by determining the minimum and maximum values of each feature within the classes. In contrast, F3 estimates the individual efficiency of each feature in separating the classes and considers the maximum value found among all features. Additionally, F4 builds on F3 and provides an overview of how the features work together.

Regarding measures of linearity, L1 assesses the linearity of data by computing the sum of the distances between misclassified instances and the linear boundary used for their classification. To obtain the linear classifier, we employ a linear SVM, as recommended by Orriols-Puig et al. \cite{orriols2010documentation}. If L1 yields zero, the problem is linearly separable and considered simpler than problems requiring a non-linear boundary. Furthermore, L2 calculates the error rate of the linear SVM classifier. Let $h\!\left(\mathbf{x}_i\right)\!$ represent the linear classifier, and L2 can be calculated as follows:
$$ \begin{aligned}
L2=\frac{\sum_{i=1}^n I\!\left(h\!\left(\mathbf{x}_i\right)\! \neq y_i\right)}{n},
\end{aligned} $$
Greater L2 values indicate more errors, implying that the data cannot be linearly separated, and hence the problem is more complex. On the other hand, L3 measures the non-linearity of a linear classifier. The procedure commences by generating a new dataset via interpolation of paired training instances belonging to the same class. Then, a linear classifier is trained on the original data, and L3 is determined by the error rate measured on the new data points.

Neighborhood measures aim to capture the shape of the decision boundary and characterize the class overlap by analyzing local neighborhoods of the data points. We take the simplest neighborhood measure N3 as an example, N3 is the error rate of a 1-Nearest-Neighbor (NN) classifier estimated by using a leave-one-out procedure. It is denoted as:
$$ \begin{aligned}
N3=\frac{\sum_{i=1}^n I\!\left(NN\!\left(\mathbf{x}_i\right)\! \neq y_i\right)}{n},
\end{aligned} $$
where $\mathbf{x}_i$ represents an instance from a defect dataset with $n$ total instances, and $y_i \in \left\{0, 1\right\}$ is its corresponding label, where $y_i = 1$ indicates that the code snippet contains one or more bugs, while $y_i = 1$ indicates a non-defective code snippet. The $NN\!\left(\mathbf{x}_i\right)$ term is the prediction made by a 1-NN classifier trained on all the other instances. N4 is similar to L3, but employs a NN classifier instead of a linear classifier. N1 calculates the percentage of vertices associated with edges linking instances of opposite classes in a minimum spanning tree derived from the data. N2 calculates the ratio of two sums: the sum of distances between each instance and its corresponding intra-class, and the sum of distances between each instance and its corresponding extra-class. T1 is the ratio between the number of remaining hyperspheres and the total number of instances in the dataset, and an alternative implementation based on the distance matrix between all examples has also been used \cite{lorena2019complex}. Lastly, the local set average cardinality measure (LSC) is defined based on a concept of Local Set (LS). The LS of an instance $\mathbf{x}_i$ in a dataset $T$ is the set of points from $T$ whose distance to $\mathbf{x}_i$ is smaller than the distance from $\mathbf{x}_i$ to $\mathbf{x}_i$'s nearest enemy:
$$ \begin{aligned}
LS\left(\mathbf{x}_i\right)=\left\{\mathbf{x}_j \mid d\left(\mathbf{x}_i, \mathbf{x}_j\right)<d\left(\mathbf{x}_i, ne\left(\mathbf{x}_i\right)\right)\right\},
\end{aligned} $$
where $ne\left(\mathbf{x}_i\right)$ is the nearest enemy from instance $\mathbf{x}_i$. In that case, LSC is calculated as:
$$ \begin{aligned}
LSC=1-\frac{1}{n^2} \sum_{i=1}^n\left|LS\left(\mathbf{x}_i\right)\right|,
\end{aligned} $$
where $\left|LS\left(\mathbf{x}_i\right)\right|$ is the cardinality of the LS for instance $\mathbf{x}_i$. Higher values indicate more complex datasets, where each instance is closer to an enemy than to others from the same class.

There are also network, dimensionality, and class imbalance measures. For network measures, Density considers the number of edges retained in the dataset's graph, normalized by the maximum number of edges between $n$ data point pairs. ClsCoef evaluates the grouping tendency of graph vertices by monitoring how close neighboring vertices are to forming cliques. Hubs assigns a score to each node based on its number of connections to other nodes, weighted by those neighbors' connection count. As for dimensionality measures, T2 divides the number of instances in the dataset by their dimensionality, here we take its inverse to yield higher values for more complex datasets. T3 is defined using Principal Component Analysis (PCA) of the dataset, while T4 represents the ratio of PCA dimension to the original dimension, providing a rough measure of the proportion of relevant dimensions for the dataset. Regarding class imbalance measures, C1 is defined as the entropy of class proportions, while C2 is derived from the imbalance ratio.

In this paper, we employ a total of 23 dataset complexity measures. Out of these, 22 measures are derived from \cite{lorena2019complex}. Additionally, considering the widespread class imbalance in defect prediction \cite{tantithamthavorn2018impact1}, we adopt the Bayes imbalance impact index (abbreviated BI3), which estimates the degree of deterioration caused solely by imbalance across the entire dataset, rather than just considering the imbalance ratio \cite{lu2019bayes}. Table \ref{table2} summarizes these measures along with essential information.

\begin{table*}[htbp]
\caption{A summary of all the dataset complexity measures adopted in this paper}
\label{table2}
\centering
\begin{threeparttable}
\footnotesize
\renewcommand\arraystretch{1.}
\setlength{\tabcolsep}{1.6mm}{
\begin{tabular}{cclccl}

\toprule
\textbf{Acronym}               & \textbf{Category}             & \textbf{Measure Name}                                      & \textbf{Min}      & \textbf{Max}      & \textbf{References} \\
\midrule
F1                             & Feature-based                 & Maximum Fisher’s discriminant ratio                        & $\approx 0$       & 1                 & \cite{ho2002complexity, lorena2019complex} \\
F1v                            & Feature-based                 & Directional vector maximum Fisher’s discriminant ratio     & $\approx 0$       & 1                 & \cite{lorena2019complex} \\
F2                             & Feature-based                 & Volume of overlapping region                               & 0                 & 1                 & \cite{ho2002complexity, lorena2019complex} \\
F3                             & Feature-based                 & Maximum individual feature efficiency                      & 0                 & 1                 & \cite{ho2002complexity, lorena2019complex} \\
F4                             & Feature-based                 & Collective feature efficiency                              & 0                 & 1                 & \cite{lorena2019complex} \\ \midrule
L1                             & Linearity                     & Sum of the error distance by linear programming            & 0                 & $\approx 1$       & \cite{ho2002complexity, lorena2019complex} \\
L2                             & Linearity                     & Error rate of linear classifier                            & 0                 & 1                 & \cite{ho2002complexity, lorena2019complex} \\
L3                             & Linearity                     & Non linearity of linear classifier                         & 0                 & 1                 & \cite{ho2002complexity, lorena2019complex} \\ \midrule
N1                             & Neighborhood                  & Faction of borderline points                               & 0                 & 1                 & \cite{ho2002complexity, lorena2019complex} \\
N2                             & Neighborhood                  & Ratio of intra/extra class NN distance                     & 0                 & $\approx 1$       & \cite{ho2002complexity, lorena2019complex} \\
N3                             & Neighborhood                  & Error rate of NN classifier                                & 0                 & 1                 & \cite{ho2002complexity, lorena2019complex} \\
N4                             & Neighborhood                  & Non linearity of NN classifier                             & 0                 & 1                 & \cite{ho2002complexity, lorena2019complex} \\
T1                             & Neighborhood                  & Fraction of hyperspheres covering data                     & 0                 & 1                 & \cite{ho2002complexity, lorena2019complex} \\
LSC                            & Neighborhood                  & Local set average cardinality                              & 0                 & $1-\tfrac{1}{n}$  & \cite{lorena2019complex} \\ \midrule
Density                        & Network                       & Density                                                    & 0                 & 1                 & \cite{lorena2019complex} \\
ClsCoef                        & Network                       & Clustering Coefficient                                     & 0                 & 1                 & \cite{lorena2019complex} \\
Hubs                           & Network                       & Hubs                                                       & 0                 & 1                 & \cite{lorena2019complex} \\ \midrule
T2                             & Dimensionality                & Average number of features per points                      & $\approx 0$       & $m$               & \cite{ho2002complexity, lorena2019complex} \\
T3                             & Dimensionality                & Average number of PCA dimensions per points                & $\approx 0$       & $m$               & \cite{lorena2019complex} \\
T4                             & Dimensionality                & Ratio of the PCA dimension to the original dimension       & 0                 & 1                 & \cite{lorena2019complex} \\ \midrule
C1                             & Class Balance                 & Entropy of classes proportions                             & 0                 & 1                 & \cite{lorena2019complex} \\
C2                             & Class Balance                 & Imbalance ratio                                            & 0                 & 1                 & \cite{lorena2019complex} \\
BI3                         & Class Balance                 & Bayes Imbalance Impact Index                               & 0                 & 1                 & \cite{lu2019bayes}       \\
\bottomrule
\end{tabular}
}
\begin{tablenotes}
\footnotesize
\item $n$ stands for the number of instances in a dataset; $m$ corresponds to its number of features.
\end{tablenotes}
\end{threeparttable}
\end{table*}

\section{Experimental methodology}
\label{sect:experiment}
This section begins by introducing four research questions to be addressed in this study. We then provide an overview of the experimental methodology employed to investigate these questions, which includes benchmark datasets, baseline classification algorithms, model training and hyper-parameter tuning, as well as correlation analysis and significance testing methods.

\subsection{Research Questions}
\label{sect:RQs}

The objective of this paper is to conduct an empirical investigation into data complexity in the context of defect prediction, examining it from both the dataset-level and instance-level perspectives, as well as exploring the application of data complexity information during the training process. To achieve this, we therefore formulate four research questions (RQs) as follows:

\begin{itemize}
\item \textbf{RQ1}: \emph{How are hardness values distributed among instances in the defective and non-defective classes, as well as across different datasets?}
\item \textbf{RQ2}: \emph{What is the manifestation of data complexity in defect prediction when viewed from the instance-level perspective?}
\item \textbf{RQ3}: \emph{How does dataset complexity manifest in defect prediction and what is the effect of data preprocessing methods on it?}
\item \textbf{RQ4}: \emph{How can we integrate data complexity information into the learning process and how effective is it when compared to baseline methods?}
\end{itemize}

\textbf{RQ1} investigates the proportions of instances that are easiest or hardest to classify and their hardness distribution across defective and non-defective classes. \textbf{RQ2} examines the classification difficulty for defective/non-defective instances from an instance-level perspective. This involves calculating instance hardness values and measures for all instances, exploring primary causes of hardness for defective/non-defective instances through correlation analysis, and determining the average of instance hardness measures for each dataset to analyze exhibited data complexities. Subsequently, \textbf{RQ3} calculates dataset complexity measures for all datasets, conducts correlation analysis to investigate reasons for classification difficulty from a dataset-level perspective, and examines the impact of data preprocessing operations on dataset complexity. Lastly, \textbf{RQ4} explores approaches for incorporating instance hardness measures and dataset complexity measures into the training process and validates their effectiveness through comparison experiments.


\subsection{Benchmark Defect Datasets}
\label{sect:datasets}

In selecting the studied defect datasets, we identified three important criteria:

\begin{itemize}
\item \textbf{Criterion 1}: To promote the research reproducibility of our experiments, we choose the benchmark defect datasets hosted in publicly available software repositories.
\item \textbf{Criterion 2}: To combat potential bias in our conclusions, the defect datasets consist of different types of metrics with different granularity. Moreover, all defect datasets are chosen from various corpora and have been widely used in the defect prediction literature.
\item \textbf{Criterion 3}: To avoid inadequate learning of the model, the sample size of each dataset is required to be more than 100. Moreover, if the dataset has more than one version, only the latest version is selected, since different versions of a project may be similar.
\end{itemize}

Finally, we select 36 datasets that satisfy our criteria. They are from 4 corpora (i.e., AEEEM, NASA, PROMISE and ReLink) and have been widely used in prior work \cite{li2020understanding, tantithamthavorn2018impact2, tantithamthavorn2016automated}. Specifically, we use the AEEEM corpus as provided by \cite{d2010extensive}, the cleaned version of the NASA corpus as provided by \cite{shepperd2013data}, the PROMISE corpus as provided by \cite{jureczko2010towards}, the ReLink corpus provided by \cite{wu2011relink}. Table \ref{table3} provides an overview of defect datasets, with their corpora, number of metrics, type of metrics, total number of instances, number/ratio of defective instances, and prediction granularity.

\begin{table*}[htbp]
\caption{An overview of the benchmark defect datasets}
\label{table3}
\centering
\begin{threeparttable}
\footnotesize
\renewcommand\arraystretch{1.}
\setlength{\tabcolsep}{1.mm}{
\begin{tabular}{p{1.3cm}p{2cm}p{0.8cm}p{0.8cm}p{1.4cm}p{1.2cm}p{1.2cm}p{1.4cm}p{1.4cm}}

\toprule
\textbf{Corpora}   & \textbf{Dataset}   &\textbf{Item}    & \textbf{\#M}   & \textbf{\#T} & \textbf{\#I}  & \textbf{\#D}  & \textbf{\%D}  & \textbf{Granularity} \\
\midrule

\multirow{5}{*}{AEEEM}    & EQ          &1         & 61         & CMs, PMs & 324        & 129                  & 39.81                & Class          \\
                          & JDT         &2           & 61         & CMs, PMs & 997        & 206                  & 20.66                & Class          \\
                          & LC          &3              & 61         & CMs, PMs & 691        & 64                   & 9.26                 & Class          \\
                          & ML          &4            & 61         & CMs, PMs & 1862       & 245                  & 13.16                & Class          \\
                          & PDE         &5        & 61         & CMs, PMs & 1497       & 209                  & 13.96                & Class          \\ \midrule
\multirow{12}{*}{NASA}    & CM1         &6             & 37         & CMs      & 327        & 42                   & 12.84                & Function        \\
                          & JM1         &7             & 21         & CMs      & 7720       & 1612                 & 20.88                & Function        \\
                          & KC1         &8             & 21         & CMs      & 1162       & 294                  & 25.30                & Function        \\
                          & KC3         &9             & 39         & CMs      & 194        & 36                   & 18.56                & Function        \\
                          & MC1         &10             & 38         & CMs      & 1952       & 36                   & 1.84                 & Function        \\
                          & MC2         &11            & 39         & CMs      & 124        & 44                   & 35.48                & Function        \\
                          & MW1         &12             & 37         & CMs      & 250        & 25                   & 10.00                & Function        \\
                          & PC1         &13             & 37         & CMs      & 679        & 55                   & 8.10                 & Function        \\
                          & PC2         &14             & 36         & CMs      & 722        & 16                   & 2.22                 & Function        \\
                          & PC3         &15             & 37         & CMs      & 1053       & 130                  & 12.35                & Function        \\
                          & PC4         &16             & 37         & CMs      & 1270       & 176                  & 13.86                & Function        \\
                          & PC5         &17             & 38         & CMs      & 1694       & 458                  & 27.04                & Function        \\ \midrule
\multirow{15}{*}{PROMISE} & ant-1.7     &18         & 20         & CMs      & 745        & 166                  & 22.28                & Class          \\
                          & arc         &19             & 20         & CMs      & 234        & 27                   & 11.54                & Class          \\
                          & camel-1.6   &20       & 20         & CMs      & 965        & 188                  & 19.48                & Class          \\
                          & ivy-2.0     &21         & 20         & CMs      & 352        & 40                   & 11.36                & Class          \\
                          & jedit-4.3   &22       & 20         & CMs      & 492        & 11                   & 2.24                 & Class          \\
                          & log4j-1.2   &23       & 20         & CMs      & 205        & 189                  & 92.20                & Class          \\
                          & lucene-2.4  &24      & 20         & CMs      & 340        & 203                  & 59.71                & Class          \\
                          & poi-3.0     &25         & 20         & CMs      & 442        & 281                  & 63.57                & Class          \\
                          & prop-6      &26          & 20         & CMs      & 660        & 66                   & 10.00                & Class          \\
                          & redaktor    &27        & 20         & CMs      & 176        & 27                   & 15.34                & Class          \\
                          & synapse-1.2 &28     & 20         & CMs      & 256        & 86                   & 33.59                & Class          \\
                          & tomcat      &29          & 20         & CMs      & 858        & 77                   & 8.97                 & Class          \\
                          & velocity-1.6&30    & 20         & CMs      & 229        & 78                   & 34.06                & Class          \\
                          & xalan-2.7   &31       & 20         & CMs      & 909        & 898                  & 98.79                & Class          \\
                          & xerces-1.4  &32      & 20         & CMs      & 588        & 437                  & 74.32                & Class          \\ \midrule
\multirow{5}{*}{ReLink}   & Apache      &33          & 26         & CMs      & 194        & 98                   & 50.52                & File           \\
                          & Eclipse34-debug &34   & 17         & CMs, PMs & 1065       & 263                  & 24.69                & File           \\
                          & Eclipse34-swt &35   & 17         & CMs, PMs & 1485       & 653                  & 43.97                & File           \\
                          & Zxing       &36           & 26         & CMs      & 399        & 118                  & 29.57                & File           \\
\bottomrule
\end{tabular}}
\begin{tablenotes}
\footnotesize
\item \#M denotes the number of metrics; \#T denotes the types of metrics; \#I denotes the number of instances; \#D denotes the number of defective instances; \%D denotes the proportion of defective instances.
\end{tablenotes}
\end{threeparttable}
\end{table*}

\subsection{Representative Classification Algorithms}
\label{sect:algorithms}

Given that most defect prediction studies focus on within-project settings using supervised learning approaches \cite{nam2014survey, li2018progress}, this paper examines the data complexity of within-project defect prediction using supervised learning approaches. Moreover, accurate estimation of instance hardness relies on the appropriate selection of classification algorithms. To ensure a good representation of $\mathcal{H}$ and a reasonable estimation of instance hardness, we bias the selection of representative algorithms to those that (1) are commonly employed in defect prediction literature, (2) representative of different learning paradigms, and (3) exhibit different algorithmic biases, thereby maintaining diversity in prediction outputs. To achieve this, we first summarize relevant literature \cite{herbold2018comparative, tantithamthavorn2018impact2, li2020understanding} to identify candidate classification algorithms. Following this, we employ the Classifier Output Difference (COD) \cite{peterson2005estimating} to choose a diverse set of representative algorithms from the candidate pool.

\subsubsection{Candidate Classification Algorithms}

The candidate classification algorithms considered in this paper can be divided into 7 categories. Each category is briefly described as follows.

\begin{itemize}
\item \textbf{Statistical Methods}: Statistical techniques rely on probability-based models \cite{kotsiantis2007supervised} to find meaningful patterns and build predictive models \cite{berson2004overview}. Naive Bayes (NB) and generalized linear model are two common statistical methods. For generalized linear regression models, Ridge classifier (Ridge) and Logistic Regression (LR) are implemented in this paper. Ridge uses l2-norm regularization, while LR uses both l1-norm and l2-norm regularization.


\item \textbf{Lazy learning Methods}: Lazy learning (a.k.a., Nearest Neighbours) methods do not require an explicit training process, resulting in shorter training times but longer testing times \cite{kotsiantis2007supervised}. In this paper, we implement the K-Nearest Neighbor (KNN) algorithm, which classify test instances to the majority class among the K most similar training instances.

\item \textbf{Decision Tree Methods}: Decision trees use a recursive and greedy method to partition the feature space, constructing an inverted tree with root, internal, and leaf nodes. This tree can be transformed into a set of rules by tracing paths from the root node to each leaf node \cite{kotsiantis2007supervised}. In this paper, we implement the Classification And Regression Tree (CART), which splits the dataset into a tree by optimizing the sub-node homogeneity using the Gini index \cite{breiman2017classification}.


\item \textbf{Support Vector Machine Methods}: Support Vector Machines (SVMs) seek a hyperplane in high-dimensional space that maximize the margin, i.e., the maximum distance between samples from both classes. SVMs utilize support vectors, the samples closest to the hyperplane that impact its position and orientation, to optimize the margin. While the original SVM algorithm is a non-probabilistic linear model, kernel SVMs can achieve flexible hyperplanes by implicitly mapping inputs into a high-dimensional space using kernel tricks.

\item \textbf{Neural Network Methods}: A neural network consists of input, hidden, and output layers with connected neurons (nodes). Its functionality depends on the neuron model, network structure, and learning algorithm, which adjusts neuron weights to generate outputs that increasingly resemble target outputs \cite{singh2009neural}. In this paper, we implement the Multi-Layer Perceptron (MLP), a widely-used neural network for defect prediction studies \cite{ghotra2015revisiting}.
\item \textbf{Ensemble Learning Methods}: Ensemble learning combines predictions from multiple diverse models to improve predictive performance \cite{zhou2012ensemble}. Bagging and Boosting are two common types of ensemble Learning solutions. Bagging reduces variance and helps prevent overfitting by training individual models on different bootstrap samples, while Boosting builds a strong classifier from multiple weak classifiers by sequentially and adaptively learning each classifier to correct the errors of its predecessor. Ensemble learning can be homogeneous, consisting of models with a single-type learning algorithm, or heterogeneous, consisting of models with different algorithms. For example, Random Forest (RF) and Extremely randomized Trees (ExtraTrees) belong to Bagging and homogeneous ensembles. AdaBoost and Gradient Boosted Decision Trees (GBDT) belong to Boosting and homogeneous ensembles. A voting classifier with different base learners belongs to heterogeneous ensembles.
\item \textbf{Rule-based Methods}: Rule-based classifiers utilize a set of IF-THEN rules for classification. There are two primary strategies for combining multiple rules: rule lists and rule sets. Both strategies address the problem of overlapping rules in different ways. A rule list establishes an order for the decision rules, while a rule set resembles a democracy of rules, with some potentially having greater voting power \cite{molnar2020interpretable}. In this paper, we implement two rule-based techniques, one for rule sets, Boosted Rule Set (Boosted-RS), and one for rule lists, Greedy rule list (Greedy-RL). Specifically, Boosted-RS sequentially fits a set of rules with the Adaboost algorithm, while Greedy-RL fits a list with the CART algorithm \cite{singh2021imodels}.

\end{itemize}

Finally, we implement 25 candidate classification algorithms with default hyper-parameters as set in a machine learning library named Scikit-Learn \cite{scikit-learn}, including NB, Ridge, LR, KNN, CART, linSVM (SVMs without a kernel), rbfSVM (SVMs with a rbf kernel), MLP1 (MLPs with a single hidden layer of 10 neurons), MLP2 (MLPs with two hidden layers of 10 neurons), RF, ExtraTrees, GBDT, Voting (a Voting classifier with 5 base learners, NB, LR, SVM, DT, and MLP), Bagging with different base learners (including NBBagging, LRBagging, SVMBagging, DTBagging, and MLPBagging), Boosting with different base learners (including NBBoosting, LRBoosting, SVMBoosting, DTBoosting, and MLPBoosting), and two rule-based classifiers (Boosted-RS, Greedy-RL). They are representative of learning paradigms in machine learning applications and defect prediction. Data normalization, feature selection and data re-sampling are considered in the training phase of representative algorithms. In fact, NBBoosting is excluded first and does not appear in the subsequent dendrogram because it fails to fit on certain datasets due to the base classifier being worse than random.



\subsubsection{Representative Classification Algorithms}

Considering that ensemble learning algorithms accounts for more than half of the candidate algorithms, we have to select a diverse set of representative classification algorithms for unbiased estimating the instance hardness values.

\begin{figure}[htbp]
\centering
\includegraphics[width=3.3in,keepaspectratio]{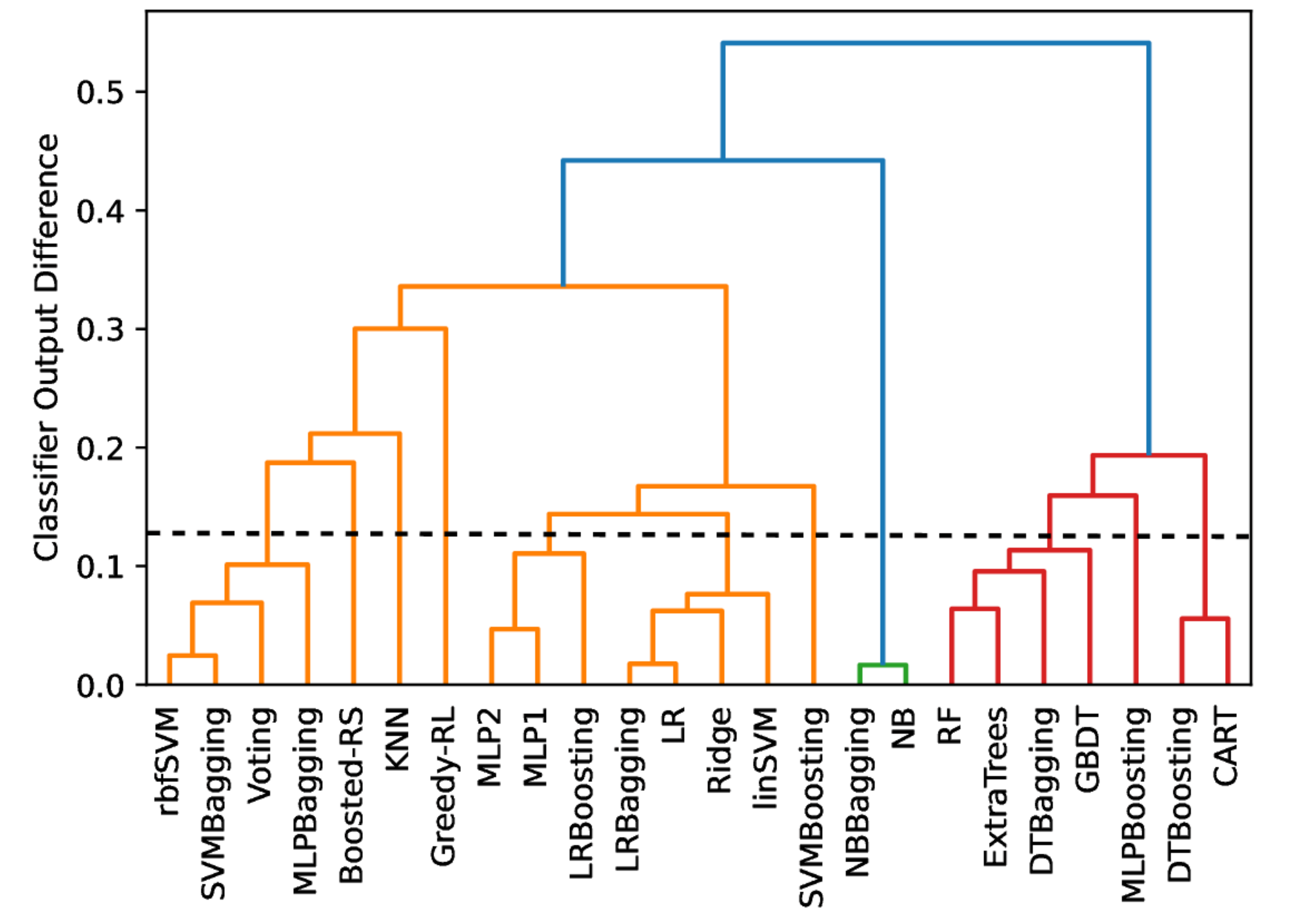}
\caption{Dendrogram of candidate classification algorithms clustered based on the COD scores}
\label{Fig2}
\end{figure}

Specifically, we adopt a concept in unsupervised meta-learning \cite{lee2011metric}, namely Classifier Output Difference (COD), to measure the diversity between candidate algorithms. COD measures the distance between two classification algorithms as the probability that the classification algorithms make different predictions. Unsupervised meta-learning then clusters these algorithms based on their COD scores with hierarchical agglomerative clustering. Hierarchical clustering is a common cluster analysis approach which aims to build a hierarchy of clusters, and agglomerative clustering is the most common type of hierarchical clustering \cite{mullner2011modern}. More specifically, agglomerative clustering works in a ``bottom-up'' manner, which means each object (it represents a classification algorithm here) is initially considered as a single-element cluster (leaf). At each step, two most similar clusters will be combined into a new bigger cluster (nodes). The above procedure is iterated until clusters have been merged into one big cluster containing all objects (root). Finally, the result is a tree-based representation of the objects, which is named dendrogram and shown in Fig. \ref{Fig2}.

As shown in Fig. \ref{Fig2}, the height of the line connecting two clusters corresponds to the distance (COD value) between them. A cut-point of 0.13 was chosen and a representative algorithm from each cluster was utilized to create $\mathcal{L}$. To limit the proportion of ensemble learning methods in the candidate algorithms, we preferentially select a base algorithm rather than an ensemble learning algorithm in a cluster. Moreover, we have to balance 7 categories of classification algorithms. Finally, a set of 11 representative classification algorithms are selected as shown in Table \ref{table4}. Thereinto, SVM represents a support vector machine without specifying a kernel type, and MLP represents a multi-layer perceptron without specifying the number of hidden layers and neurons.

\begin{table*}[htbp]
\centering
\caption{A set of 11 representative classification algorithms selected in 25 candidate algorithms}
\label{table4}
\begin{threeparttable}
\footnotesize
\renewcommand\arraystretch{1.1}
\setlength{\tabcolsep}{3.3mm}{
\begin{tabular}{ll}
\toprule
\textbf{Representative Classification Algorithms} \\
\midrule
K Nearest Neighbors (KNN)     & Naive Bayes (NB) \\
Classification And Regression Tree (CART) & Logistic Regression (LR)  \\
Support Vector Machines (SVM) & Multi-Layer Perceptron (MLP)  \\
Boosted Rule Sets (Boosted-RS) & Greedy Rule Lists (Greedy-RL) \\
Random Forest (RF) & Boosting with Support Vector Machines (SVMBoosting) \\
Boosting with Multi-Layer Perceptrons (MLPBoosting)\\
\bottomrule
\end{tabular}
}
\end{threeparttable}
\end{table*}

It is noted that instance hardness could be calculated with either more specific or boarder sets of representative classification algorithms, and each set may obtain different results. In general, more classification algorithms may obtain a more accurate estimate of instance hardness, we expect to select a relatively small and diverse set of representative classification algorithms to achieve a trade-off between efficiency and accuracy. Moreover, as the development of ML-based defect prediction techniques, the set of candidate algorithms is constantly growing and evolving and hence no exact solution is possible. Even so, we can still fairly select a set of representative algorithms from more candidate algorithms and simply adjust $\mathcal{L}$ by the way described above.


%

\subsection{Model Training and Hyper-parameter Tuning}
\label{sect:training}

For calculating the instance hardness of each instance in a dataset $D$, we have to compute the probability that a instance is misclassified when a prediction model is trained on the other instances from the dataset. Since the leave-one-out procedure is computationally prohibitive, a 5 by 5-fold cross-validation procedure is conventional and adopted to conduct the experiments. Specifically, the dataset $D$ is randomly shuffled 5 times. In each time, the shuffled dataset is divided into 5 folds, 4 folds are used for training, the left-out fold is left for testing. The process is repeated 5 times, using a different fold for testing each time. And in the training process, we also need to divide the training data into training set and validation set, the purpose of validation set is to provide an unbiased evaluation of a model during training. Since our datasets are imbalanced, the division uses stratification to maintain similar class distribution over different folds, and data re-sampling (SMOTE) is applied only to the training set. In addition, standard normalization, correlation-based feature selection (CFS) are utilized in the data preprocessing stage. After that, a prediction model is trained on the preprocessed training data, and then it is applied to each test instance to predict the existence of defects. Therefore, each instance is tested 5 times by the same classification algorithm, and whether these algorithms classify the instances correctly in each time is recorded.

\begin{figure}[htbp]
\centering
\includegraphics[width=4.8in,keepaspectratio]{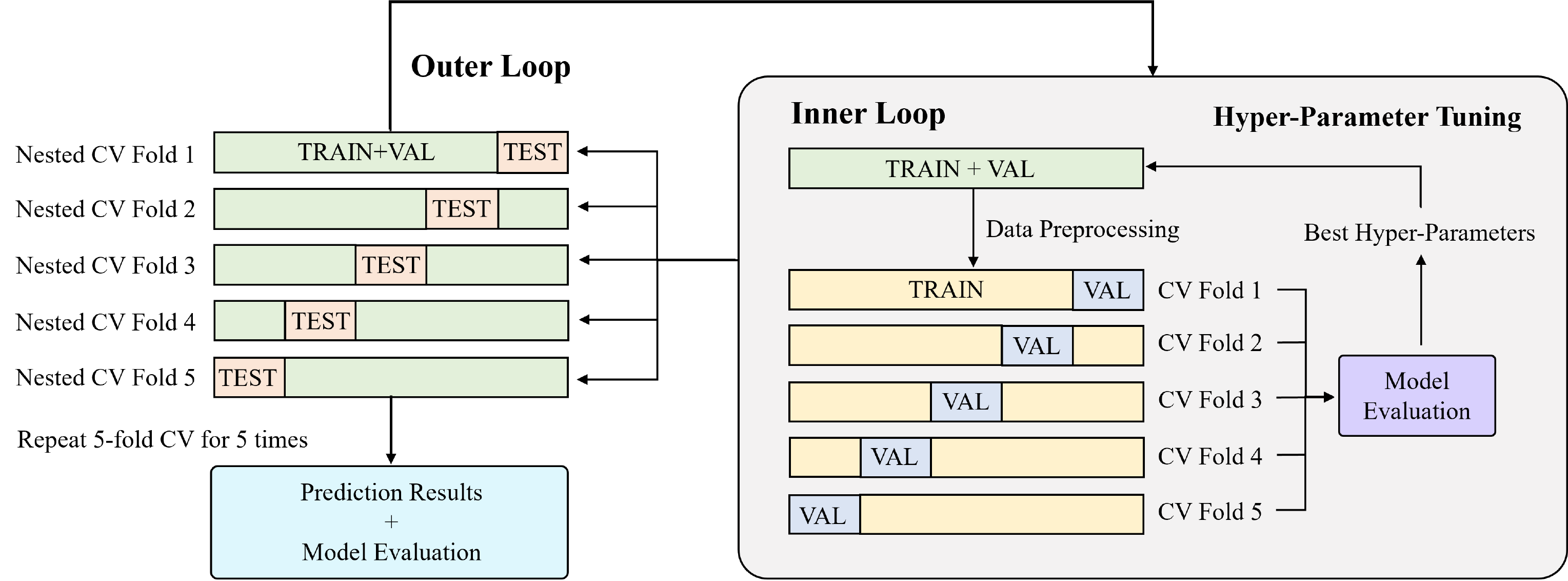}
\caption{A nested cross-validation process for training and testing the defect prediction models}
\label{Fig3}
\end{figure}

\begin{table*}[htbp]
\caption{The hyper-parameter space of each classification algorithm considered in our experiments}
\label{table5}
\centering
\begin{threeparttable}
\scriptsize
\renewcommand\arraystretch{1}
\setlength{\tabcolsep}{0.5mm}{
\begin{tabular}{clll}
\toprule
\multirow{2}{*}{\textbf{Algorithms}} & \multicolumn{3}{c}{\textbf{Hyper-parameters} } \\
\cmidrule(l){2-4}
& \multicolumn{1}{c}{\textbf{Name}} & \multicolumn{1}{c}{\textbf{Description}} & \multicolumn{1}{c}{\textbf{Range}} \\
\midrule
\multirow{4}{*}{KNN}
& n\_neighbors              & Number of neighbors to use for kneighbors queries [R]             & [1, 10]                              \\
& weights                   & Weight function used in prediction [C]                            & \{`uniform', `distance'\}              \\
& p                         & Power parameter for the Minkowski metric [R]                      & [1, 5]                               \\
& algorithm                 & Nearest neighborhood search algorithm [C]               & \{`auto', `ball\_tree', `kd\_tree'\}  \\
\midrule
\multirow{1}{*}{NB}
& var\_smoothing            & Portion of the largest variance of features added to variances [R]    & [$e^{-10}$, $e^{-1}$]  \\
\midrule
\multirow{4}{*}{CART}
& criterion                 & The function to measure the quality of a split [C]                    & \{`gini', `entropy'\}         \\
& max\_features             & The number of features to consider at each split [C]                  & \{`auto', `sqrt', `log2'\}    \\
& min\_samples\_split       & The minimum number of samples required to split the node [N]  & [2, 5]                        \\
& min\_samples\_leaf        & The minimum number of samples required to be at a leaf node [N]       & [1, 5]                            \\
\midrule
\multirow{4}{*}{LR}
& l1\_ratio                 & The Elastic-Net mixing parameter [R]                                  & [0, 1]         \\
& C                         & Inverse of regularization strength [R]                                & [$e^{-10}$, $e^{10}$]            \\
& penalty                   & The norm of the penalty, here both L1 and L2 terms are added [C]      & \{`elasticnet'\} \\
& solver                    & Optimization algorithm, which depends on the penalty chosen [C]       & \{`saga'\}    \\
\midrule
\multirow{4}{*}{SVM}
& kernel                    & The kernel type to be used in the algorithm [C]                       & \{`linear', `poly', `rbf'\} \\
& C                         & Regularization parameter to control the regularization strength [R]   & [$e^{-1}$, $e^{10}$] \\
& gamma                     & Kernel coefficient for the specified kernel [C]                       & \{0.01$\times$(range(10)+1), `auto'\}   \\
& max\_iter                 & Maximum number of iterations [C]                                      & \{10$\times$range(5, 21)\}    \\
\midrule
\multirow{4}{*}{MLP}
& alpha                     & Strength of the L2 regularization term [R]                            & [$e^{-10}$, $e^{-1}$]         \\
& max\_iter                 & Maximum number of iterations [C]                                      & \{10$\times$range(5, 21)\}    \\
& learning\_rate            & Learning rate schedule for weight updates [C]                         & \{`constant', `invscaling', `adaptive'\} \\
& hidden\_layer\_sizes      & The number of neurons in each hidden layer [C]                        &  \{(i,) or (i, j) for i, j in [5, 7, 9, 11]\} \\
\midrule
\multirow{4}{*}{Boosted-RS}
& n\_estimators             & The maximum number of estimators [C]                                  & \{10$\times$(range(5)+1)\}    \\
& criterion                 & The function to measure the quality of a split [C]                    & \{`gini', `entropy'\}         \\
& max\_depth                & The maximum depth of the tree [N]                                     & [1, 3]                        \\
& max\_features             & The number of features to consider at each split [C]                  & \{`auto', `sqrt', `log2'\}    \\
\midrule
\multirow{2}{*}{Greedy-RL}
& criterion                 & The function to measure the quality of a split [C]                    & \{`gini', `entropy'\}         \\
& max\_depth                & The maximum depth of the tree [N]                                     & [1, 5]                        \\
\midrule
\multirow{4}{*}{RF}
& n\_estimators             & The number of trees in the forest [C]                                 & \{10$\times$(range(5)+1)\}         \\
& criterion                 & The function to measure the quality of a split [C]                    & \{`gini', `entropy'\}              \\
& max\_features             & The number of features to consider at each split [C]                  & \{`sqrt', `log2', 0.7\}            \\
& min\_samples\_split       & The minimum number of samples required to split the node [N]          & [2, 5]                             \\
\midrule
\multirow{5}{*}{SVMBoosting}
& n\_estimators             & The maximum number of estimators [C]                                 & \{10$\times$(range(5)+1)\}         \\
& kernel                    & The kernel type to be used in the algorithm [C]                       & \{`linear', `poly', `rbf'\} \\
& C                         & Regularization parameter to control the regularization strength [R]   & [$e^{-1}$, $e^{10}$] \\
& gamma                     & Kernel coefficient for the specified kernel [C]                       & \{0.01$\times$(range(10)+1), `auto'\}   \\
& max\_iter                 & Maximum number of iterations [C]                                      & \{10$\times$range(5, 21)\}    \\
\midrule
\multirow{5}{*}{MLPBoosting}
& n\_estimators             & The maximum number of estimators [C]                                 & \{10$\times$(range(5)+1)\}         \\
& alpha                     & Strength of the L2 regularization term [R]                            & [$e^{-10}$, $e^{-1}$]         \\
& max\_iter                 & Maximum number of iterations [C]                                      & \{10$\times$range(5, 21)\}    \\
& learning\_rate            & Learning rate schedule for weight updates [C]                         & \{`constant', `invscaling', `adaptive'\} \\
& hidden\_layer\_sizes      & The number of neurons in each hidden layer [C]                        &  \{(i,) or (i, j) for i, j in [5, 7, 9, 11]\} \\
\bottomrule
\end{tabular}}
\begin{tablenotes}
\footnotesize
\item[] [N] denotes an integer value from range; [R] denotes real value from range; [C] denotes a choice from categories.
\end{tablenotes}
\end{threeparttable}
\end{table*}

Since hyper-parameter optimization can impact the experimental results \cite{fu2016tuning, tantithamthavorn2016automated, qu2018impact, tantithamthavorn2018impact2, li2020understanding}, we incorporate a hyper-parameter optimization procedure in our experimental setup, acting as an inner loop for each training fold of the outer cross-validation. The nested cross-validation process is illustrated in Fig. \ref{Fig3}. Within the inner loop, a candidate set of hyper-parameters is evaluated through 5-fold cross-validation upon the training folds from the outer loop. In addition, we employe a Bayesian optimization technique, provided by an open-source hyper-parameter optimization library named Hyperopt \cite{bergstra2013making}, to determine the hyper-parameter values for each classification algorithm. The objective is to get closer to the optimal prediction performance achievable for the given instances and classification algorithms. A candidate tuple of hyper-parameters is evaluated within the inner loop. MCC is adopted to assess defect prediction performance, as recommended by recent studies \cite{yao2020assessing, moussa2022use}. MCC is perfectly symmetric and accounts for all four situations in the confusion matrix, capturing overall performance even with imbalanced data. The MCC value ranges from -1 to 1, where -1 indicates perfect misclassification, 1 indicates perfect classification, and 0 represents a random guess classifier. To address the impracticality of assessing all hyper-parameter values, we establish a candidate value space based on relevant studies  \cite{tantithamthavorn2016automated, tantithamthavorn2018impact2, li2020understanding}, as presented in Table \ref{table5}. For unmentioned hyper-parameters, we adopt the recommended values in Scikit-Learn. Our experiments are conducted on Ubuntu 20.04 using two NVIDIA GTX 3090 GPUs, four 10-core CPUs with Intel(R) Xeon(R) Silver 4210R@2.40GHz, and 256GB DDR4 memory.



\subsection{Correlation Analysis and Significance Test}
\label{sect:significance}

As we employ so many measures to analyze the data complexity, it is crucial to examine if there's any correlation between these measures through correlation coefficient analysis. In statistics, there are several types of correlation coefficients, such as Pearson's $r$, Spearman's rho ($r_{s}$), and Kendall's Tau ($\tau$) \cite{boddy2009statistical}. For our study, we adopt Spearman's $r_{s}$ \cite{spearman1961proof} to perform correlation analysis of complexity measures. Spearman's $r_{s}$ is a nonparametric rank statistical measure of the strength and the direction of the arbitrary monotonic association between two ranked variables or one ranked variable and one measurement variable. Therefore, it does not require to make any assumptions about the frequency distribution and the linear relationship between the two variables, nor measured on interval scale. Spearman's $r_{s}$ ranges from -1.0 (a perfect negative correlation) to 1.0 (a perfect positive correlation),with an $r_{s}$ of 0 indicating no correlation. The strength of the correlation can be verbally described using a guide for the absolute value of $r_{s}$, as presented in Table \ref{table6} \cite{evans1996straightforward}.


\begin{table*}[htbp]
\centering
\caption{Interpretation of the Spearman's correlation coefficients}
\label{table6}
\begin{threeparttable}
\footnotesize
\renewcommand\arraystretch{1.1}
\setlength{\tabcolsep}{10.5mm}{
\begin{tabular}{cc}
\toprule
\textbf{Absolute value of $r_{s}$}      & \textbf{Strength of relationship} \\
\midrule
$0.0 \leq \vert r_{s}\vert < 0.2$       & Very weak correlation  \\
$0.2 \leq \vert r_{s}\vert < 0.4$       & Weak correlation \\
$0.4 \leq \vert r_{s}\vert < 0.6$       & Moderate correlation \\
$0.6 \leq \vert r_{s}\vert < 0.8$       & Strong correlation \\
$0.8 \leq \vert r_{s}\vert \leq 1.0$    & Very strong correlation \\
\bottomrule
\end{tabular}
}
\end{threeparttable}
\end{table*}

Additionally, a significance test of correlations between different measures was conducted by using the Wilcoxon rank test ($p$ values). It aims to determine whether the null hypothesis (a general statement that there is no relationship between the two measured variables) should be accepted or rejected. If a $p$ value is less than a pre-specified significance level (usually $\alpha = 0.05$), the hypothesis test is statistically significant, and the null hypothesis should be rejected. Hence, the $p$ value denotes the likelihood that the strength of the monotonic association is a chance occurrence.

\section{Experimental Results}
\label{sect:results}

In this section, we answer our research questions based on the experimental results.

\subsection{RQ1: How are hardness values distributed among instances in the defective and non-defective classes, as well as across different datasets?}
\label{rq1}

To answer \textbf{RQ1}, we first calculate the hardness of each instance using prediction results from 11 representative models trained on 36 defect datasets. We then explore the differences in the distribution of instance hardness values between defective and non-defective classes, as well as the potential variations in the distribution of instance hardness across datasets. To estimate the probability of an instance $x$ being misclassified when the model is trained on other instances from the dataset, we adopt a 5-by-5-fold cross-validation procedure. Each instance is independently predicted 55 times (11 models $\times$ 5 repeat times) to determine the frequency of an instance being misclassified, as an estimate of its hardness. Notably, our analysis encompasses over 33,000 instances, and we produce a total of 9,900 models from 11 algorithms trained on 36 datasets using 5-by-5-fold cross-validation. With a significant volume and diversity of datasets and models, our results provide extensive insights into the distribution of instances across varying levels of hardness.

\begin{figure}[htbp]
\centering
\includegraphics[width=4.5in,keepaspectratio]{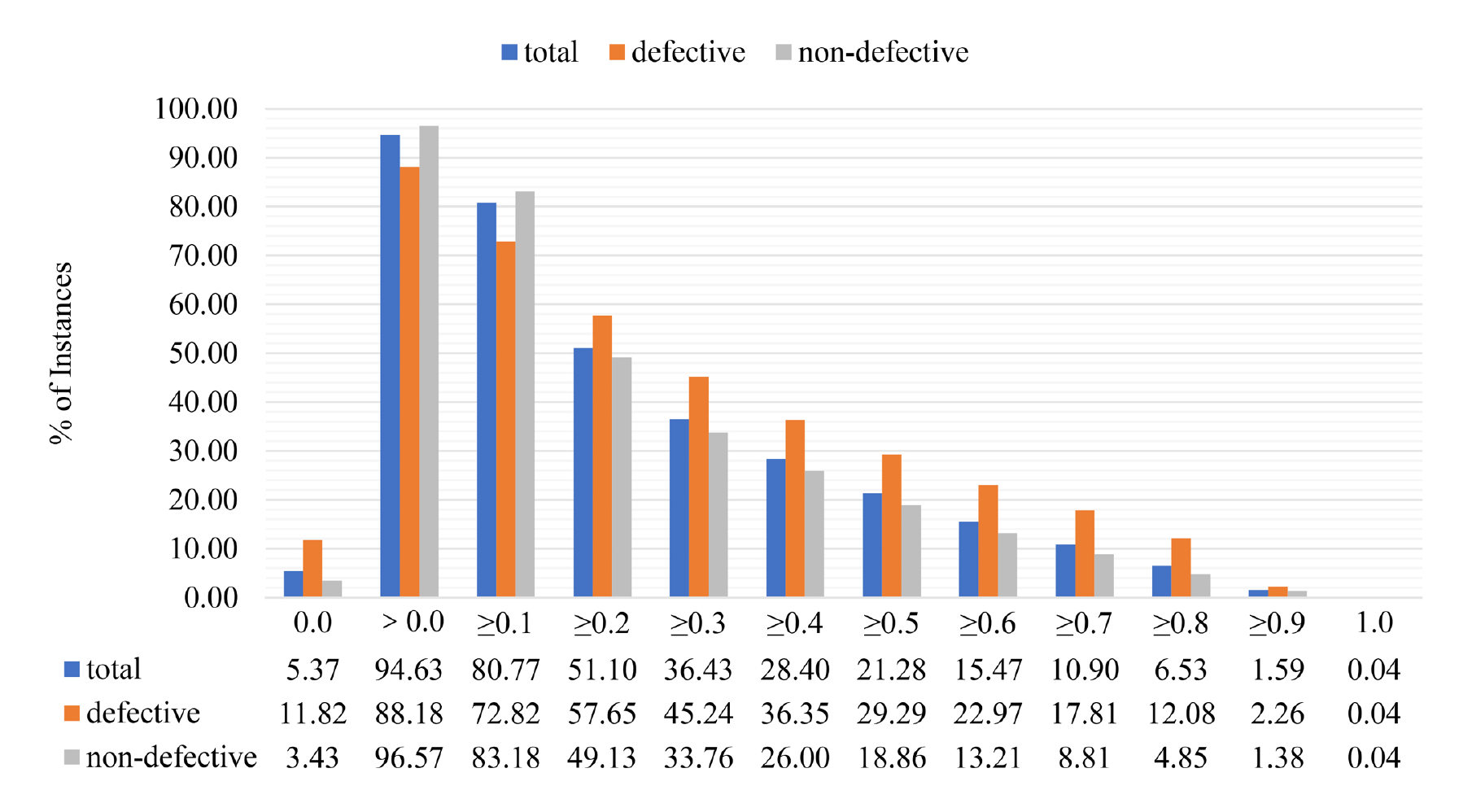}
\caption{The cumulative percentage of instances distributed in different hardness value intervals}
\label{Fig4}
\end{figure}

Fig. \ref{Fig4} depicts a cumulative frequency histogram of instances distributed across different hardness value intervals. All instances are grouped into intervals of 0.1 based on their estimated hardness values, which range from 0 to 1. The horizontal axis in Fig. \ref{Fig4} denotes the value interval, while the vertical axis represents the cumulative percentage of instances falling within each respective interval. The first column of the histogram presents the percentage of instances with a hardness value of 0, representing correct prediction by all models across multiple trials. The last column indicates the percentage of instances with a hardness value of 1, indicating incorrect prediction by any model throughout all trials. To explore differences in the distribution of instance hardness values between defective and non-defective classes, we calculate the cumulative percentage of the two classes separately. As illustrated in Fig. \ref{Fig4}, the blue section presents the cumulative distribution of hardness values in all instances, while the orange and gray sections represent the cumulative distribution of hardness values in two classes, respectively. Moreover, a table below the figure showcases the cumulative percentages of instances distributed across various intervals.

As illustrated in Fig. \ref{Fig4}, the distribution of instance hardness values is right-skewed, characterized by a pattern of ``low at both ends, high in the middle, and higher on the left than the right''. Instances that are either exceptionally easy or hard to classify are relatively rare, with hardness values of 0 and 1 constituting 5.37\% and 0.04\% of all instances, respectively. The majority of instances have hardness values within the range of $(0, 0.4)$, constituting 66.23\% ($94.63-28.40=66.23\%$) of the total. Furthermore, the interval with the largest proportion falls between $[0.1, 0.2)$, comprising 29.67\% ($80.77-51.10=29.67\%$) of all instances. The proportion of instances within subsequent intervals declines sharply, with those in $[0.5, 1]$ making up only 21.28\% of the total.

\begin{center}
\begin{tcolorbox}
\textbf{Finding 1}: Regardless of the defective or non-defective class, instance hardness across all datasets exhibits a pronounced right-skewed distribution, characterized by a pattern of low ends, a high middle, and a higher left than right.
\end{tcolorbox}
\end{center}

In addition, examining the distribution of both classes separately reveals disparities between defective and non-defective instances. Non-defective instances exhibit a similar trend to the overall distribution, with the majority falling within $(0, 0.3)$, making up 62.81\% ($96.57-33.76=62.81\%$) of the total. In addition, as hardness values increase, the number of non-defective instances decreases significantly, particularly outside these intervals. In contrast, the distribution of hardness values for defective instances is more dispersed, with the largest proportion of instances in $(0, 0.1)$, constituting 15.36\% ($88.18-72.82=15.36\%$) of all defective instances, followed by $[0.1, 0.2)$ with a proportion of 15.17\% ($72.82-57.65=15.17\%$). Interestingly, the proportion of defective instances that are extremely easy to classify (with a hardness value of 0) reaches 11.82\%, which is significantly higher than both the overall distribution and the non-defective class. Nevertheless, this does not necessarily mean that defective instances are more easily classified, as they are evenly distributed across multiple intervals, implying a relatively large proportion of defective instances with high hardness values. While some defective instances can be easily classified by all models, others remain difficult to classify correctly. For example, the proportion of defective instances with hardness values exceeding 0.6 and 0.7 is nearly twice that of non-defective instances.

\begin{center}
\begin{tcolorbox}
\textbf{Finding 2}: Overall, the defective class has a more scattered distribution of instance hardness, with a smaller proportion of instances having low hardness values and a larger proportion of instances having high hardness values compared to the non-defective class.
\end{tcolorbox}
\end{center}

\begin{figure*}[htbp]
\centering    
\begin{minipage}{\linewidth}
\centering
\includegraphics[width=0.92\textwidth]{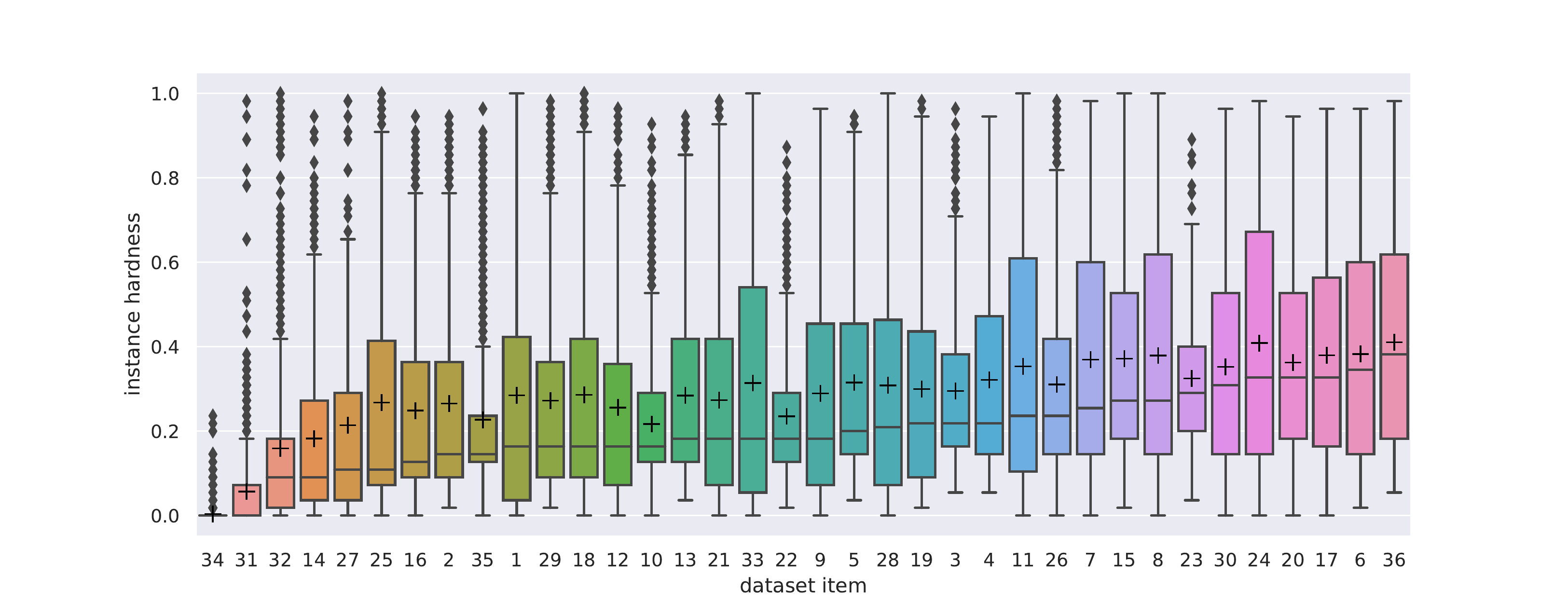}
\vspace{-5pt}
\caption{Distribution of instance hardness values in the whole dataset}
\label{Fig5a}
\end{minipage}\hfill
\begin{minipage}{\linewidth}
\centering
\includegraphics[width=0.92\textwidth]{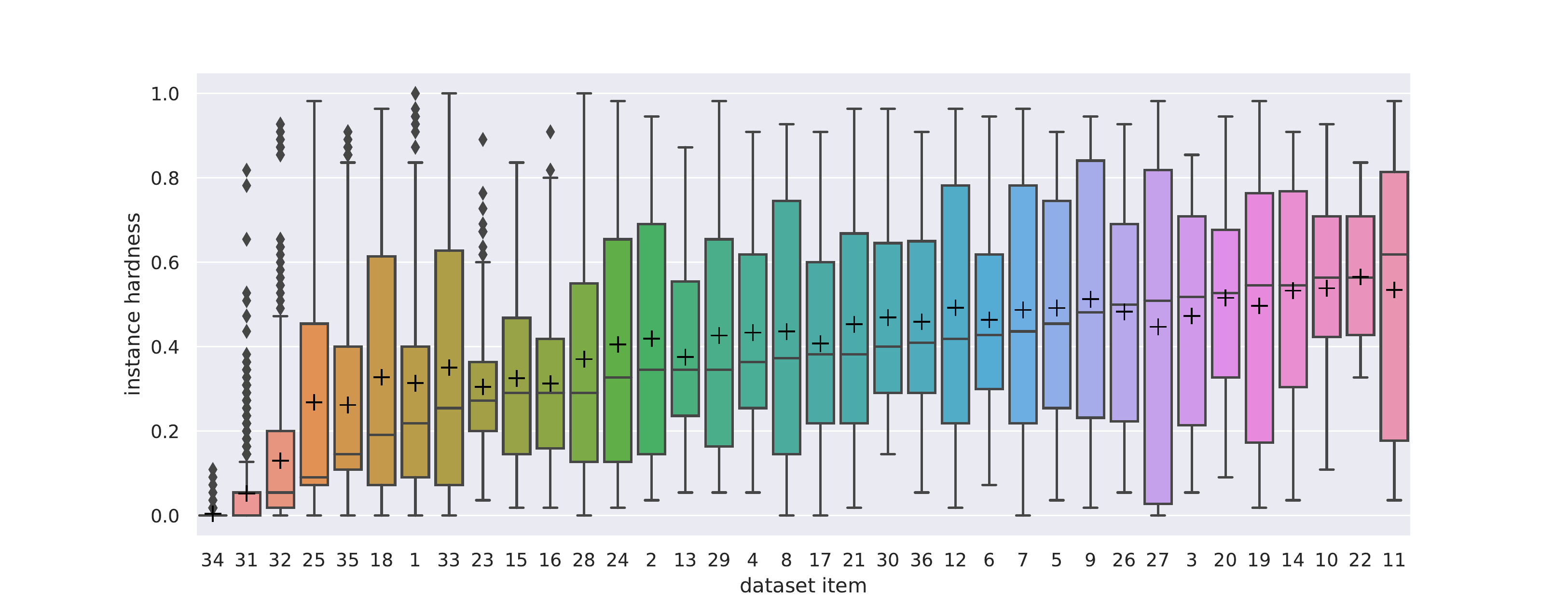}
\vspace{-5pt}
\caption{Distribution of instance hardness values in the defective class}
\label{Fig5b}
\end{minipage}\hfill
\begin{minipage}{\linewidth}
\centering
\vspace{0.1cm} 
\includegraphics[width=0.92\textwidth]{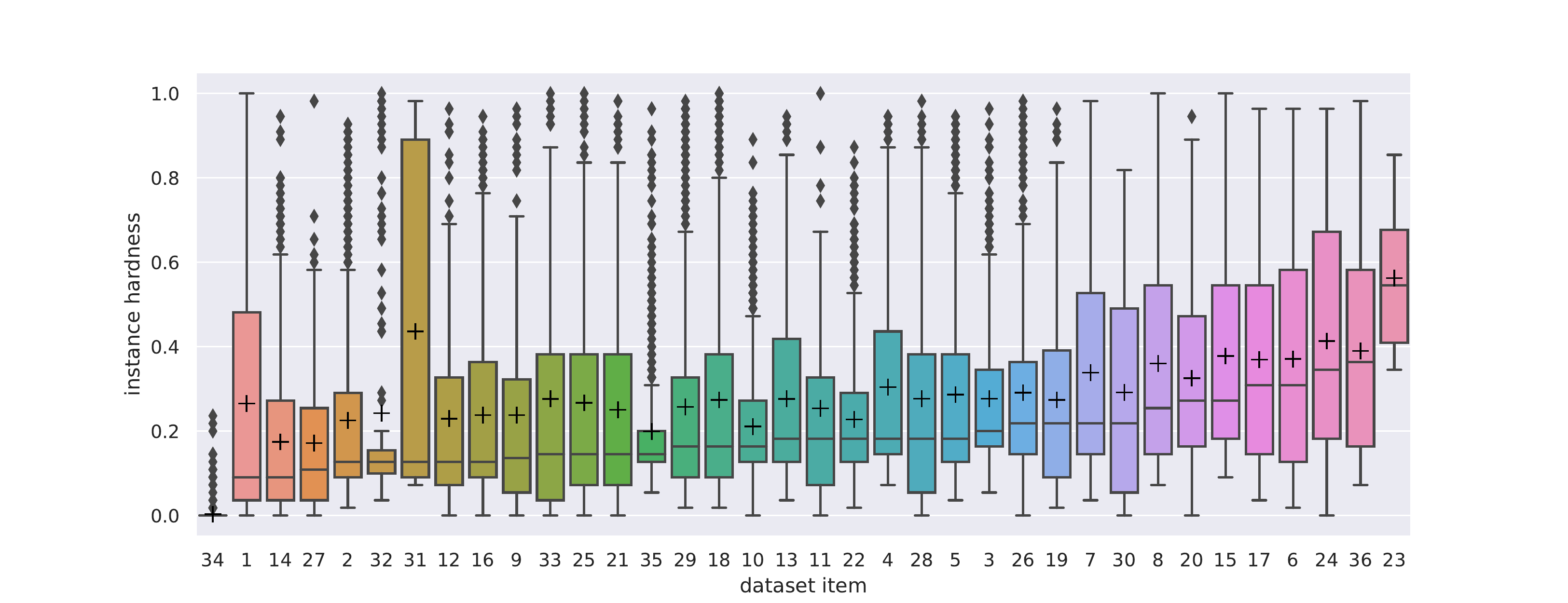}
\vspace{-5pt}
\caption{Distribution of instance hardness values in the non-defective class}
\label{Fig5c}
\end{minipage}
\end{figure*}

Fig. \ref{Fig5a} shows the instance hardness distribution for each dataset using boxplots. The horizontal axis denotes the dataset number (with corresponding dataset names found in Section \ref{sect:datasets}), while the vertical axis indicates the instance hardness value. Each boxplot consists of five horizontal lines representing five non-outlier sample statistics: minimum, 25th, 50th (median), 75th percentiles, and maximum. All datasets are arranged from left to right in ascending order based on the median of instance hardness values. Additionally, the mean of instance hardness values is marked with a ``+'' symbol within each boxplot. Considering the prevalent issue of class imbalance in defect prediction, separate boxplots have been provided for the distribution of instance hardness values within the defective (Fig. \ref{Fig5b}) and non-defective (Fig. \ref{Fig5c}) classes. As depicted in Fig. \ref{Fig5a}, \ref{Fig5b}, and \ref{Fig5c}, nearly all datasets exhibit a right-skewed distribution of instance hardness values, indicating a higher frequency of lower hardness values and fewer instances with higher hardness values. This finding supports the analysis results in Fig. \ref{Fig4} and holds true for the entire dataset as well as the defective and non-defective classes individually. However, several datasets, including redaktor (27), LC (3), arc (19), and MC2 (11), reveal a subtle left-skewed distribution of hardness values for defective instances, implying a marginally higher frequency of elevated hardness values, which in turn results in poor predictive accuracy for ML models on these datasets. Furthermore, as demonstrated in Fig. \ref{Fig5a}, Eclipse34-debug (34), xalan-2.7 (31), xerces-1.4 (32), xerces-1.4 (14), and redaktor (27) exhibit the lowest mean instance hardness values, indicating relatively low classification difficulty for these datasets. In other words, most models can attain relatively high average prediction accuracy on these datasets. Conversely, lucene-2.4 (24), camel-1.6 (20), PC5 (17), CM1 (6), and Zxing (36) possess the highest instance hardness values, indicating that achieving satisfactory predictive performance on these datasets is challenging for most models. Therefore, it can be concluded that the distribution of instance hardness exhibits notable variation across diverse datasets and classes.


\begin{center}
\begin{tcolorbox}
\textbf{Finding 3}: The distribution of instance hardness exhibits significant variation among different datasets and classes. This is illustrated by the fact that the majority of datasets exhibit a right-skewed distribution of instance hardness for both classes, while a few datasets display a slightly left-skewed distribution of instance hardness for the defective class.

%


\end{tcolorbox}
\end{center}



\vspace{5pt} 
\begin{tcolorbox}[title = {Summary of answers to RQ1 and their implications}:]
Although the distribution of instance hardness usually exhibits a right-skewed tendency, it exhibits considerable variation when examining specific classes or datasets. Recognizing these differences is crucial when examining data complexity in a defect prediction task.

\end{tcolorbox}

\subsection{RQ2: What is the manifestation of data complexity in defect prediction when viewed from the instance-level perspective?}
\label{rq2}

To answer \textbf{RQ2}, we calculate hardness measures for instances across 36 datasets and analyze their correlations to identify overlaps and investigate the relationship between instance hardness and the measures. We also examine the differences in hardness measures between defective/non-defective and easy/hard instances, as well as across datasets. Specifically, 15 hardness measures are calculated for defective/non-defective instances, easy/hard instances, and different datasets.

\begin{figure}[htbp]
\centering
\includegraphics[width=4.3in,keepaspectratio]{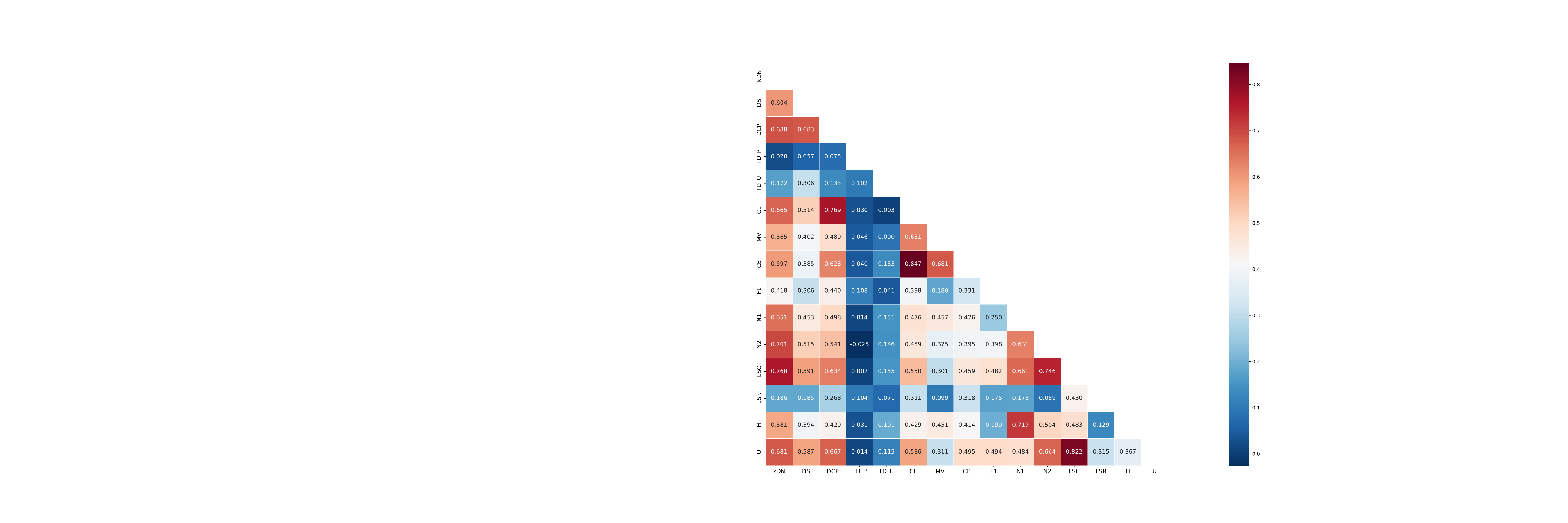}
\caption{Spearman correlation matrix for the instance hardness measures}
\label{Fig8}
\end{figure}

Fig. \ref{Fig8} presents a pairwise comparison of instance hardness measures using the Spearman correlation. CB and CL exhibit the strongest correlation with $r_{s}=0.847 > 0.8$, implying a strong relationship between them. CB measures the skewness of the instance's belonging class, while CL provides a global measure of overlap. The strong correlation suggests that the presence of class imbalance may worsen the instance overlap issue, and minority class instances are more likely to overlap. This is because minority classes typically have less distinct boundaries compared to majority classes, making accurate classification challenging. LSC and U measures also have a very strong correlation with $r_{s}=0.82 > 0.8$, as they both analyze the distribution and density of neighborhood sets to assess the difficulty of classifying an instance. Additionally, CL and DCP ($r_{s}=0.775$), kDN and LSC ($r_{s}=0.768$), LSC and N2 ($r_{s}=0.746$), kDN and N2 ($r_{s}=0.701$), and H and N2 ($r_{s}=0.719$) exhibit strong correlations with each other and capture different aspects of class overlap, such as feature overlap, structural overlap, instance overlap, and multiresolution overlap \cite{santos2022joint}. One notable exception is TD, where both pruned (TD\_P) and unpruned (TD\_U) versions of the measure have extremely low or even negligible correlation with other measures ($r_{s}<0.2$). TD estimates the difficulty of classifying an instance by estimating the description length required for classification, indicating that class overlap and class imbalance may not directly affect the description length. In addition, most measures exhibit either strong or weak positive correlations with each other, but no extreme correlations are observed ($r_{s}>0.9$). This suggests that the considered measures can measure the hardness of an instance based on different properties without redundancy.

\begin{center}
\begin{tcolorbox}
\textbf{Finding 4}: The 15 instance hardness measures in this paper can effectively measure the hardness of an instance , taking into account distinct properties without redundancy.
\end{tcolorbox}
\end{center}


After conducting the aforementioned analysis, a more intriguing question arises: how does instance hardness relate to the considered instance hardness measures? Table \ref{table7} displays the Spearman correlation coefficients between instance hardness and these measures for both defective and non-defective classes. To thoroughly analyze this correlation, we examine instance hardness estimated by all classifiers (referred to as IH) and by individual classifiers (denoted by the name of each classifier). The last column of Table \ref{table7}, labeled ``Lin'', presents correlation coefficients for a linear regression model predicting instance hardness using the considered measures. The strongest correlated measure in each row (excluding the combined measure in the last column) is underlined, and the top three measures with the highest correlation coefficients are highlighted in bold. The penultimate row of the table gives the average correlation coefficient for a total of 12 cases, while the last row gives the frequency of each measure ranking first or in the top three for correlation strength per row. Furthermore, the first section of the table displays the Spearman correlation coefficients for instance hardness measures in the defective class, while the second section presents the results for the non-defective class. A significance test is performed on each correlation coefficient, and those failing to pass the test with a $p$-value above 0.05 are marked in gray.

As indicated in Table \ref{table7}, DCP and kDN have the highest correlation with instance hardness for the defective class, with DCP being the strongest in 8 of 12 cases and kDN in 3 cases, while kDN has the strongest correlation in 3 cases. Both consistently rank among the top three measures in almost all cases. The situation is slightly different when it comes to the non-defective class. While DCP and kDN are still regarded as the measures with the strongest correlation with instance hardness, their complete dominance has been somewhat diminished, with DCP being strongest in 6 cases and kDN in 3. The ranking of these two measures is primarily challenged by DS and LSC, with DS being in the top three in 6 cases and LSC in 5 cases. Notably, the correlation between some measures and instance hardness can vary significantly between classes. For example, MV and CB show a moderate correlation in the defective class but are nearly negligible in the non-defective class. TD\_P and TD\_U may not accurately reflect classification difficulty due to weak correlations with instance hardness ($r_{s}<0.2$) for both classes in most cases. In addition, the last column labeled ``Lin'' in Table \ref{table7} suggests a linear combination of all measures correlates better with instance hardness than any individual measure, which illustrates that identifying hard instances in defect prediction datasets can be challenging, as the hardness may arise from multiple factors.

When comparing instance hardness measures, DCP and kDN exhibit the strongest correlation across all datasets, regardless of whether all representative learning algorithms or a specific one is considered. These results suggest that class overlap may be the primary factor contributing to instance hardness in defect prediction. For defective instances, only TD\_P, TD\_U, MV, CB, and F1 exhibit weak correlations with instance hardness on average ($r_{s} < 0.4$), and other measures have moderate or strong correlation with instance hardness on average. This suggests that diverse factors contribute to the difficulty of classifying defective instances. In contrast, the factors leading to difficulties in classifying non-defective instances are more concentrated. For example, DCP exhibits only a weak correlation with instance hardness in terms of the mean correlation coefficient, while TD\_P, TD\_U, MV, CB, and LSR demonstrate almost no correlation, and other measures show only weak correlation. The scarcity of defective instances results in a concentration of data complexities such as multiple concepts, overlapping, and small disjuncts within these instances. These issues are intricate, interrelated, and mutually influential, often manifesting simultaneously and complicating instance classification. This explains the strong correlation between many measures and instance hardness. In contrast, non-defective instances are more abundant, and data complexities are not concentrated, resulting in a weaker correlation between each measure and instance hardness. However, no single instance hardness measure can fully capture the data complexity.

\begin{center}
\begin{tcolorbox}
\textbf{Finding 5}: Instance hardness may arise from various factors, however, a single instance hardness measure can only characterize the data complexity from a particular perspective, rendering it inadequate to fully encompass the data complexity as a whole.
\end{tcolorbox}
\end{center}

\begin{figure*}[htbp]
\centering
\begin{minipage}{0.48\linewidth}
\centering
\includegraphics[width=\textwidth]{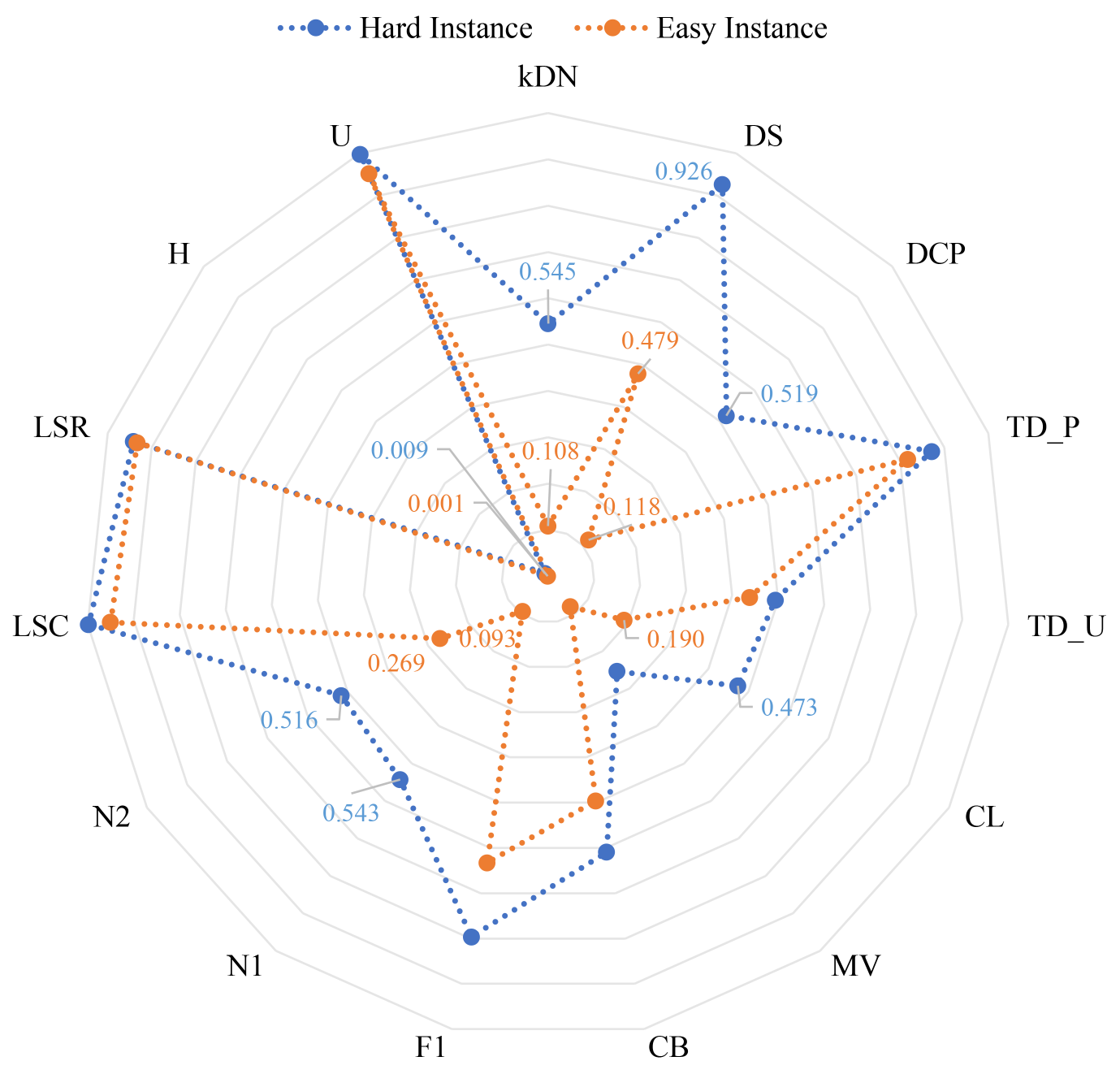}
\label{Fig9a}
\end{minipage}
\begin{minipage}{0.48\linewidth}
\centering
\includegraphics[width=\textwidth]{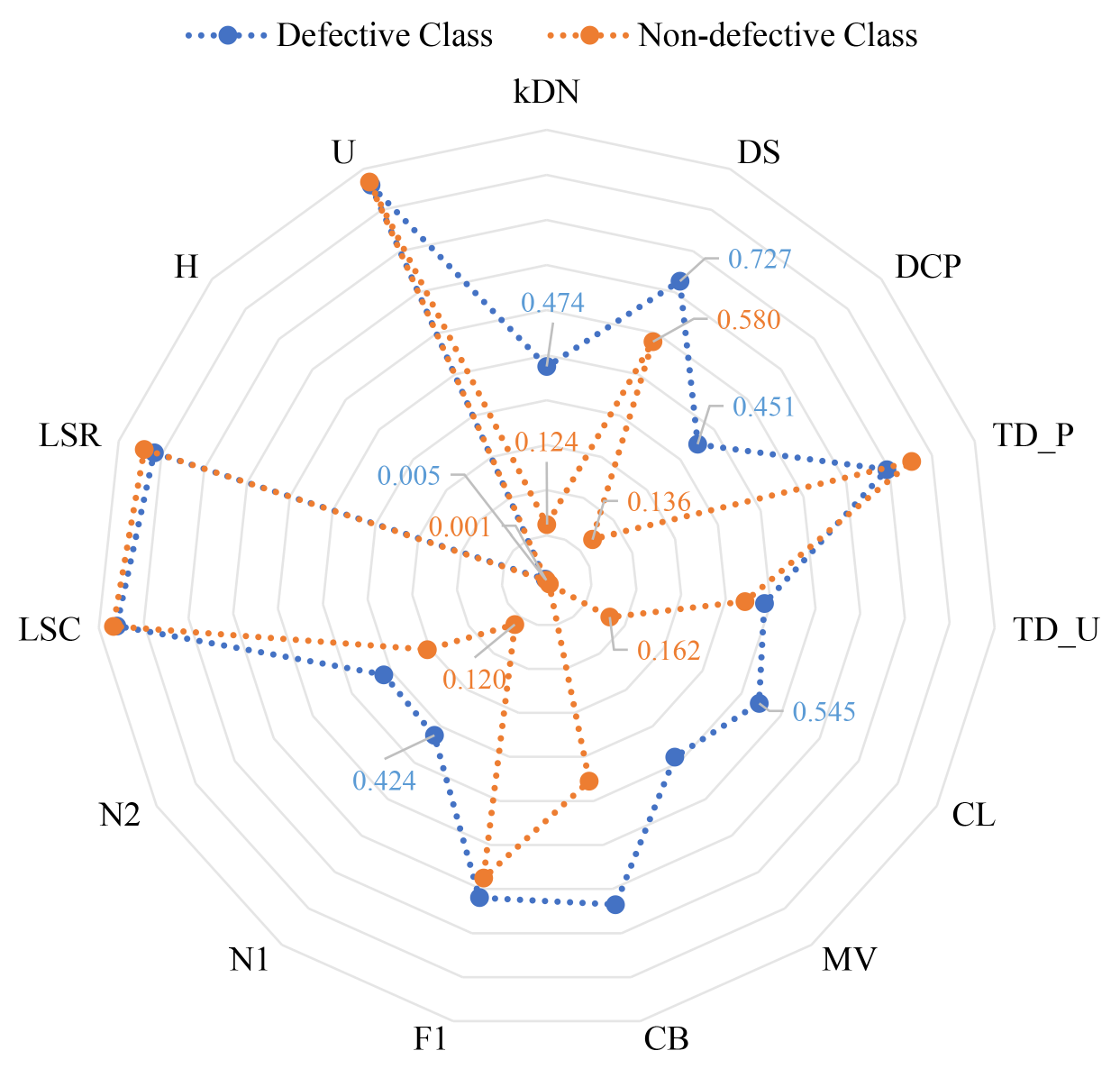}
\label{Fig9b}
\end{minipage}
\vspace{-15pt}
\caption{The average of instance hardness measures of defective/non-defective and hard/easy instances}
\label{Fig9}
\end{figure*}

Furthermore, we conduct a comprehensive analysis of hardness measures for defective/non-defective instances and easy/hard instances. Here easy instances refer to those with an instance hardness value lower than 0.3, while hard instances refer to those with an instance hardness value higher than 0.7. Fig. \ref{Fig9} presents a radar chart that illustrates the average of all instance hardness measures and some representative measures are highlighted to facilitate further analysis.

\begin{landscape}
\begin{table*}[htbp]
\centering
\caption{Spearman correlation between instance hardness measures and instance hardness for defective/non-defective instances across different classifiers}
\label{table7}
\begin{threeparttable}
\footnotesize
\renewcommand\arraystretch{1.05}
\setlength{\tabcolsep}{2.2mm}{
\begin{longtable}{ccccccccccccccccc}
\toprule
\multicolumn{1}{c}{Defective Class} & kDN   & DS    & DCP   & TD\_P & TD\_U  & CL    & MV    & CB    & F1   & N1    & N2    & LSC   & LSR   & H     & U     & Lin \\
\midrule
IH                & \textbf{0.736}       & 0.663          & \textbf{\uline{0.768}} & 0.252 & 0.168                        & \textbf{0.695}       & 0.422 & 0.451          & 0.162  & 0.599 & 0.584          & 0.637          & 0.575          & 0.540 & 0.635          & 0.854 \\
KNN               & \textbf{\uline{0.664}} & 0.563          & \textbf{0.627}       & 0.180 & 0.187                          & 0.554                & 0.367 & 0.388          & 0.130  & 0.562 & \textbf{0.590} & \textbf{0.590} & 0.423          & 0.506 & 0.573          & 0.731 \\
NB                & 0.440       & 0.390          & \textbf{0.510}       & 0.307 & 0.080                          & \textbf{0.567} & 0.183 & 0.212          & 0.200  & 0.362 & 0.295          & 0.388          & \textbf{\uline{0.699}}          & 0.315 & 0.380          & 0.772 \\
CART              & \textbf{\uline{0.734}} & 0.648          & \textbf{0.713}       & 0.230 & 0.180                          & 0.647                & 0.512 & 0.533          & 0.252  & 0.607 & 0.650          & \textbf{0.657} & 0.408          & 0.531 & 0.654          & 0.783 \\
LR                & \textbf{0.513}       & 0.474          & \textbf{\uline{0.551}} & 0.063 & 0.242                          & 0.464                & 0.176 & 0.185          & -0.042 & 0.401 & 0.344          & 0.405          & \textbf{0.493} & 0.410 & 0.407          & 0.737 \\
SVM               & \textbf{0.491}       & 0.454          & \textbf{\uline{0.521}} & 0.257 & \cellcolor[gray]{0.9}-0.009 & 0.468                & 0.391 & 0.412          & 0.278  & 0.389 & 0.410          & 0.466          & 0.356          & 0.296 & \textbf{0.484} & 0.554 \\
MLP               & \textbf{0.558}       & \textbf{0.479} & \textbf{\uline{0.582}} & 0.063 & 0.198                          & 0.460                & 0.210 & 0.231          & -0.035 & 0.437 & 0.407          & 0.430          & 0.423          & 0.445 & 0.437          & 0.704 \\
Boosted\_RS       & \textbf{0.661}       & \textbf{0.568} & \textbf{\uline{0.675}} & 0.101 & 0.195                          & 0.566                & 0.374 & 0.397          & 0.055  & 0.526 & 0.543          & 0.530          & 0.365          & 0.523 & 0.531          & 0.731 \\
Greedy\_RL        & \textbf{0.439}       & 0.371          & \textbf{\uline{0.492}} & 0.028 & 0.141                          & 0.339                & 0.105 & 0.117          & -0.069 & 0.340 & 0.309          & 0.328          & \textbf{0.382} & 0.320 & 0.325          & 0.641 \\
RF                & \textbf{\uline{0.724}} & 0.624          & \textbf{0.693}       & 0.203 & 0.196                          & \textbf{0.635}       & 0.483 & 0.498          & 0.185  & 0.603 & 0.632          & 0.629          & 0.419          & 0.530 & 0.631          & 0.770 \\
SVMBoosting       & \textbf{0.292}       & 0.287          & \textbf{\uline{0.314}} & 0.115 & \cellcolor[gray]{0.9}-0.020 & 0.290                & 0.277 & \textbf{0.308} & 0.028  & 0.226 & 0.258          & 0.270          & 0.034          & 0.208 & 0.285          & 0.460 \\
MLPBoosting       & \textbf{0.611}       & \textbf{0.535} & \textbf{\uline{0.631}} & 0.137 & 0.148                          & 0.524                & 0.340 & 0.361          & 0.023  & 0.491 &
0.485          & 0.484          & 0.375          & 0.476 & 0.487          & 0.689 \\
\midrule
Avg.             & \textbf{0.572}  & 0.505   & \textbf{\uline{0.590}}   & 0.161 &	0.142 &	\textbf{0.517} &	0.320 &	0.341 &	0.097 &	0.462 &	0.459 &	0.485 &	0.413 &	0.425 &	0.486 &	0.702 \\
Top-Count         & 3/11                 & 0/3            & 8/12                 & 0/0   & 0/0  & 0/3                  & 0/0   & 0/1            & 0/0    & 0/0   & 0/1            & 0/2            & 1/3            & 0/0   & 0/1            & --- \\
\bottomrule \\
\toprule
\multicolumn{1}{c}{Non-defective Class} & kDN   & DS    & DCP  & TD\_P  & TD\_U  & CL    & MV     & CB & F1  & N1    & N2   & LSC & LSR & H     & U    & Lin \\
\midrule
IH                    & \textbf{0.530}       & \textbf{0.541}       & \textbf{\uline{0.578}} & 0.077                          & 0.078                         & 0.333 & -0.014                        & 0.032           & 0.450                         & 0.333 & 0.514          & 0.512          & -0.092                        & 0.274 & 0.507          & 0.762 \\
KNN                   & \textbf{\uline{0.516}} & 0.418                & 0.474                & -0.025                         & 0.101                         & 0.317 & 0.022                         & 0.143           & 0.361                         & 0.351 & \textbf{0.514} & \textbf{0.482} & -0.043                        & 0.295 & 0.463          & 0.620 \\
NB                    & 0.132                & \textbf{0.170}       & 0.087                & 0.109                          & \cellcolor[gray]{0.9}0.000 & 0.022 & \cellcolor[gray]{0.9}0.002 & \textbf{-0.255} & 0.145                         & 0.092 & 0.135          & 0.022          & \textbf{\uline{-0.353}}         & 0.104 & 0.023          & 0.533 \\
CART                  & \textbf{\uline{0.519}} & 0.433                & \textbf{0.514}       & -0.021                         & 0.116                         & 0.367 & 0.048                         & 0.221           & 0.383                         & 0.340 & 0.506          & \textbf{0.509} & 0.023                         & 0.285 & \textbf{0.509} & 0.626 \\
LR                    & \textbf{0.389}       & \textbf{0.379}       & \textbf{\uline{0.390}} & 0.086                          & 0.063                         & 0.191 & \cellcolor[gray]{0.9}0.011 & -0.059          & 0.318                         & 0.240 & 0.330          & 0.292          & -0.295                        & 0.217 & 0.325          & 0.608 \\
SVM                   & 0.100                & \textbf{\uline{0.222}} & \textbf{0.185}       & \cellcolor[gray]{0.9}-0.002 & 0.050                         & 0.044 & -0.036                        & -0.104          & \cellcolor[gray]{0.9}0.011 & 0.078 & 0.154          & \textbf{0.159} & \cellcolor[gray]{0.9}0.000 & 0.066 & 0.139          & 0.335 \\
MLP                   & \textbf{0.465}       & \textbf{0.431}       & \textbf{\uline{0.480}} & 0.086                          & 0.092                         & 0.241 & \cellcolor[gray]{0.9}0.007 & 0.018           & 0.362                         & 0.287 & 0.414          & 0.376          & -0.231                        & 0.253 & 0.395          & 0.646 \\
Boosted\_RS           & \textbf{0.538}       & 0.465                & \textbf{\uline{0.577}} & 0.056                          & 0.089                         & 0.366 & 0.051                         & 0.180           & 0.400                         & 0.346 & \textbf{0.480} & 0.453          & -0.124                        & 0.301 & 0.460          & 0.661 \\
Greedy\_RL            & \textbf{0.377}       & \textbf{0.419}       & \textbf{\uline{0.444}} & 0.121                          & 0.080                         & 0.164 & -0.016                        & -0.066          & 0.359                         & 0.228 & 0.358          & 0.285          & -0.277                        & 0.200 & 0.294          & 0.676 \\
RF                    & \textbf{\uline{0.499}} & 0.402                & \textbf{0.481}       & \cellcolor[gray]{0.9}0.008  & 0.106                         & 0.353 & 0.061                         & 0.213           & 0.350                         & 0.354 & 0.465          & \textbf{0.470} & 0.046                         & 0.307 & 0.456          & 0.573 \\
SVMBoosting           & 0.197                & 0.220                & 0.276                & 0.041                          & -0.086                        & 0.263 & -0.059                        & 0.095           & \textbf{0.283}                & 0.102 & 0.165          & \textbf{0.309} & 0.275                         & 0.056 & \textbf{0.284} & 0.401 \\
MLPBoosting           & \textbf{0.514}       & 0.427                & \textbf{\uline{0.518}} & 0.075                          & 0.098                         & 0.307 & 0.036                         & 0.122           & 0.360                         & 0.336 & \textbf{0.449} & 0.426          & -0.130                        & 0.298 & 0.425          & 0.614 \\
\midrule
Avg.             & \textbf{0.398} &	\textbf{0.377} & \textbf{\uline{0.417}} &	0.051 &	0.066 &	0.247 &	0.009 &	0.045 &	0.315 &	0.257 &	0.374 &	0.358 &	-0.100 & 0.221 & 0.357 & 0.588 \\
Top-Count        & 3/9                  & 1/6                  & 6/9                  & 0/0                            & 0/0                           & 0/0   & 0/0                           & 0/1             & 0/1                           & 0/0   & 0/3            & 0/5            & 1/1                           & 0/0   & 0/2            & --- \\
\bottomrule
\end{longtable}
}
\end{threeparttable}
\end{table*}
\end{landscape}

The left chart highlights notable differences in the measure values between hard and easy instances. In particular, hard instances exhibit higher mean values for all measures, with some key measures (strongly correlated with instance hardness) showing a 2-fold or even multiple-fold difference. For example, kDN is about 5 times higher ($0.545/0.108 \approx 5.0$), DS is nearly 1.9 times higher ($0.926/0.479 \approx 1.9$), and DCP is around 4.4 times higher ($0.519/0.118 \approx 4.4$). This suggests that these measures can effectively measure the hardness level for accurate classification. The right chart presents the average of hardness measures for defective/non-defective instances. Similarly, most measures for defective instances exhibit higher average values than those for non-defective instances. For example, the average kDN for defective instances is about 3.8 times higher than that for non-defective instances ($0.473/0.124 \approx 3.8$), DCP is around 3.3 times higher ($0.451/0.136 \approx 3.3$), and CL is nearly 3.3 times higher ($0.545/0.162 \approx 3.3$). In comparison to non-defective instances, defective instances demonstrate greater data complexity across more measures, indicating that there are more factors contributing to the hardness of defective instances.

\begin{center}
\begin{tcolorbox}
\textbf{Finding 6}: The defective class  spans a more extensive range of data complexities, with numerous factors influencing instance hardness. This is evidenced by the tendency of hardness measures for defective instances to generally exhibit higher average values and stronger correlations with instance hardness when compared to their non-defective counterparts.
\end{tcolorbox}
\end{center}

Table \ref{table8} presents the correlation between instance hardness and hardness measures for instances in each dataset. The column labeled ``DSH'' denotes the dataset hardness as defined in Section \ref{sect:ih}. The most highly correlated measure in each row is underlined, while the top three measures with the highest correlation coefficients are highlighted in bold. The penultimate row of the table provides the average of correlation coefficients across 36 datasets, and the final row gives the frequency of each measure being ranked in the first or top three in terms of correlation strength. Moreover, those correlation coefficients with a $p$-value greater than 0.05 are shaded in gray. As shown in Table \ref{table8}, on average, LSC, kDN, and U exhibit the highest correlation with instance hardness, with $r_s$ values exceeding 0.56. Additionally, CL, DS, and DCP also demonstrate relatively high correlations with instance hardness, with $r_s$ values greater than 0.50. These measures belong to the class overlap measures, which include feature overlap, structural overlap, instance overlap, and multiresolution overlap. These measures of class overlap are highly correlated with each other. In contrast, measures of class imbalance (MV and CB) tend to display weak correlation with instance hardness, with $r_s$ values less than 0.20. This suggests that class imbalance may not be a significant factor in instance classification challenges. Instead, the primary difficulty in defect prediction tasks is attributed to the issue of class overlap, however, the presence of class imbalance can, to some extent, exacerbate the challenge of correctly classifying overlapping instances.

\begin{center}
\begin{tcolorbox}
\textbf{Finding 7}: Class imbalance may not be the primary cause of instance hardness; instead, class overlap could be the principal contributor. This is supported by the observation that MV and CB exhibit little to no correlation with instance hardness, whereas DCP and kDN exhibit the strongest correlation with instance hardness.
\end{tcolorbox}
\end{center}

In addition, we observe that in Table \ref{table7}, the two instance hardness measures, kDN and DCP, exhibit the strongest correlation with instance hardness, far ahead of the other measures. However, in Table \ref{table8}, while these two measures still demonstrate a strong correlation with DSH, other measures such as U, LSC, CL, and DS also show comparable or even stronger correlation with the former two measures. After a thorough analysis, we have discovered that the primary reason for this disparity is that the correlation analysis in Table \ref{table7} is conducted on the entire population of all instances, which may introduce a bias towards larger datasets. For example, JM1 (7) is the largest dataset with approximately a quarter of the total instances, and Table \ref{table8} reveals that kDN and DCP are the two measures with the strongest correlation in this dataset. Therefore, this implies that conducting a correlation analysis on the entire population of instances alone is inadequate, and it is essential to conduct separate correlation analyses for each dataset in order to obtain more generalizable conclusions regarding the extent and sources of data complexity across diverse datasets.

\begin{center}
\begin{tcolorbox}
\textbf{Finding 8}: LSC, kDN, U, CL, DS, and DCP all exhibit a relatively strong correlation with instance hardness, suggesting that various groups of class overlap measures -- encompassing feature overlap, structural overlap, instance overlap, and multiresolution overlap -- can be employed to thoroughly characterize overlapping instances.
\end{tcolorbox}
\end{center}

\vspace{5pt} 

\begin{tcolorbox}[title = {Summary of answers to RQ2 and their implications}:]
Although numerous factors can contribute to instance hardness (particularly for the defective class), class overlap, rather than class imbalance, emerges as the most significant factor among all data complexities. In order to comprehensively characterize overlapping instances, we can employ different groups of class overlap measures, including feature overlap, structural overlap, instance overlap, and multiresolution overlap.

\end{tcolorbox}

\subsection{RQ3: How does dataset complexity manifest in defect prediction and what is the effect of data preprocessing methods on it?}
\label{rq3}

To answer \textbf{RQ3}, we calculate dataset complexity measures and examine their correlations to identify potential overlap and explore the relationship between dataset complexity and the considered measures. Moreover, we calculate these measures for datasets before and after some data preprocessing methods, in order to analyze the impact of data preprocessing methods on data complexity.
\setcounter{table}{7}

\begin{landscape}
\begin{table*}[htbp]
\centering
\caption{Spearman correlation between instance hardness measures and dataset hardness across multiple datasets}
\label{table8}
\begin{threeparttable}
\footnotesize
\renewcommand\arraystretch{0.9}
\setlength{\tabcolsep}{2.3mm}{
\begin{longtable}{ccccccccccccccccc}
\toprule
\multicolumn{1}{c}{Item} & DSH & kDN   & DS    & DCP   & TD\_P  & TD\_U  & CL     & MV     & CB     & F1     & N1    & N2     & LSC   & LSR    & H     & U     \\
\midrule
1         & 0.285 & \textbf{\uline{0.745}} & \textbf{0.711}       & 0.512                      & 0.578                      & 0.544                      & 0.559                & 0.173  & 0.173  & 0.534          & 0.508 & 0.589          & 0.671                & -0.346         & 0.493 & \textbf{0.720}       \\
2         & 0.266 & \textbf{0.639}       & 0.516                & \textbf{0.597}             & -0.419                     & -0.137                     & 0.592                & 0.311  & 0.311  & 0.453          & 0.492 & 0.589          & 0.590                & -0.320         & 0.493 & \textbf{\uline{0.640}} \\
3         & 0.295 & 0.460                & 0.507                & 0.544                      & \cellcolor[gray]{0.9}0.056                      & 0.487                      & 0.488                & 0.193  & 0.193  & \textbf{0.581} & 0.401 & \textbf{0.567} & 0.541                & -0.277         & 0.410 & \textbf{\uline{0.637}} \\
4         & 0.321 & 0.484                & \textbf{0.569}       & \textbf{\uline{0.741}}       & -0.386                     & 0.090                      & 0.566                & 0.211  & 0.211  & 0.332          & 0.396 & 0.524          & 0.506                & -0.245         & 0.366 & \textbf{0.587}       \\
5         & 0.315 & 0.519                & 0.481                & \textbf{0.606}             & 0.430                      & 0.130                      & \textbf{0.608}       & 0.266  & 0.266  & 0.536          & 0.410 & 0.533          & 0.540                & -0.448         & 0.374 & \textbf{\uline{0.634}} \\
6         & 0.383 & 0.663                & 0.572                & \textbf{0.681}             & \cellcolor[gray]{0.9}NA & 0.390                      & \textbf{\uline{0.724}} & 0.139  & 0.139  & 0.584          & 0.404 & 0.590          & \textbf{0.680}       & \cellcolor[gray]{0.9}-0.067         & 0.329 & 0.602                \\
7         & 0.370 & \textbf{0.526}       & 0.405                & \textbf{\uline{0.619}}       & 0.093                      & -0.026                     & 0.500                & 0.223  & 0.223  & 0.326          & 0.377 & 0.445          & 0.437                & -0.219         & 0.393 & \textbf{0.510}       \\
8         & 0.379 & \textbf{0.556}       & 0.470                & 0.090                      & \cellcolor[gray]{0.9}NA & 0.164                      & \textbf{\uline{0.568}} & 0.090  & 0.090  & 0.334          & 0.345 & 0.460          & 0.479                & -0.190         & 0.372 & \textbf{0.520}       \\
9         & 0.289 & \textbf{0.628}       & 0.626                & 0.578                      & -0.256                     & \cellcolor[gray]{0.9}0.044                      & \textbf{\uline{0.736}} & 0.369  & 0.369  & 0.385          & 0.454 & 0.573          & 0.564                & -0.311         & 0.406 & \textbf{0.631}       \\
10        & 0.217 & 0.359                & 0.469                & 0.408                      & \cellcolor[gray]{0.9}NA & 0.310                      & \textbf{\uline{0.639}} & 0.177  & 0.177  & 0.391          & 0.256 & 0.289          & \textbf{0.558}       & 0.174          & 0.235 & \textbf{0.550}       \\
11        & 0.354 & \textbf{\uline{0.744}} & 0.572                & 0.530                      & \cellcolor[gray]{0.9}NA & 0.310                      & \textbf{0.645}       & 0.402  & 0.402  & 0.416          & 0.590 & 0.499          & \textbf{0.655}       & \cellcolor[gray]{0.9}0.119          & 0.521 & 0.552                \\
12        & 0.256 & 0.641                & \textbf{\uline{0.719}} & \cellcolor[gray]{0.9}0.061                      & 0.271                      & 0.321                      & \textbf{0.706}       & 0.273  & 0.273  & 0.465          & 0.532 & 0.619          & \textbf{0.666}       & \cellcolor[gray]{0.9}-0.074         & 0.466 & 0.625                \\
13        & 0.285 & 0.583                & 0.684                & 0.591                      & \cellcolor[gray]{0.9}NA & \textbf{0.738}             & 0.641                & 0.130  & 0.130  & 0.616          & 0.351 & 0.520          & \textbf{0.737}       & 0.167          & 0.341 & \textbf{\uline{0.739}} \\
14        & 0.182 & 0.452                & \textbf{0.745}       & 0.728                      & \cellcolor[gray]{0.9}NA & \textbf{0.744}             & \textbf{\uline{0.879}} & 0.183  & 0.183  & 0.606          & 0.296 & 0.600          & 0.697                & 0.234          & 0.250 & 0.543                \\
15        & 0.371 & 0.435                & 0.511                & 0.395                      & \cellcolor[gray]{0.9}NA & 0.507                      & \textbf{0.527}       & -0.081 & -0.081 & 0.465          & 0.236 & 0.450          & \textbf{0.581}       & 0.208          & 0.206 & \textbf{\uline{0.593}} \\
16        & 0.249 & 0.580                & \textbf{0.697}       & 0.636                      & \textbf{\uline{0.771}}       & \textbf{0.715}             & 0.626                & 0.184  & 0.184  & 0.549          & 0.374 & 0.541          & 0.614                & \cellcolor[gray]{0.9}0.006          & 0.364 & 0.605                \\
17        & 0.380 & 0.537                & 0.540                & \textbf{\uline{0.615}}       & \textbf{0.588}             & 0.511                      & \textbf{0.591}       & 0.084  & 0.084  & 0.402          & 0.388 & 0.432          & 0.502                & -0.073         & 0.408 & 0.552                \\
18        & 0.286 & 0.505                & 0.449                & 0.531                      & -0.017                     & 0.210                      & \textbf{0.589}       & \cellcolor[gray]{0.9}0.006  & \cellcolor[gray]{0.9}0.006  & 0.368          & 0.372 & 0.517          & \textbf{0.563}       & \cellcolor[gray]{0.9}-0.057         & 0.435 & \textbf{\uline{0.591}} \\
19        & 0.300 & 0.603                & \textbf{0.607}       & 0.220                      & \cellcolor[gray]{0.9}NA & 0.535                      & 0.583                & 0.220  & 0.220  & 0.383          & 0.493 & \textbf{0.620} & \textbf{\uline{0.647}} & \cellcolor[gray]{0.9}0.101          & 0.410 & 0.529                \\
20        & 0.362 & \textbf{0.577}       & 0.509                & 0.323                      & \cellcolor[gray]{0.9}NA & 0.474                      & \textbf{\uline{0.612}} & 0.323  & 0.323  & 0.335          & 0.457 & \textbf{0.588} & 0.556                & -0.129         & 0.451 & 0.553                \\
21        & 0.273 & 0.687                & 0.692                & 0.572                      & \cellcolor[gray]{0.9}NA & 0.695                      & \textbf{\uline{0.787}} & 0.246  & 0.246  & 0.559          & 0.485 & \textbf{0.730} & 0.716                & -0.128         & 0.441 & \textbf{0.757}       \\
22        & 0.235 & 0.374                & \textbf{\uline{0.507}} & 0.384                      & -0.149                     & \textbf{0.463}             & -0.216               & 0.223  & 0.223  & \textbf{0.416} & 0.314 & 0.298          & 0.189                & -0.242         & 0.290 & 0.257                \\
23        & 0.325 & 0.412                & \textbf{\uline{0.636}} & \textbf{0.625}             & 0.468                      & \textbf{0.595}             & 0.551                & 0.345  & 0.345  & 0.185          & 0.358 & 0.202          & 0.327                & 0.242          & 0.394 & 0.191                \\
24        & 0.409 & \textbf{\uline{0.658}} & 0.361                & 0.397                      & \cellcolor[gray]{0.9}0.069                      & 0.353                      & 0.496                & \cellcolor[gray]{0.9}0.018  & \cellcolor[gray]{0.9}0.018  & \cellcolor[gray]{0.9}0.078          & 0.355 & 0.483          & \textbf{0.525}       & 0.249          & 0.434 & \textbf{0.547}       \\
25        & 0.268 & \textbf{\uline{0.723}} & 0.589                & 0.643                      & \cellcolor[gray]{0.9}NA & 0.276                      & 0.394                & \cellcolor[gray]{0.9}0.046  & \cellcolor[gray]{0.9}0.046  & -0.231         & 0.546 & 0.535          & \textbf{0.688}       & 0.142          & 0.530 & \textbf{0.650}       \\
26        & 0.310 & 0.534                & 0.509                & \textbf{0.557}             & -0.393                     & \cellcolor[gray]{0.9}0.070                      & \textbf{\uline{0.611}} & 0.204  & 0.204  & 0.463          & 0.401 & 0.492          & \textbf{0.549}       & 0.163          & 0.416 & 0.476                \\
27        & 0.214 & \textbf{0.641}       & 0.395                & 0.272                      & \cellcolor[gray]{0.9}NA & 0.363                      & 0.492                & 0.182  & 0.182  & 0.331          & 0.517 & 0.448          & \textbf{\uline{0.698}} & 0.268          & 0.469 & \textbf{0.675}       \\
28        & 0.308 & \textbf{\uline{0.716}} & 0.609                & 0.554                      & \cellcolor[gray]{0.9}NA & 0.237                      & \textbf{0.645}       & 0.186  & 0.186  & 0.248          & 0.484 & 0.583          & \textbf{0.647}       & \cellcolor[gray]{0.9}0.105          & 0.502 & 0.632                \\
29        & 0.272 & 0.603                & 0.455                & 0.599                      & \cellcolor[gray]{0.9}0.028                      & 0.366                      & \textbf{\uline{0.759}} & 0.195  & 0.195  & 0.475          & 0.392 & \textbf{0.701} & 0.671                & \cellcolor[gray]{0.9}0.001          & 0.359 & \textbf{0.687}       \\
30        & 0.352 & \textbf{\uline{0.739}} & \textbf{0.635}       & 0.313                      & 0.483                      & 0.407                      & \textbf{0.697}       & 0.358  & 0.358  & 0.324          & 0.546 & 0.605          & 0.626                & \cellcolor[gray]{0.9}-0.001         & 0.466 & 0.522                \\
31        & 0.057 & 0.316                & 0.533                & 0.351                      & 0.435                      & 0.462                      & \textbf{\uline{0.681}} & 0.166  & 0.166  & -0.082         & 0.263 & -0.142         & \textbf{0.590}       & \textbf{0.636} & 0.213 & 0.333                \\
32        & 0.159 & 0.588                & \textbf{0.707}       & \textbf{\uline{0.754}}       & 0.206                      & 0.293                      & \textbf{0.716}       & 0.334  & 0.334  & -0.493         & 0.489 & 0.188          & 0.566                & 0.632          & 0.460 & 0.548                \\
33        & 0.314 & \textbf{\uline{0.762}} & 0.555                & \textbf{0.663}             & -0.164                     & 0.358                      & -0.547               & -0.147 & -0.147 & 0.148          & 0.455 & 0.487          & 0.519                & \cellcolor[gray]{0.9}0.014          & 0.517 & \textbf{0.623}       \\
34        & 0.004 & \textbf{0.251}       & 0.165                & \cellcolor[gray]{0.9}NA & \cellcolor[gray]{0.9}NA & \cellcolor[gray]{0.9}NA & \textbf{\uline{0.337}} & 0.165  & 0.165  & 0.032          & 0.161 & 0.216          & 0.233                & -0.112         & 0.175 & \textbf{0.278}       \\
35        & 0.227 & 0.544                & \textbf{0.575}       & \textbf{\uline{0.650}}       & \cellcolor[gray]{0.9}NA & 0.255                      & 0.413                & -0.055 & -0.055 & 0.232          & 0.469 & 0.320          & 0.529                & \cellcolor[gray]{0.9}0.009          & 0.459 & \textbf{0.583}       \\
36        & 0.411 & \textbf{\uline{0.559}} & 0.445                & 0.500                      & \cellcolor[gray]{0.9}NA & 0.199                      & \textbf{0.553}       & 0.133  & 0.133  & 0.299          & 0.401 & 0.468          & 0.515                & -0.314         & 0.386 & \textbf{0.542}       \\
\midrule
Avg.       & 0.286 & \textbf{0.565}       & 0.548                & 0.510                      & 0.135                      & 0.356                      & 0.549                & 0.180  & 0.180  & 0.335          & 0.410 & 0.477          & \textbf{\uline{0.566}} & -0.002         & 0.395 & \textbf{0.561}       \\
Top-Count & ---    & 8/15                 & 3/11                 & 5/11                       & 1/2                        & 0/5                        & 11/21                & 0/0    & 0/0    & 0/2            & 0/0   & 0/5            & 2/14                 & 0/1            & 0/0   & 6/21 \\
\bottomrule
\end{longtable}
}
\end{threeparttable}
\end{table*}
\end{landscape}

\begin{figure}[htbp]
\centering
\includegraphics[width=4.3in,keepaspectratio]{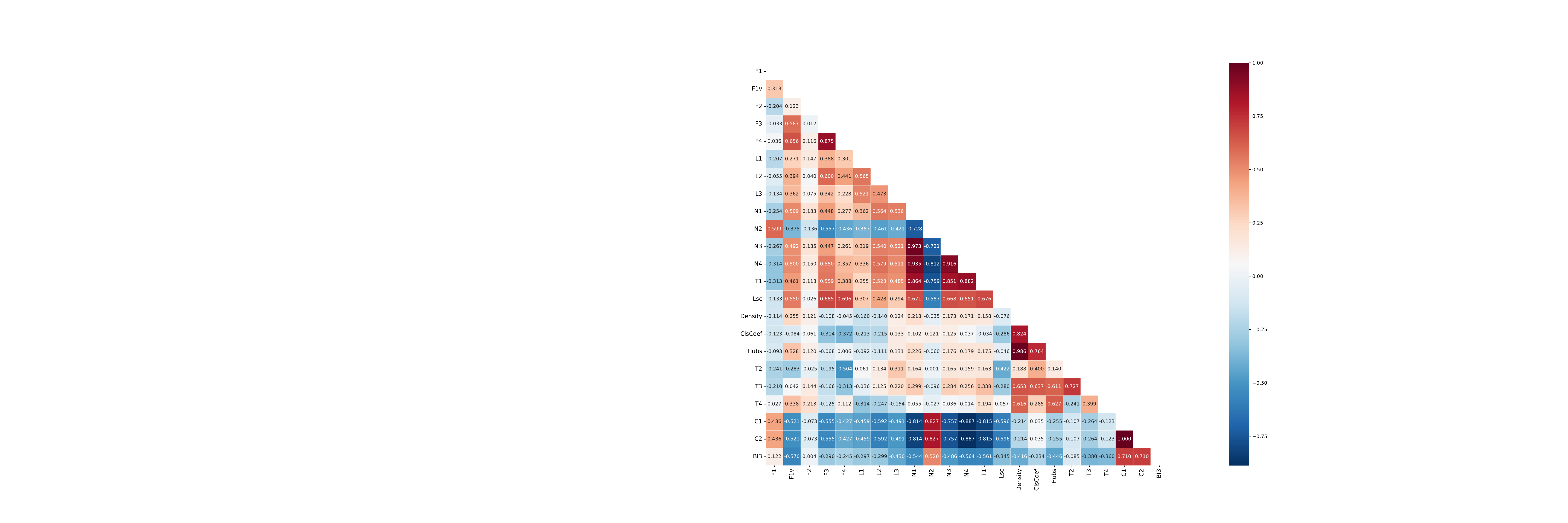}
\vspace{-10pt}
\caption{Spearman correlation matrix for the dataset complexity measures}
\label{Fig10}
\end{figure}

Fig. \ref{Fig10} presents pairwise comparisons of dataset complexity measures using the Spearman correlation. C1 and C2 are perfectly positively correlated ($r_{s}=1.0$), indicating significant redundancy in measuring data characteristics, particularly for class imbalance. Density and Hubs are approximately perfectly positively correlated with $r_{s}=0.986$, while N1, N3, and N4 are approximately perfectly positively correlated with $r_{s}$ values of 0.973, 0.946, and 0.917, respectively, indicating significant redundancy between these measures. F2 has little to no correlation with almost all other measures, with the highest correlation observed between T4 and F2 with $r_{s}=0.216$. Further analysis reveals that the low correlation is mainly due to F2 taking a value of 0 on the vast majority of datasets, making it insufficient to adequately describe the data characteristics. Additionally, three network-based measures, Density, ClsCoef, and Hubs, have low correlation with other measures, but their coefficient of variation is not low, indicating that other data complexities do not directly affect the graph of a dataset. Hence, F2, C2, N3, N4, and Hubs are removed in subsequent analysis. The remaining measures exhibit either strong or weak positive correlation with each other, but no extreme correlation ($r_{s}>0.9$) is observed, suggesting that they complement each other and provide a comprehensive description of dataset complexity from various perspectives.

\begin{center}
\begin{tcolorbox}
\textbf{Finding 9}: Besides F2, C2, N3, N4, and Hubs, the considered dataset complexity measures contribute to a thorough understanding of dataset complexity from multiple perspectives.
\end{tcolorbox}
\end{center}

After conducting the aforementioned analysis, two intriguing questions emerge: how about the data complexities of different datasets in terms of different dataset complexity measures, and what is the relationship between the imbalanced dataset hardness (defined in Section \ref{sect:ih} and abbreviated as IDSH) and the considered dataset complexity measures? Table \ref{table9} presents 18 dataset complexity measures for each datset, along with the Spearman correlation coefficients relating the IDSH value to the these measures. The highest value in each column (represents the highest value in the corresponding measure) is underlined, and the top three highest values are highlighted in bold. The penultimate row of the table gives the average value for all 36 datasets, while the last row provides the Spearman correlation coefficients relating the IDSH value to the considered dataset complexity measures. Additionally, a significance test is performed on each correlation coefficient, and those failing to pass the test with a $p$-value greater than 0.05 are marked in gray.

As shown in Table \ref{table9}, jedit-4.3 (22), log4j-1.2 (23), and CM1 (6) are the most difficult datasets, with IDSH values over 0.88, implying an average predictive accuracy (measured by MCC) of all classifiers on these datasets below 0.12. Other difficult datasets include PC2 (14), MC1 (10), camel-1.6 (20), and Zxing (36), all with IDSH values above 0.85, thereby demonstrating their high classification difficulty. Further analysis reveals that these datasets tend to exhibit high values across multiple dataset complexity measures. For example, jedit-4.3 (22) has the highest values in the N2 and BI3 measures, while log4j-1.2 (23) has the highest values in the L3, ClsCoef, and T3 measures, and ranks in the top three or ties for the highest values in the F1 and T4 measures. Furthermore, Eclipse34-debug (34) is the easiest dataset with an IDSH value of 0.009, indicating an average MCC value of 0.991 and nearly perfect prediction. Actually, we observe that the F1 complexity measure for this dataset is the lowest among all of the datasets, which suggests that there are few instances of class overlap in this dataset. Moreover, analyzing the correlation between IDSH values and various measures reveals significant correlations only with F1, N2, C1, and BI3. Specifically, F1 shows the strongest correlation with IDSH, with an $r_s$ over 0.7. The remaining measures have moderate correlations with IDSH, with $r_s$ above 0.4. These results indicate that even when examining at the dataset level, class overlap remains the most significant factor contributing to the complexity of a defect dataset, particularly, feature overlap and structural overlap are worth our attention.

\begin{center}
\begin{tcolorbox}
\textbf{Finding 10}: Feature overlap and structural overlap serve as the primary contributors to defect dataset complexity, as evidenced by the strong correlation of F1 with the overall dataset hardness and the moderate correlation of N2.
\end{tcolorbox}
\end{center}

Moreover, two measures of class imbalance exhibit moderate correlations with the IDSH value, seemingly contradicting the conclusions in Section \ref{rq2}. In that section, we have concluded that measures of class imbalance display weak correlations with the dataset hardness (DSH). Upon further analysis, we realize that the main reason for this discrepancy lies in the fact that dataset hardness quantification in Section \ref{rq2} is determined by averaging instance hardness across the entire dataset. Unfortunately, this approach neglects to account for the class imbalance issue prevalent in defect prediction tasks, resulting in an overemphasis on non-defective instances (i.e., majority class instances) when characterizing dataset hardness. Since the MCC is a perfectly symmetric evaluation metric, it penalizes misclassification of different class instances equally. As a result, compared to the DSH value in Section \ref{rq2}, the IDSH value, as calculated by the MCC, considers the hardness of minority class instances to a greater extent. The difference in the definition of dataset hardness consequently leads the imbalance ratio to significantly impact the IDSH value. In addition, in defect prediction tasks, accurately predicting the defective class typically holds greater importance than predicting the non-defective class. Consequently, relying solely on the overall instance hardness distribution fails to accurately assess the predictive difficulty of a dataset. For example, PC2 (14) is one of the top 5 datasets with the lowest median (or average) instance hardness value as shown in Fig. \ref{Fig5a}. However, in fact, it is among the top 5 datasets with the highest dataset hardness in terms of IDSH values, indicating that most models struggle to accurately predict defects on PC2 (14). Additionally, PC5 (17), one of the top 5 datasets with the highest median (or average) instance hardness value, does not belong to the 10 datasets with the highest IDSH values. Therefore, even if the majority of instances possess low hardness values, a defect prediction model may still fail to achieve satisfactory predictive performance in terms of MCC. In other words, class imbalance does not substantially influence the hardness of individual instances. However, when assessing data complexity at the dataset level, it is necessary to introduce a novel definition of dataset hardness (such as the IDSH in this paper) that considers the misclassification costs for both defective and non-defective instances, where class imbalance may inevitably have an impact.

\begin{center}
\begin{tcolorbox}
\textbf{Finding 11}: A novel definition of dataset hardness is necessary in defect prediction since averaging instance hardness disregards different misclassification costs of two classes. However, this definition will inevitably relate to class imbalance or imbalance ratio.
\end{tcolorbox}
\end{center}

\begin{figure*}[htbp]
\centering
\begin{minipage}{0.47\linewidth}
\centering
\includegraphics[width=0.9\textwidth]{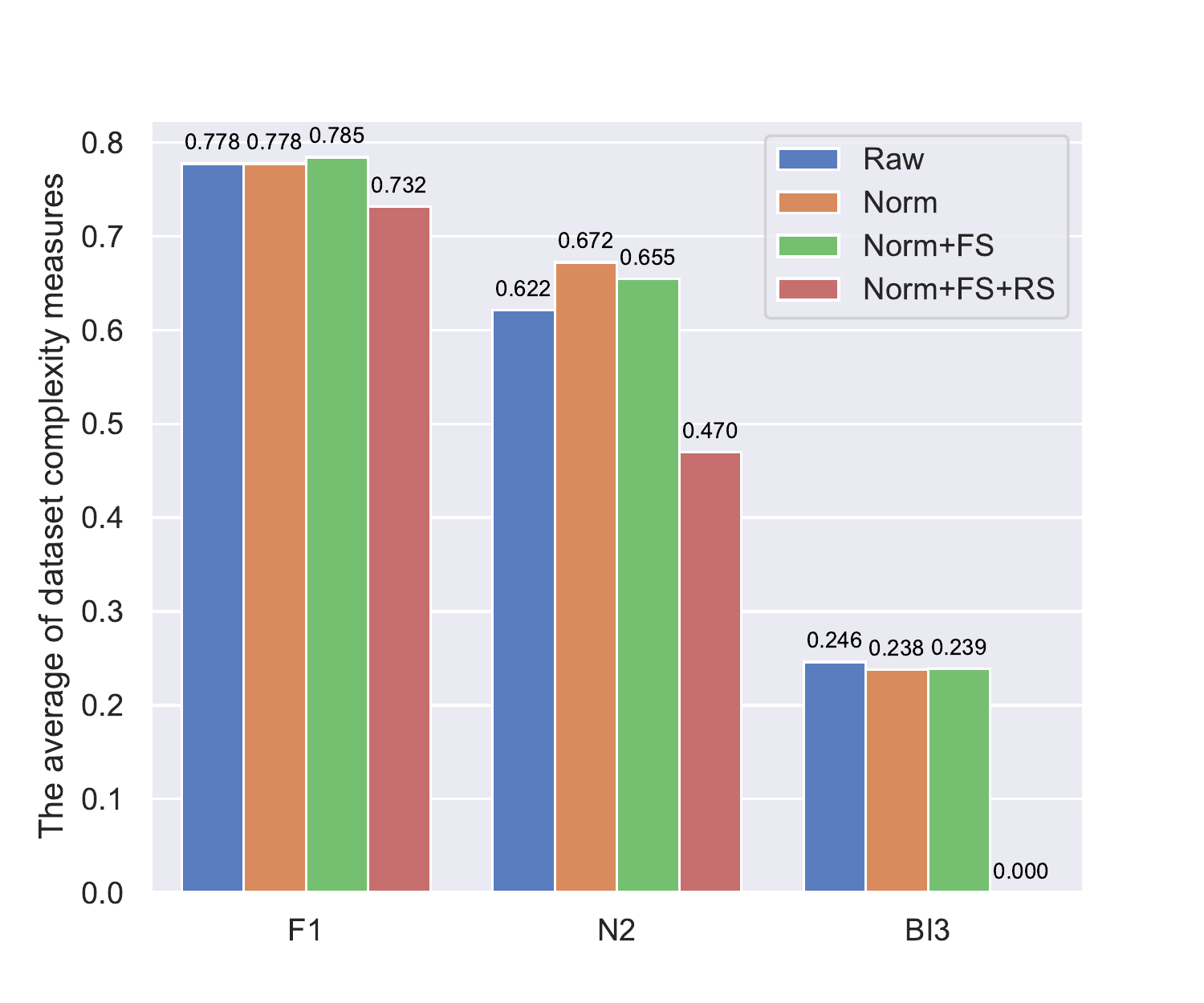}
\caption{Avg. dataset complexity measures across 36 datasets with data preprocessing}
\label{Fig11a}
\end{minipage}\hfill
\begin{minipage}{0.47\linewidth}
\centering
\includegraphics[width=0.9\textwidth]{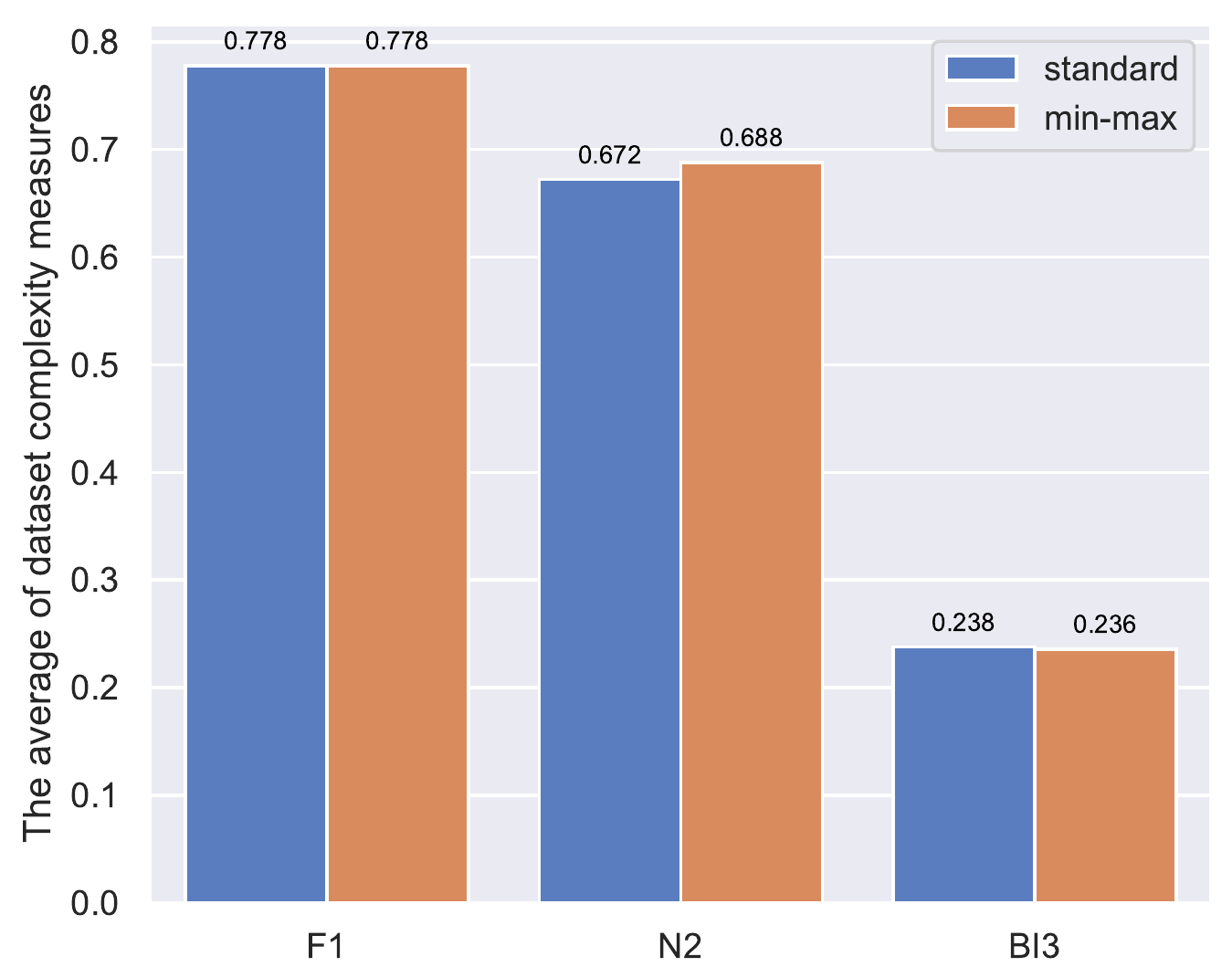}
\caption{Avg. dataset complexity measures across 36 datasets with data normalization}
\label{Fig11b}
\end{minipage}

\vspace{10pt}

\begin{minipage}{0.47\linewidth}
\centering
\includegraphics[width=0.9\textwidth]{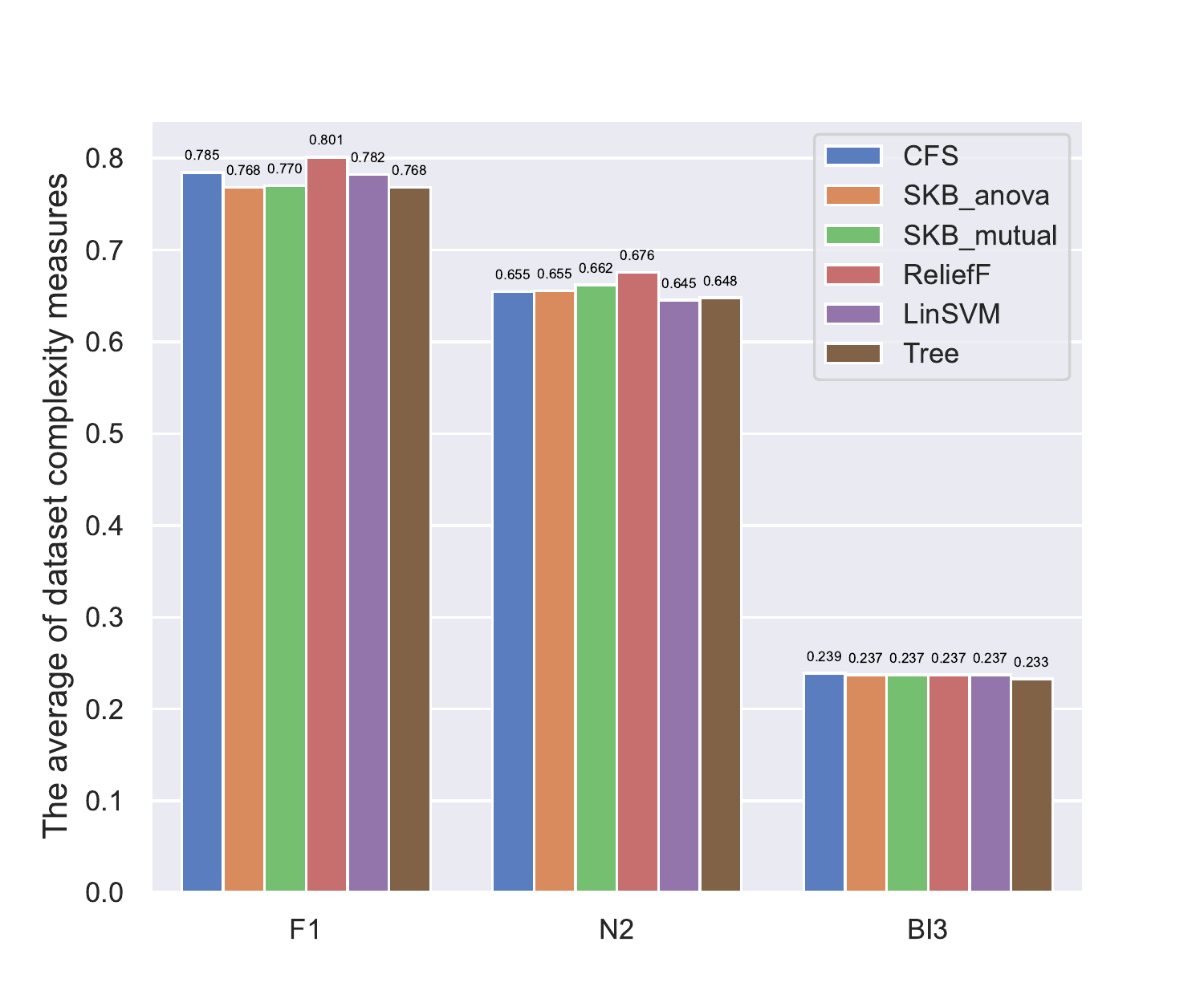}
\caption{Avg. dataset complexity measures across 36 datasets with feature selection}
\label{Fig11c}
\end{minipage}\hfill
\begin{minipage}{0.47\linewidth}
\centering
\includegraphics[width=0.9\textwidth]{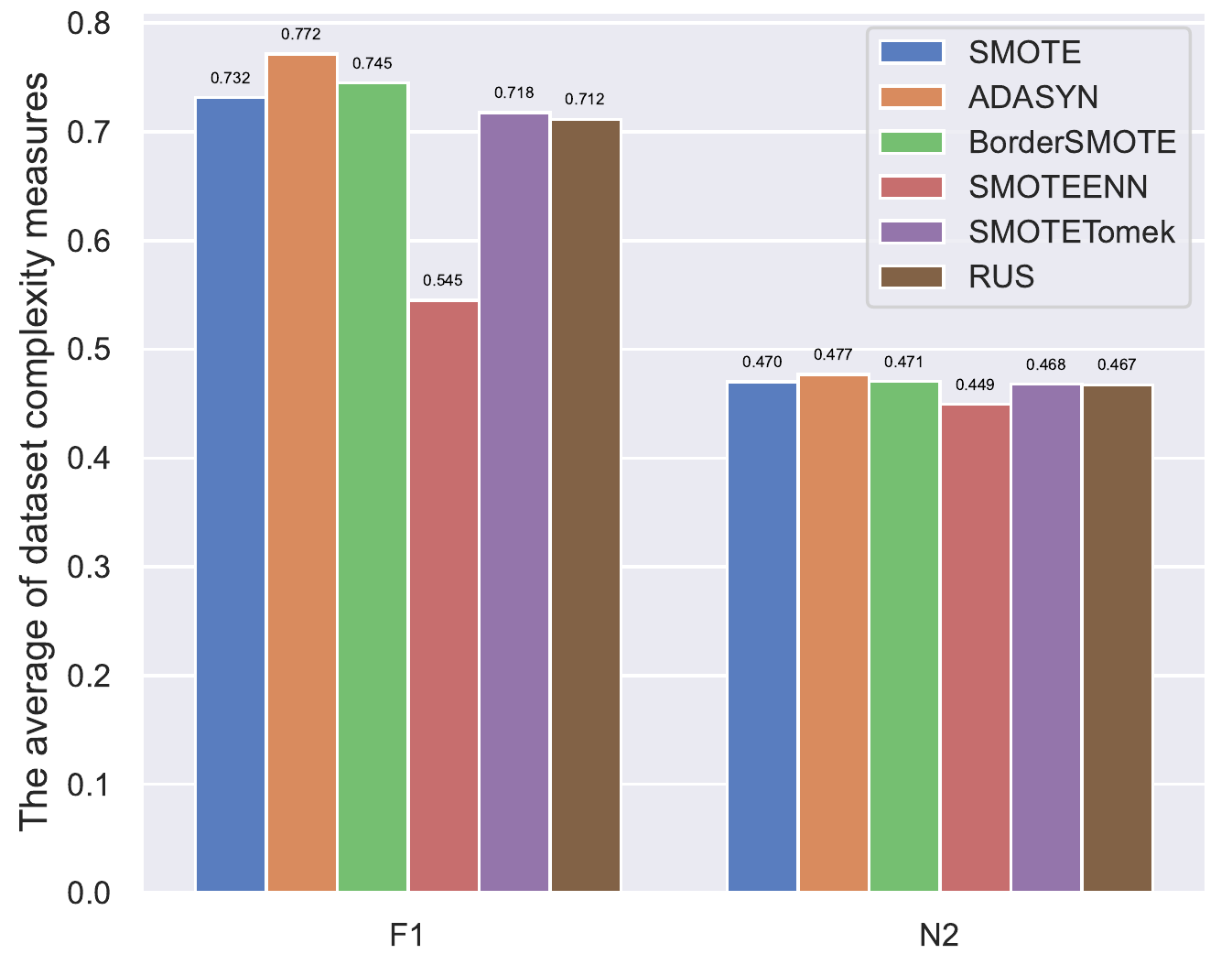}
\caption{Avg. dataset complexity measures across 36 datasets with data re-sampling}
\label{Fig11d}
\end{minipage}
\end{figure*}

Furthermore, Fig. \ref{Fig11a} presents the average of complexity measures for datasets processed using data normalization, feature selection, and data re-sampling. ``Raw'' denotes raw data, ``Norm'' refers to data normalization (standard normalization), ``Norm+FS'' signifies data with feature selection (CFS) applied after normalization, and ``Norm+FS+RS'' refers to data with re-sampling (SMOTE) executed following normalization and feature selection. Subsequently, more data preprocessing techniques are implemented to investigate their impacts on the data complexity in terms of dataset complexity measures. Specifically, standard and min-max normalization are implemented for the study of data normalization, 6 feature selection methods are implemented for the study of feature selection, including 2 filter-based subset selection methods (CFS, ReliefF), 2 filter-based feature ranking methods (SKB\_anova, SKB\_mutual), and 2 wrapper-based subset selection methods (LinSVM, Tree), which select features according to the feature importance given by the linear SVM and ExtraTree classifier, respectively \cite{xu2016impact}. Moreover, 6 data re-sampling methods are implemented for the study of data re-sampling, including 3 oversampling method (SMOTE, ADASYN, and BorderSMOTE), 1 undersampling method (RUS) and 2 hybrid sampling method (SMOTETomek, SMOTEENN) \cite{JMLR:v18:16-365}. The average of complexity measures for datasets processed using data preprocessing methods are depicted in Fig. \ref{Fig11b}, Fig. \ref{Fig11c}, and Fig. \ref{Fig11d}. In this subsection, we employ three dataset complexity measures: F1, N2, and BI3, which represent feature overlap, structural overlap, and class imbalance, respectively. As mentioned earlier, these measures exhibit the strongest correlation with IDSH values. Notably, C1 has been excluded from consideration due to its inability to estimate the deterioration caused solely by class imbalance in the entire dataset \cite{lu2019bayes}.

As depicted in Fig. \ref{Fig11a}, standard normalization does not affect the extent of feature overlap and can help mitigate the effects of class imbalance. However, it is essential to exercise caution when introducing data normalization in certain datasets, as it may inadvertently worsen the issue of structural overlap, complicating model training and ultimately impairing the test performance of defect prediction models. Therefore, careful consideration is necessary when applying data normalization techniques in defect prediction. On the other hand, feature selection reduces the feature space, potentially increasing feature overlap while reducing structural overlap. Lastly, data re-sampling decreases both feature and structural overlap, as well as alleviates the effect of class imbalance on instance classification. For re-sampled datasets, the class imbalance issue is addressed (i.e., each class constitutes 50\%), leading to C1 and BI3 measure values being equal to 0.

\setcounter{table}{8}

\begin{landscape}
\begin{table*}[htbp]
\centering
\caption{Dataset complexity measures for each dataset and their Spearman correlation coefficients for the imbalanced dataset hardness}
\label{table9}
\begin{threeparttable}
\footnotesize
\renewcommand\arraystretch{0.87}
\setlength{\tabcolsep}{1.6mm}{

\begin{longtable}{cccccccccccccccccccc}
\toprule
\multicolumn{1}{l}{Item} & IDSH                  & F1                   & F1v                           & F3                             & F4                             & L1                             & L2                             & L3                            & N1                             & N2                   & T1                             & Lsc                            & Dens.                        & ClsC.                        & T2                            & T3                             & T4                             & C1                   & BI3                  \\
\midrule
1                                & 0.577                & 0.598                & 0.201                         & 0.938                          & 0.562                          & 0.203                          & 0.302                          & 0.262                         & 0.136                          & 0.410                & 0.639                          & 0.952                          & 0.642                          & 0.152                          & \textbf{0.188}                & 0.006                          & 0.033                          & 0.030                & 0.076                \\
2                                & 0.666                & 0.532                & 0.186                         & 0.966                          & 0.779                          & 0.182                          & 0.212                          & 0.150                         & 0.089                          & 0.378                & 0.465                          & 0.985                          & 0.533                          & 0.181                          & 0.061                         & 0.001                          & 0.016                          & 0.265                & 0.198                \\
3                                & 0.816                & NA                   & 0.139                         & 0.915                          & 0.637                          & 0.255                          & 0.080                          & \textbf{0.609}                & 0.064                          & 0.681                & 0.417                          & 0.963                          & 0.218                          & 0.044                          & 0.088                         & 0.003                          & 0.033                          & 0.555                & 0.434                \\
4                                & 0.792                & 0.804                & 0.294                         & \textbf{0.987}                 & \textbf{0.905}                 & 0.212                          & 0.134                          & 0.161                         & 0.095                          & 0.669                & 0.481                          & 0.991                          & 0.339                          & 0.074                          & 0.033                         & 0.001                          & 0.033                          & 0.438                & 0.346                \\
5                                & 0.818                & 0.782                & 0.286                         & 0.973                          & 0.850                          & 0.271                          & 0.169                          & 0.235                         & 0.089                          & 0.658                & 0.556                          & 0.992                          & 0.429                          & 0.106                          & 0.041                         & 0.003                          & 0.066                          & 0.417                & 0.350                \\
6                                & \textbf{0.881}       & 0.833                & 0.267                         & 0.758                          & 0.190                          & 0.455                          & 0.128                          & \textbf{0.535}                & 0.104                          & 0.710                & 0.300                          & 0.972                          & 0.474                          & 0.144                          & 0.113                         & 0.003                          & 0.027                          & 0.447                & 0.374                \\
7                                & 0.843                & 0.897                & \textbf{\uline{0.610}}          & \textbf{\uline{0.998}}           & \textbf{\uline{0.989}}           & \textbf{\uline{0.957}}           & 0.224                          & 0.251                         & 0.151                          & 0.430                & 0.526                          & \textbf{\uline{0.999}}           & 0.333                          & 0.002                          & 0.003                         & 0.000                          & 0.048                          & 0.261                & 0.257                \\
8                                & 0.811                & 0.832                & 0.494                         & 0.957                          & 0.904                          & \textbf{0.940}                 & 0.336                          & 0.254                         & 0.168                          & 0.497                & 0.510                          & \textbf{0.994}                 & 0.490                          & 0.074                          & 0.018                         & 0.001                          & 0.048                          & 0.184                & 0.211                \\
9                                & 0.775                & 0.776                & 0.210                         & 0.897                          & 0.021                          & 0.735                          & 0.186                          & 0.222                         & 0.144                          & 0.576                & 0.428                          & 0.973                          & 0.737                          & 0.310                          & \textbf{0.201}                & 0.005                          & 0.026                          & 0.308                & 0.284                \\
10                               & 0.858                & 0.928                & 0.118                         & 0.876                          & 0.608                          & 0.384                          & 0.019                          & 0.025                         & 0.016                          & \textbf{0.937}       & 0.057                          & 0.952                          & 0.264                          & 0.132                          & 0.019                         & 0.001                          & 0.026                          & \textbf{0.867}       & \textbf{0.574}       \\
11                               & 0.779                & 0.719                & 0.227                         & 0.855                          & 0.024                          & \textbf{0.955}                 & 0.363                          & 0.500                         & \textbf{\uline{0.194}}           & 0.397                & 0.540                          & 0.976                          & 0.746                          & 0.310                          & \textbf{\uline{0.315}}          & 0.008                          & 0.026                          & 0.062                & 0.101                \\
12                               & 0.799                & 0.732                & 0.165                         & 0.760                          & 0.008                          & 0.267                          & 0.128                          & 0.132                         & 0.086                          & 0.756                & 0.252                          & 0.940                          & 0.743                          & \textbf{0.342}                 & 0.148                         & 0.004                          & 0.027                          & 0.531                & 0.441                \\
13                               & 0.757                & 0.748                & 0.177                         & 0.632                          & 0.249                          & 0.213                          & 0.362                          & 0.206                         & 0.070                          & 0.679                & 0.216                          & 0.951                          & 0.350                          & 0.127                          & 0.054                         & 0.001                          & 0.027                          & 0.594                & 0.452                \\
14                               & 0.866                & \textbf{0.955}       & 0.107                         & 0.370                          & 0.000                          & 0.114                          & 0.044                          & 0.032                         & 0.020                          & 0.904                & 0.066                          & 0.803                          & 0.142                          & 0.053                          & 0.050                         & 0.001                          & 0.028                          & \textbf{0.847}       & \textbf{\uline{0.639}} \\
15                               & 0.782                & 0.793                & 0.275                         & 0.745                          & 0.508                          & 0.218                          & 0.128                          & 0.138                         & 0.094                          & 0.816                & 0.307                          & 0.982                          & 0.397                          & 0.115                          & 0.035                         & 0.001                          & 0.027                          & 0.461                & 0.385                \\
16                               & 0.633                & 0.703                & 0.196                         & 0.909                          & 0.722                          & 0.293                          & 0.139                          & 0.141                         & 0.107                          & 0.659                & 0.398                          & 0.993                          & 0.745                          & 0.292                          & 0.029                         & 0.001                          & 0.027                          & 0.419                & 0.351                \\
17                               & 0.783                & 0.802                & 0.447                         & \textbf{0.979}                 & 0.904                          & 0.299                          & 0.270                          & 0.275                         & \textbf{0.175}                 & 0.423                & 0.540                          & \textbf{0.995}                 & 0.562                          & 0.142                          & 0.022                         & 0.001                          & 0.026                          & 0.158                & 0.192                \\
18                               & 0.637                & 0.586                & 0.287                         & 0.907                          & 0.820                          & 0.460                          & 0.195                          & 0.291                         & 0.126                          & 0.442                & 0.501                          & 0.985                          & 0.866                          & 0.297                          & 0.027                         & 0.003                          & \textbf{0.100}                 & 0.235                & 0.224                \\
19                               & 0.829                & 0.755                & 0.240                         & 0.756                          & 0.380                          & 0.194                          & 0.115                          & 0.115                         & 0.090                          & 0.679                & 0.380                          & 0.962                          & 0.857                          & 0.234                          & 0.085                         & \textbf{0.009}                 & \textbf{0.100}                 & 0.484                & 0.353                \\
20                               & 0.855                & 0.931                & \textbf{0.520}                & 0.951                          & 0.841                          & 0.271                          & \textbf{0.418}                 & 0.307                         & 0.150                          & 0.560                & 0.585                          & 0.992                          & 0.842                          & 0.207                          & 0.021                         & 0.002                          & \textbf{0.100}                 & 0.289                & 0.290                \\
21                               & 0.776                & 0.676                & 0.206                         & 0.710                          & 0.401                          & 0.198                          & 0.116                          & 0.105                         & 0.087                          & 0.608                & 0.344                          & 0.950                          & 0.798                          & 0.273                          & 0.057                         & 0.006                          & \textbf{0.100}                 & 0.489                & 0.381                \\
22                               & \textbf{0.895}       & 0.926                & 0.195                         & 0.591                          & 0.075                          & 0.167                          & 0.024                          & 0.043                         & 0.016                          & \textbf{0.913}       & 0.085                          & 0.829                          & 0.803                          & 0.301                          & 0.041                         & 0.002                          & 0.050                          & 0.846                & \textbf{0.574}       \\
23                               & \textbf{\uline{0.909}} & \textbf{0.942}       & 0.319                         & 0.659                          & 0.195                          & 0.170                          & 0.088                          & \textbf{\uline{0.610}}          & 0.078                          & 0.881                & 0.307                          & 0.949                          & 0.850                          & \textbf{\uline{0.385}}           & 0.098                         & \textbf{\uline{0.010}}           & \textbf{0.100}                 & 0.605                & -0.184               \\
24                               & 0.820                & 0.860                & 0.417                         & 0.938                          & 0.765                          & 0.241                          & 0.309                          & 0.312                         & \textbf{0.185}                 & 0.544                & \textbf{\uline{0.738}}           & 0.991                          & \textbf{\uline{0.919}}           & \textbf{0.338}                 & 0.059                         & 0.006                          & \textbf{0.100}                 & 0.027                & -0.064               \\
25                               & 0.541                & 0.769                & 0.357                         & 0.977                          & 0.876                          & 0.500                          & 0.235                          & 0.457                         & 0.129                          & 0.559                & 0.484                          & 0.985                          & 0.821                          & 0.255                          & 0.045                         & 0.005                          & \textbf{0.100}                 & 0.054                & -0.057               \\
26                               & 0.837                & 0.888                & 0.430                         & 0.956                          & 0.826                          & 0.154                          & 0.105                          & 0.100                         & 0.070                          & 0.756                & 0.262                          & 0.973                          & \textbf{0.911}                 & 0.309                          & 0.032                         & 0.003                          & 0.095                          & 0.531                & 0.423                \\
27                               & 0.657                & NA                   & 0.159                         & 0.795                          & 0.330                          & 0.160                          & 0.119                          & 0.148                         & 0.097                          & 0.744                & 0.392                          & 0.955                          & \textbf{0.893}                 & 0.317                          & 0.119                         & 0.006                          & 0.048                          & 0.382                & 0.300                \\
28                               & 0.654                & 0.687                & 0.329                         & 0.914                          & 0.723                          & 0.302                          & \textbf{0.422}                 & 0.199                         & 0.168                          & 0.466                & 0.625                          & 0.978                          & 0.886                          & 0.317                          & 0.078                         & 0.008                          & \textbf{0.100}                 & 0.079                & 0.118                \\
29                               & 0.778                & NA                   & 0.211                         & 0.833                          & 0.683                          & 0.199                          & 0.091                          & 0.193                         & 0.071                          & 0.641                & 0.296                          & 0.966                          & 0.753                          & 0.252                          & 0.023                         & 0.002                          & \textbf{0.100}                 & 0.564                & 0.425                \\
30                               & 0.763                & 0.740                & 0.334                         & 0.917                          & 0.616                          & 0.655                          & 0.358                          & 0.279                         & \textbf{0.179}                 & 0.416                & 0.638                          & 0.978                          & 0.867                          & 0.295                          & 0.087                         & \textbf{0.009}                 & \textbf{0.100}                 & 0.075                & 0.131                \\
31                               & 0.665                & \textbf{\uline{0.988}} & 0.327                         & 0.340                          & 0.024                          & 0.007                          & 0.009                          & 0.017                         & 0.006                          & \textbf{\uline{0.984}} & 0.050                          & 0.864                          & 0.826                          & 0.308                          & 0.022                         & 0.002                          & \textbf{0.100}                 & \textbf{\uline{0.906}} & -0.021               \\
32                               & 0.388                & 0.784                & 0.311                         & 0.813                          & 0.706                          & 0.274                          & 0.112                          & 0.114                         & 0.094                          & 0.679                & 0.355                          & 0.983                          & 0.880                          & 0.253                          & 0.034                         & 0.003                          & \textbf{0.100}                 & 0.178                & -0.084               \\
33                               & 0.619                & 0.750                & 0.432                         & 0.933                          & 0.510                          & 0.849                          & \textbf{\uline{0.454}}           & 0.289                         & 0.160                          & 0.501                & \textbf{0.722}                 & 0.982                          & 0.659                          & 0.112                          & 0.134                         & 0.005                          & 0.038                          & 0.000                & -0.003               \\
34                               & 0.009                & 0.294                & 0.126                         & 0.000                          & 0.000                          & 0.000                          & 0.000                          & 0.000                         & 0.141                          & 0.454                & 0.562                          & 0.992                          & 0.569                          & 0.157                          & 0.016                         & 0.002                          & \textbf{\uline{0.118}}           & 0.194                & 0.210                \\
35                               & 0.443                & 0.739                & 0.433                         & 0.950                          & \textbf{0.910}                 & 0.217                          & 0.191                          & 0.292                         & 0.073                          & 0.427                & 0.302                          & 0.974                          & 0.553                          & 0.042                          & 0.011                         & 0.001                          & 0.059                          & 0.011                & 0.015                \\
36                               & 0.855                & 0.915                & \textbf{0.608}                & 0.932                          & 0.774                          & 0.343                          & 0.296                          & 0.283                         & 0.160                          & 0.541                & \textbf{0.704}                 & 0.988                          & 0.626                          & 0.151                          & 0.065                         & 0.005                          & 0.077                          & 0.124                & 0.162                \\
\midrule
Avg.                   & 0.727                & 0.778                & 0.296                         & 0.816                          & 0.537           & 0.342
& 0.191                & 0.230                & 0.108                & 0.622                         & 0.418                          & 0.964           & 0.648
& 0.206                & 0.069                & 0.004                & 0.062                         & 0.359                          & 0.246                \\
$r_{s}$                              & ---                  & 0.724                & \cellcolor[gray]{0.9}0.052 & \cellcolor[gray]{0.9}-0.043 & \cellcolor[gray]{0.9}-0.058 & \cellcolor[gray]{0.9}-0.040 & \cellcolor[gray]{0.9}-0.217 & \cellcolor[gray]{0.9}0.070 & \cellcolor[gray]{0.9}-0.207 & 0.445                & \cellcolor[gray]{0.9}-0.235 & \cellcolor[gray]{0.9}-0.178 & \cellcolor[gray]{0.9}-0.211 & \cellcolor[gray]{0.9}-0.076 & \cellcolor[gray]{0.9}0.026 & \cellcolor[gray]{0.9}-0.058 & \cellcolor[gray]{0.9}-0.153 & 0.503                & 0.446 \\
\bottomrule
\end{longtable}
}
\end{threeparttable}
\end{table*}
\end{landscape}

%

Following this, we investigate the effects of data normalization, feature selection, and data re-sampling on dataset complexity, respectively. First, Fig. \ref{Fig11b} illustrates the impact of data normalization on data complexity. After applying two data normalization methods, the datasets exhibit almost no noticeable differences in F1 and BI3 measures. However, when utilizing min-max normalization, the dataset demonstrates a marginally higher N2 measure (naturally surpassing that of the raw data). Consequently, it is crucial to recognize that although data normalization approaches can mitigate the effects of class imbalance to some extent, they may exacerbate structural overlaps (especially in the case of min-max normalization) in defect prediction. Second, the impact of feature selection on data complexity is illustrated in Fig. \ref{Fig11c}. Overall, Tree surpasses other methods by consistently achieving low scores across all measures, showcasing its capacity to reduce feature and structural overlaps while mitigating class imbalance effects on prediction accuracy. In contrast, LinSVM, a linear model-based approach, falls short due to its limited ability to fit non-linear classification boundaries. On the other hand, ExtraTree, an ensemble learning model, excels at learning non-linear boundaries. This allows for a more accurate feature selection, ultimately reducing dataset complexity, particularly regarding class overlap. Additionally, both SKB\_anova and SKB\_mutual effectively decrease feature and structural overlaps, primarily due to their reliance on ANOVA F-value or mutual information statistics for feature selection, which helps identify and eliminate redundant or irrelevant features. Third, the impact of data re-sampling on data complexity is illustrated in Fig. \ref{Fig11d}. Although data re-sampling methods can alleviate class imbalance issues (at least by balancing the number of instances in different classes), some methods (such as ADASYN) may introduce negative side effects, such as worsening feature overlap and structural overlap, which in turn substantially affect the predictive performance. On the other hand, some superior re-sampling methods not only tackle class imbalance but also further reduce class overlap, leading to a decrease in dataset complexity. Generally, hybrid sampling methods perform better than other methods, with SMOTEENN being the best among six re-sampling methods. In fact, SMOTEENN integrates SMOTE with Edited Nearest Neighbor (ENN) to generate synthetic samples and eliminate misclassified instances \cite{batista2003balancing}, while SMOTETomek incorporates SMOTE alongside Tomek Links to produce synthetic samples and remove borderline majority class instances \cite{batista2004study}. Considering that ENN and Tomek are designed to remove noise and ambiguous instances from the majority class, it is reasonable to conclude that these two data cleaning techniques help in reducing data complexity. Furthermore, we also notice that dataset processed with RUS exhibits lower complexity in terms of F1 and N2 measures. However, given that RUS eliminates a substantial number of majority class instances, resulting in a smaller dataset, its measure values may be less dependable. In conclusion, different data preprocessing operations have varying impacts on dataset complexity.

\begin{center}
\begin{tcolorbox}
\textbf{Finding 12}: Data normalization may not always be beneficial for model training, as it can worsen structural overlap issues. On the other hand, feature selection can decrease structural overlap at the expense of increased feature overlap. In contrast, hybrid sampling techniques can address class imbalance through oversampling while also minimizing both feature and structural overlap through data cleaning.
\end{tcolorbox}
\end{center}

\vspace{5pt} 

\begin{tcolorbox}[title = {Summary of answers to RQ3 and their implications}:]
Class overlap (measured by feature overlap and structural overlap) stands as the primary factor contributing to dataset complexity. Yet, when considering varying misclassification costs for each class, the imbalance ratio becomes critical. Moreover, various data preprocessing methods yield differing effects on dataset complexity, with no universal solution for all datasets, and inappropriate operations can worsen data complexity. Thus, it is essential to choose appropriate methods customized to the characteristics of each dataset.

\end{tcolorbox}

\subsection{RQ4: How can we integrate data complexity information into the learning process and how effective is it when compared to baseline methods?}
\label{rq4}

To answer RQ4, we explore how to integrate data complexity information into the learning process and thereby improve the test performance of defect prediction models. On the one hand, we inspire from \cite{walmsley2018ensemble} and propose a novel algorithm named HMSMOTEBagging for the generation of pools of classifiers based on the SMOTEBagging algorithm \cite{wang2009diversity}, in which the probability of an instance being selected during the ensemble generation process is inversely proportional to its instance hardness measures. On the other hand, we propose AdaptivePreprocessing, an adaptive data preprocessing procedure based on dataset complexity measures. Therefore, in this section, we will present the fundamental concepts and detailed procedures of both HMSMOTEBagging and AdaptivePreprocessing, followed by conducting experiments to validate their effectiveness.

\subsubsection{Integrating instance hardness measures into the ensemble generation process}
\label{application1}

Traditional ensemble learning presumes an equal likelihood of selecting each instance during the sampling process. This can lead to bootstrapped training sets containing a high proportion of overlapping instances. When faced with such data, some classifiers may either overfit these overlapping instances or fail to learn at all, consequently compromising the generalization performance of the final ensemble model. To address this challenge, HMSMOTEBagging selectively removes overlapping instances probabilistically during the ensemble generation process, building upon the Bagging ensemble. By identifying and eliminating these overlapping instances, the training process will become more stable, resulting in enhanced generalization performance. Section \ref{rq2} emphasizes that various class overlap measures, such as LSC, kDN, U, CL, DS, and DCP, can be employed to identify overlapping instances. Thus, the core idea of this algorithm involves refining the process of generating bootstrapped training sets within the Bagging framework. The selection probability of an instance is now designed to be inversely proportional to its instance hardness measures. This aims to reduce the likelihood of selecting overlapping instances while still allowing for the inclusion of hard instances potentially located along the class boundaries, thereby preserving essential class boundary information. The pseudocode of HMSMOTEBagging is presented in \textbf{Algorithm 1}.

\begin{algorithm}[htbp]
  \caption{Integrating instance hardness measures into the ensemble generation process}
  \label{alg:Framwork}
  \footnotesize
  \begin{algorithmic}[1]

    \REQUIRE
    Training set $\mathcal{D}$;
    Hardness measure $\mathcal{HM}$;
    Base classifier $f$;
    Number of base classifier $n$.

    \ENSURE
    Final ensemble $F(x)$.

    \STATE \textbf{begin}
    \STATE $\mathcal{P}$ $\Leftarrow$ minority instances in $\mathcal{D}$, $\mathcal{N}$ $\Leftarrow$ majority instances in $\mathcal{D}$;
    \STATE Calculate the hardness measure for all instances: $h(x)=\mathcal{HM}(x), \forall x\in D$;
    \FOR {$x \in \mathcal{P}$}
    \STATE Calculate the selection probability $p_{+}(x)=normalize(x, h(x), \mathcal{P}) $;
    \ENDFOR
    \FOR {$x \in \mathcal{N}$}
    \STATE Calculate the selection probability $p_{-}(x)=normalize(x, h(x), \mathcal{N}) $;
    \ENDFOR
    \FOR {$i$ = 1 to $n$}
    \STATE Initialize the training set $\mathcal{D}_{i}$ as the empty set;
    \FOR {$j$ from 1 to $\left|\mathcal{P}\right|$}
    \STATE Add an instance $x_{j} \in \mathcal{P}$ to $\mathcal{D}_{i}$, sampled with replacement according to $p_{+}$;
    \ENDFOR
    \FOR {$j$ from 1 to $\left|\mathcal{N}\right|$}
    \STATE Add an instance $x_{j} \in \mathcal{N}$ to $\mathcal{D}_{i}$, sampled with replacement according to $p_{-}$;
    \ENDFOR
    \STATE The bootstrapped set $\mathcal{D}_{i}$ is oversampled with SMOTE to obtain $\mathcal{D}^{\prime}_{i}$;
    \STATE Train the base classifier $f_{i}$ using the balanced bootstrapped set $\mathcal{D}^{\prime}_{i}$;
    \ENDFOR
    \RETURN Final ensemble $F(x)=\frac{1}{n} \sum_{i=1}^{n} f_{i}(x)$
  \end{algorithmic}
\end{algorithm}

Line 3 of \textbf{Algorithm 1} shows the process of calculating the hardness of all instances. Lines 4 to 9 describe the procedure for calculating the selection probability of each instance from majority and minority classes, which returns a selection probability which is inversely proportional to the hardness measure. Let $n$ be the number of instances in the class to which an instance $x_{i} \in \mathcal{D}$ belongs, and the function $f(x_{i})$ of the instance $x_{i}$ is defined as follows:
$$
\begin{aligned}
f(x_{i})=\frac{1}{n}+(1-\mathcal{HM}(x_{i})),
\end{aligned}
$$
The first term of the equation ensures that every instance belonging to the same class as $x_{i}$ has an equal chance of being selected. This makes sure that even instances with a hardness measure value of 1 are given consideration, avoiding the chance of them being ignored completely. This arrangement is put in place to preserve any hard instances that may be situated at class boundaries. Meanwhile, the second term makes it less likely for instances with higher hardness measures to be selected. As demonstrated in lines 5 and 8, the value of $f(x_{i})$ is normalized by dividing it by the sum of all instances in $\mathcal{P}$ or $\mathcal{N}$ to calculate the probability $p(x_{i})$ of selecting instance $x_{i}$:
$$
\begin{aligned}
p\left(x_i\right)=\frac{f\left(x_i\right)}{\sum_{i=1}^n f\left(x_i\right)}
\end{aligned}
$$


Normalization ensures that $p$ is a valid probability distribution, which is crucial for the proper implementation of the choosing-with-replacement procedure. Next, $n$ bootstrapped training sets are generated by drawing with replacement from the calculated probability distribution (lines 10 to 20). Bearing in mind the issue of class imbalance in defect prediction tasks, the generated training set is then re-sampled using the SMOTE algorithm (line 18). $n$ base classifiers are trained on the balanced training set $\mathcal{D}^{\prime}_{i}$ (line 19), and ultimately, an ensemble classifier is obtained. By making the selection probability inversely proportional to the hardness measure, our goal is to pick hard instances with a lower probability than easy instances during the bootstrapping process.

To verify the effectiveness of HMSMOTEBagging, we utilize kDN, DS, DCP, CL, LSC, and U as instance hardness measures and build models using the algorithm, and the SMOTEBagging algorithm is implemented as a baseline for comparison. For the sake of fairness, we utilize a decision tree as the base classifier for all ensemble learning algorithms, with a total of 50 base classifiers and all other hyper-parameters set to their default values\footnote{\url{https://scikit-learn.org/stable/modules/generated/sklearn.tree.DecisionTreeClassifier.html}}. Additionally, we carry out a 5 by 5-fold cross-validation process on each dataset and record the average of the 25 MCC values obtained from the cross-validation. The average test performance of the HMSMOTEBagging algorithms based on 6 instance hardness measures, alongside the baseline method, SMOTEBagging, is presented in Table \ref{table9} (excluding the last three columns). The top three MCC values among the 7 algorithms are highlighted in bold, and the highest MCC value is underlined. Lastly, we calculate the percentage improvement in test performance (Imp. (\%)) for each HMSMOTEBagging algorithm relative to the baseline method, and those with positive improvement percentages are denoted in gray. As depicted in Table \ref{table9}, 6 versions of the HMSMOTEBagging algorithm employing instance hardness measures outperform the baseline method on average regarding MCC, with average improvement percentages ranging from 4.488\% to 16.444\%. However, we also note that the inclusion of instance hardness measures may yield negative effects on certain datasets, such as CM1 (6) and MC1 (10), where HMSMOTEBagging does not achieve the expected performance enhancement. We attribute this to two primary factors: first, these datasets are extremely difficulty to classify, leading all algorithms to yield low MCC values, which in turn leads to the improvement of prediction performance being considerably impacted by random elements. Second, some datasets exhibit multiple sources of data complexity, with class overlap representing only one aspect; while all hardness measures we employ solely characterize the degree of class overlap, they may not accurately represent the classification difficulty of instances within the dataset. Nonetheless, on the whole, integrating instance hardness measures does contribute to enhancing the learning capacity of the SMOTEBagging algorithm.


\subsubsection{Integrating dataset complexity measures into the data preprocessing process}
\label{application2}

As mentioned in Section \ref{rq3}, there is no universal preprocessing method suitable for every dataset; however, some dataset complexity measures, such as F1 and N2, can assist in identifying suitable preprocessing methods for a given dataset. Notably, data normalization may be harmful for model training, as it can intensify structural overlap as measured by N2, feature selection can amplify feature overlap as indicated by F1, and hybrid sampling can effectively minimize both feature and structural overlap. Drawing from these insights, we propose utilizing the N2 measure to ascertain whether to employ data normalization and utilizing the F1 measure to handpick the optimal operation from the pool of feature selection and data re-sampling methods. This selection process for data preprocessing operations enables the customization of preprocessing strategies to accommodate the dataset characteristics. The procedure of AdaptivePreprocessing is presented in \textbf{Algorithm 2}.

\begin{algorithm}[htbp]
  \caption{Integrating dataset complexity measures into the data preprocessing process}
  \label{alg:Framwork}
  \footnotesize
  \begin{algorithmic}[1]

    \REQUIRE
    Training set $\mathcal{D}$;

    \ENSURE
    Final training set $\mathcal{D}^{\prime}$.
    \STATE \textbf{begin}
    \FOR {$norm\_type \in [Standard, None]$}
    \STATE Calculate the N2 measure of the dataset after normalization with $norm\_type$;
    \ENDFOR
    \STATE Normalize the raw dataset using the method with the lowest N2 value;
    \FOR {$fs\_type \in [SKB\_anova, SKB\_mutual, LinSVM, Tree, None]$}
    \STATE Calculate the f1 measure of the dataset processed by feature selection with $fs\_type$;
    \ENDFOR
    \STATE Select features from normalized dataset using the method with the lowest f1 value;
    \FOR {$rs\_type \in [SMOTE, BorderSMOTE, SMOTETomek, SMOTEENN]$}
    \STATE Calculate the f1 measure of the dataset processed by re-sampling method $rs\_type$;
    \ENDFOR
    \STATE Re-sample the feature-selected dataset using the method with the lowest f1 value.
    \RETURN Final training set $\mathcal{D}^{\prime}$ preprocessed by optimal data preprocessing operations
  \end{algorithmic}
\end{algorithm}

To validate the effectiveness of AdaptivePreprocessing, we use the default operation mentioned in Section \ref{sect:training} as the baseline method. We employ random forest as the learning algorithm, with all hyper-parameters set to default values\footnote{\url{https://scikit-learn.org/stable/modules/generated/sklearn.ensemble.RandomForestClassifier.html}}. We conduct a 5 by 5-fold cross-validation process on each dataset and record average MCC values. The MCC values of the trained models using both preprocessing methods are exhibited in the last three columns of Table \ref{table9}. The final row shows the percentage improvement in test performance (Imp. (\%)) for AdaptivePreprocessing compared to the baseline method, and those with positive improvement percentages are marked in gray. As depicted in Table \ref{table9}, AdaptivePreprocessing outperforms the baseline method in 22 out of 36 datasets, boasting an average improvement of 9.732\%. However, it is necessary to recognize that our approach exhibits a noticeable negative impact on 5 datasets, including KC3 (9), MC1 (10), MC2 (11), PC1 (13), and jedit-4.3 (22). After careful analysis, we deduce that these datasets either exhibit serious class imbalance or suffer from insufficient training data. Serious class imbalance may imply that multiple factors contribute to dataset complexity, with class overlap being just one aspect. Conversely, insufficient training data may result in unreliable measure values, leading the adaptive strategy based on dataset complexity measures to stumble. However, overall, incorporating dataset complexity measures facilitates the customization of preprocessing strategies for specific datasets, ultimately enhancing the predictive performance of defect prediction models.


\vspace{5pt} 
\begin{tcolorbox}[title = {Summary of answers to RQ4 and their implications}:]
By utilizing instance hardness measures to refine the ensemble generation process and leveraging dataset complexity measures for preprocessing method selection, data complexity information serves to boost an algorithm's learning capacity. Nevertheless, they may fail for datasets with inadequate training data or an extremely high degree of class imbalance.
\end{tcolorbox}

\setcounter{table}{9}

\begin{landscape}
\begin{table*}[htbp]
\centering
\caption{Experimental results for comparing the performance of HMSMOTEBagging and AdaptivePreprocessing with baseline methods}
\label{table10}
\begin{threeparttable}
\footnotesize
\renewcommand\arraystretch{0.92}
\setlength{\tabcolsep}{1.90mm}{
\begin{longtable}{cccccccccccccc|ccc}
\toprule
Item & Baseline & kDN   & Imp. (\%)                      & DS    & Imp. (\%)                      & DCP   & Imp. (\%)                      & CL    & Imp. (\%)                      & LSC   & Imp. (\%)                      & U     & \multicolumn{1}{c}{Imp. (\%)}                      & Baseline & Adaptive & Imp. (\%)                      \\
\midrule
1    & 0.497                & \textbf{\uline{0.551}} & \cellcolor[gray]{0.9}10.847 & \textbf{0.540}       & \cellcolor[gray]{0.9}8.650  & 0.531                & \cellcolor[gray]{0.9}6.906  & 0.536                & \cellcolor[gray]{0.9}7.854  & 0.522                & \cellcolor[gray]{0.9}5.042  & \textbf{0.537}       & \cellcolor[gray]{0.9}8.076                      & 0.535   & 0.500    & -6.639                         \\
2    & 0.423                & \textbf{0.464}       & \cellcolor[gray]{0.9}9.525  & 0.443                & \cellcolor[gray]{0.9}4.706  & \textbf{\uline{0.471}} & \cellcolor[gray]{0.9}11.215 & \textbf{0.464}       & \cellcolor[gray]{0.9}9.527  & 0.449                & \cellcolor[gray]{0.9}6.126  & 0.444                & \cellcolor[gray]{0.9}5.006                      & 0.431   & 0.475    & \cellcolor[gray]{0.9}10.240 \\
3    & 0.224                & \textbf{0.310}       & \cellcolor[gray]{0.9}38.265 & 0.136                & -39.548                        & \textbf{\uline{0.324}} & \cellcolor[gray]{0.9}44.411 & \textbf{\uline{0.324}} & \cellcolor[gray]{0.9}44.230 & 0.218                & -2.710                         & 0.238                & \cellcolor[gray]{0.9}6.083                      & 0.225   & 0.271    & \cellcolor[gray]{0.9}20.576 \\
4    & 0.284                & \textbf{\uline{0.318}} & \cellcolor[gray]{0.9}11.741 & 0.253                & -11.120                        & 0.309                & \cellcolor[gray]{0.9}8.731  & \textbf{0.314}       & \cellcolor[gray]{0.9}10.345 & 0.293                & \cellcolor[gray]{0.9}2.934  & \textbf{0.317}       & \cellcolor[gray]{0.9}11.493                     & 0.288   & 0.379    & \cellcolor[gray]{0.9}31.706 \\
5    & 0.203                & 0.228                & \cellcolor[gray]{0.9}12.248 & 0.214                & \cellcolor[gray]{0.9}5.666  & 0.223                & \cellcolor[gray]{0.9}10.080 & \textbf{0.248}       & \cellcolor[gray]{0.9}22.019 & \textbf{0.234}       & \cellcolor[gray]{0.9}15.194 & \textbf{\uline{0.262}} & \cellcolor[gray]{0.9}28.850                     & 0.212   & 0.276    & \cellcolor[gray]{0.9}30.263 \\
6    & 0.071                & 0.034                & -52.858                        & \textbf{0.118}       & \cellcolor[gray]{0.9}65.419 & 0.058                & -19.001                        & 0.045                & -37.145                        & \textbf{\uline{0.137}} & \cellcolor[gray]{0.9}92.563 & \textbf{0.115}       & \cellcolor[gray]{0.9}61.289                     & 0.110   & 0.169    & \cellcolor[gray]{0.9}52.962 \\
7    & 0.157                & \textbf{0.186}       & \cellcolor[gray]{0.9}18.555 & 0.173                & \cellcolor[gray]{0.9}9.742  & \textbf{\uline{0.194}} & \cellcolor[gray]{0.9}23.282 & 0.177                & \cellcolor[gray]{0.9}12.656 & 0.177                & \cellcolor[gray]{0.9}12.299 & \textbf{0.183}       & \cellcolor[gray]{0.9}16.646                     & 0.142   & 0.245    & \cellcolor[gray]{0.9}72.627 \\
8    & 0.203                & \textbf{0.234}       & \cellcolor[gray]{0.9}15.160 & 0.197                & -2.924                         & 0.217                & \cellcolor[gray]{0.9}7.135  & \textbf{\uline{0.235}} & \cellcolor[gray]{0.9}15.676 & 0.214                & \cellcolor[gray]{0.9}5.332  & \textbf{0.228}       & \cellcolor[gray]{0.9}12.433                     & 0.190   & 0.218    & \cellcolor[gray]{0.9}14.925 \\
9    & 0.180                & 0.324                & \cellcolor[gray]{0.9}79.921 & \textbf{0.339}       & \cellcolor[gray]{0.9}88.740 & 0.285                & \cellcolor[gray]{0.9}58.594 & \textbf{\uline{0.353}} & \cellcolor[gray]{0.9}96.285 & 0.314                & \cellcolor[gray]{0.9}74.772 & \textbf{0.332}       & \cellcolor[gray]{0.9}84.859                     & 0.239   & 0.214    & -10.528                        \\
10   & \textbf{0.281}       & 0.194                & -30.983                        & 0.212                & -24.538                        & \textbf{0.274}       & -2.268                         & \textbf{\uline{0.287}} & \cellcolor[gray]{0.9}2.141  & 0.213                & -24.280                        & 0.213                & -23.944                                            & 0.260   & 0.220    & -15.340                        \\
11   & 0.266                & \textbf{\uline{0.313}} & \cellcolor[gray]{0.9}17.797 & 0.244                & -8.152                         & \textbf{0.292}       & \cellcolor[gray]{0.9}10.116 & \textbf{0.311}       & \cellcolor[gray]{0.9}17.203 & 0.285                & \cellcolor[gray]{0.9}7.458  & 0.289                & \cellcolor[gray]{0.9}8.913                      & 0.227   & 0.189    & -16.422                        \\
12   & 0.190                & 0.264                & \cellcolor[gray]{0.9}38.938 & \textbf{\uline{0.334}} & \cellcolor[gray]{0.9}76.151 & 0.228                & \cellcolor[gray]{0.9}20.311 & 0.211                & \cellcolor[gray]{0.9}11.235 & \textbf{0.273}       & \cellcolor[gray]{0.9}43.682 & \textbf{0.283}       & \cellcolor[gray]{0.9}49.246                     & 0.261   & 0.276    & \cellcolor[gray]{0.9}5.999  \\
13   & 0.293                & 0.281                & -3.966                         & 0.291                & -0.627                         & \textbf{0.308}       & \cellcolor[gray]{0.9}5.255  & \textbf{0.294}       & \cellcolor[gray]{0.9}0.476  & \textbf{\uline{0.310}} & \cellcolor[gray]{0.9}5.610  & 0.288                & -1.693                                             & 0.363   & 0.295    & -18.703                        \\
14   & 0.105                & 0.099                & -5.975                         & \textbf{0.170}       & \cellcolor[gray]{0.9}61.134 & 0.065                & -38.090                        & 0.087                & -17.424                        & \textbf{0.182}       & \cellcolor[gray]{0.9}72.263 & \textbf{\uline{0.220}} & \cellcolor[gray]{0.9}108.899                    & 0.140   & 0.177    & \cellcolor[gray]{0.9}26.488 \\
15   & 0.252                & 0.245                & -2.839                         & \textbf{0.308}       & \cellcolor[gray]{0.9}22.424 & 0.268                & \cellcolor[gray]{0.9}6.394  & 0.233                & -7.531                         & \textbf{0.311}       & \cellcolor[gray]{0.9}23.635 & \textbf{\uline{0.323}} & \cellcolor[gray]{0.9}28.086                     & 0.280   & 0.357    & \cellcolor[gray]{0.9}27.819 \\
16   & \textbf{\uline{0.516}} & 0.479                & -7.196                         & \textbf{0.488}       & -5.295                         & \textbf{0.504}       & -2.264                         & 0.455                & -11.830                        & 0.473                & -8.313                         & 0.478                & -7.223                                             & 0.510   & 0.492    & -3.684                         \\
17   & 0.272                & 0.303                & \cellcolor[gray]{0.9}11.437 & \textbf{0.304}       & \cellcolor[gray]{0.9}11.732 & 0.294                & \cellcolor[gray]{0.9}8.108  & 0.272                & \cellcolor[gray]{0.9}0.110  & \textbf{0.308}       & \cellcolor[gray]{0.9}13.252 & \textbf{\uline{0.311}} & \cellcolor[gray]{0.9}14.278                     & 0.277   & 0.332    & \cellcolor[gray]{0.9}20.050 \\
18   & 0.456                & 0.467                & 2.481                          & \textbf{0.472}       & \cellcolor[gray]{0.9}3.437  & \textbf{\uline{0.484}} & \cellcolor[gray]{0.9}6.081  & 0.463                & \cellcolor[gray]{0.9}1.461  & 0.464                & \cellcolor[gray]{0.9}1.682  & \textbf{0.476}       & \cellcolor[gray]{0.9}4.332                      & 0.462   & 0.457    & -0.974                         \\
19   & 0.187                & \textbf{\uline{0.283}} & \cellcolor[gray]{0.9}51.239 & \textbf{0.277}       & \cellcolor[gray]{0.9}47.990 & 0.260                & \cellcolor[gray]{0.9}39.070 & 0.199                & \cellcolor[gray]{0.9}6.336  & \textbf{0.263}       & \cellcolor[gray]{0.9}40.795 & 0.255                & \cellcolor[gray]{0.9}36.182                     & 0.224   & 0.219    & -2.262                         \\
20   & 0.219                & 0.218                & -0.551                         & 0.206                & -5.954                         & \textbf{0.228}       & \cellcolor[gray]{0.9}3.840  & \textbf{\uline{0.247}} & \cellcolor[gray]{0.9}12.621 & 0.224                & \cellcolor[gray]{0.9}2.336  & \textbf{0.243}       & \cellcolor[gray]{0.9}10.705                     & 0.225   & 0.221    & -1.845                         \\
21   & 0.210                & 0.265                & \cellcolor[gray]{0.9}26.026 & \textbf{0.288}       & \cellcolor[gray]{0.9}36.838 & 0.210                & -0.026                         & 0.250                & \cellcolor[gray]{0.9}18.691 & \textbf{0.280}       & \cellcolor[gray]{0.9}33.313 & \textbf{\uline{0.317}} & \cellcolor[gray]{0.9}50.917                     & 0.267   & 0.339    & \cellcolor[gray]{0.9}26.693 \\
22   & \textbf{0.274}       & \textbf{\uline{0.291}} & \cellcolor[gray]{0.9}5.954  & 0.180                & -34.332                        & 0.110                & -60.063                        & 0.229                & -16.347                        & \textbf{0.248}       & -9.565                         & 0.238                & -13.124                                            & 0.247   & 0.103    & -58.274                        \\
23   & 0.190                & 0.131                & -31.302                        & \textbf{0.214}       & \cellcolor[gray]{0.9}12.580 & 0.177                & -7.165                         & 0.148                & -22.124                        & \textbf{\uline{0.215}} & \cellcolor[gray]{0.9}12.839 & \textbf{0.205}       & \cellcolor[gray]{0.9}7.750                      & 0.204   & 0.303    & \cellcolor[gray]{0.9}48.929 \\
24   & \textbf{0.216}       & 0.198                & -8.552                         & 0.188                & -13.155                        & 0.198                & -8.544                         & \textbf{\uline{0.228}} & \cellcolor[gray]{0.9}5.350  & 0.203                & -6.252                         & \textbf{0.205}       & -5.300                                             & 0.215   & 0.308    & \cellcolor[gray]{0.9}42.880 \\
25   & 0.553                & 0.555                & \cellcolor[gray]{0.9}0.400  & 0.557                & \cellcolor[gray]{0.9}0.703  & 0.557                & \cellcolor[gray]{0.9}0.808  & \textbf{0.564}       & \cellcolor[gray]{0.9}1.996  & \textbf{0.563}       & \cellcolor[gray]{0.9}1.776  & \textbf{\uline{0.573}} & \cellcolor[gray]{0.9}3.670                      & 0.569   & 0.579    & \cellcolor[gray]{0.9}1.650  \\
26   & \textbf{0.268}       & 0.218                & -18.576                        & 0.199                & -25.636                        & 0.196                & -26.885                        & 0.227                & -15.271                        & \textbf{0.258}       & -3.610                         & \textbf{\uline{0.280}} & \cellcolor[gray]{0.9}4.788                      & 0.251   & 0.266    & \cellcolor[gray]{0.9}6.261  \\
27   & 0.455                & \textbf{0.510}       & \cellcolor[gray]{0.9}12.211 & 0.436                & -4.128                         & \textbf{\uline{0.535}} & \cellcolor[gray]{0.9}17.708 & \textbf{0.480}       & \cellcolor[gray]{0.9}5.655  & 0.425                & -6.550                         & 0.425                & -6.586                                             & 0.436   & 0.369    & -15.430                        \\
28   & 0.402                & 0.410                & \cellcolor[gray]{0.9}1.799  & 0.408                & \cellcolor[gray]{0.9}1.305  & \textbf{0.419}       & \cellcolor[gray]{0.9}4.022  & 0.404                & \cellcolor[gray]{0.9}0.486  & \textbf{0.432}       & \cellcolor[gray]{0.9}7.320  & \textbf{\uline{0.434}} & \cellcolor[gray]{0.9}7.724                      & 0.446   & 0.457    & \cellcolor[gray]{0.9}2.458  \\
29   & 0.203                & 0.248                & \cellcolor[gray]{0.9}22.536 & \textbf{0.273}       & \cellcolor[gray]{0.9}34.532 & 0.227                & \cellcolor[gray]{0.9}12.047 & 0.252                & \cellcolor[gray]{0.9}24.136 & \textbf{0.303}       & \cellcolor[gray]{0.9}49.387 & \textbf{\uline{0.335}} & \cellcolor[gray]{0.9}65.450                     & 0.279   & 0.332    & \cellcolor[gray]{0.9}19.150 \\
30   & 0.322                & \textbf{0.362}       & \cellcolor[gray]{0.9}12.483 & \textbf{0.349}       & \cellcolor[gray]{0.9}8.415  & \textbf{\uline{0.365}} & \cellcolor[gray]{0.9}13.422 & 0.294                & -8.572                         & 0.334                & \cellcolor[gray]{0.9}3.807  & 0.337                & \cellcolor[gray]{0.9}4.717                      & 0.301   & 0.352    & \cellcolor[gray]{0.9}16.969 \\
31   & 0.582                & \textbf{\uline{0.636}} & \cellcolor[gray]{0.9}9.230  & 0.441                & -24.178                        & \textbf{0.606}       & \cellcolor[gray]{0.9}4.226  & 0.589                & \cellcolor[gray]{0.9}1.186  & 0.560                & -3.730                         & \textbf{0.591}       & \cellcolor[gray]{0.9}1.564                      & 0.532   & 0.556    & \cellcolor[gray]{0.9}4.536  \\
32   & 0.754                & \textbf{0.803}       & \cellcolor[gray]{0.9}6.542  & \textbf{0.800}       & \cellcolor[gray]{0.9}6.157  & \textbf{\uline{0.812}} & \cellcolor[gray]{0.9}7.681  & 0.761                & \cellcolor[gray]{0.9}0.939  & 0.778                & \cellcolor[gray]{0.9}3.198  & 0.792                & \cellcolor[gray]{0.9}5.054                      & 0.748   & 0.768    & \cellcolor[gray]{0.9}2.696  \\
33   & \textbf{0.425}       & 0.417                & -1.790                         & 0.416                & -1.949                         & \textbf{0.425}       & \cellcolor[gray]{0.9}0.210  & \textbf{\uline{0.452}} & \cellcolor[gray]{0.9}6.362  & 0.418                & -1.569                         & \textbf{0.427}       & \cellcolor[gray]{0.9}0.530                      & 0.427   & 0.412    & -3.360                         \\
34   & \textbf{\uline{1.000}} & \textbf{\uline{1.000}} & 0.000                          & \textbf{\uline{1.000}} & 0.000                          & \textbf{\uline{1.000}} & 0.000                          & \textbf{\uline{1.000}} & 0.000                          & \textbf{\uline{1.000}} & 0.000                          & \textbf{\uline{1.000}} & 0.000                                              & 1.000   & 1.000    & 0.000                          \\
35   & \textbf{\uline{0.785}} & \textbf{0.776}       & -1.150                         & 0.755                & -3.796                         & 0.772                & -1.589                         & \textbf{0.786}       & \cellcolor[gray]{0.9}0.132  & 0.757                & -3.476                         & 0.757                & -3.463                                             & 0.786   & 0.784    & -0.158                         \\
36   & 0.212                & 0.208                & -1.476                         & \textbf{\uline{0.227}} & \cellcolor[gray]{0.9}7.191  & 0.207                & -2.189                         & 0.187                & -11.365                        & \textbf{0.214}       & \cellcolor[gray]{0.9}0.939  & \textbf{0.224}       & \cellcolor[gray]{0.9}5.792                      & 0.189   & 0.224    & \cellcolor[gray]{0.9}18.079 \\
\midrule
Avg. & 0.337                & \textbf{0.356}       & \cellcolor[gray]{0.9}6.892  & 0.346                & \cellcolor[gray]{0.9}8.561  & 0.351                & \cellcolor[gray]{0.9}4.488  & 0.350                & \cellcolor[gray]{0.9}5.208  & \textbf{0.357}       & \cellcolor[gray]{0.9}12.986 & \textbf{\uline{0.366}} & \multicolumn{1}{c}{\cellcolor[gray]{0.9}16.444} & 0.347   & 0.365    & \cellcolor[gray]{0.9}9.732 \\
\bottomrule
\end{longtable}
}
\end{threeparttable}
\end{table*}
\end{landscape}

\section{Threats to Validity}
\label{sect:validity}

Threats to validity refer to the extent to which a study matches what the authors want to investigate \cite{siegmund2015views}. Threats to validity usually include threats to external, internal, and construct validity. As in other empirical studies, there are also several threats to the validity of our study.

\subsection{External Validity}
\label{ext-val}

An external threat to validity concerns the generalization of our results and conclusions to other defect datasets and defect prediction models. Our empirical study in this paper is conducted on a limited number of models and datasets, which may restrict the generalizability of our findings to newer models and datasets. However, we have examined 36 datasets from different corpora (i.e., AEEEM, NASA, PROMISE, and ReLink) and 11 models with varying biases (i.e., LR, SVM, DT, RF, AdaBoost, and NN). This diversity of datasets and models helps to ensure that our experimental results can, to some extent, satisfy the requirements of generalization and generalizability. Additionally, we have provided a detailed description of our experimental setup, which can be useful for researchers to replicate our study on a wider range of datasets and models.


\subsection{Internal Validity}
\label{int-val}

An internal threat to validity concerns the correctness of our implementations. To ensure the correctness of our implementations to the greatest extent possible, we attempt to utilize established open-source libraries wherever feasible. Specifically, we implement instance hardness measures and dataset complexity measures using PyHard and Problexity, as previously noted in Section \ref{subsect:dc}. Additionally, we implement classification algorithms using Scikit-Learn \cite{scikit-learn}, Bayesian optimization techniques using Hyperopt \cite{bergstra2013making}, and data preprocessing techniques using Scikit-Learn \cite{scikit-learn} and imbalanced-learn \cite{JMLR:v18:16-365}. For the remaining portions, we borrow source code made available in relevant studies \cite{li2020understanding, shrikanth2021early}. Another internal validity threat concerns the potential impact of the randomness inherent in the data splitting approach. To mitigate this threat, we conduct all experiments using a 5 by 5-fold cross-validation procedure. Furthermore, we utilize the Wilcoxon rank test to determine the practical significance of our results, thus limiting the impact of this threat on our study.



\subsection{Construct Validity}
\label{con-val}

An external threat to validity concerns the training process of defect prediction models. We employ a Bayesian optimization technique to determine the hyper-parameter values of each classification algorithm. Considering using different hyper-parameter values may lead to different model performance, we define a candidate value space for key hyper-parameters in each algorithm based on relevant studies \cite{tantithamthavorn2016automated, tantithamthavorn2018impact2, li2020understanding}, while other hyper-parameters use the recommended values in Scikit-Learn. Despite our efforts to adhere to common practices in relevant studies, we cannot guarantee that variations in hyper-parameter spaces and candidate value spaces will not affect the testing performance of the model or the implications of the experimental conclusions. Nonetheless, we have made all experimental data and source code in this paper freely available to enable future researchers to replicate our experiments and conduct further investigations.

\section{Conclusions and Future Work}
\label{sect:conclusion}

Data complexity is a fundamental concept in ML research that encompasses instance hardness and dataset complexity. Data complexity analysis provides us with a deeper understanding of the factors that make learning tasks challenging. For ML-based defect prediction, a comprehensive understanding of data complexity can provide valuable perspectives that can further improve defect prediction techniques. Despite its critical importance, however, there has been a noticeable lack of comprehensive empirical studies on data complexity in the area of defect prediction.

In this paper, we estimate the hardness of over 33,000 instances within 36 defect datasets using 11 representative algorithms chosen from a pool of 25 candidates. Additionally, we employ 15 instance hardness measures and 23 dataset complexity measures to characterize data complexity in defect prediction tasks. Building upon this, we use Spearman's correlation analysis method to investigate the sources of data complexity at both the instance and dataset levels, as well as to analyze the effects of data preprocessing operations on dataset complexity. We find that (1) despite the right-skewed overall distribution of instance hardness for both classes, significant disparities across datasets and classes require consideration when analyzing data complexity in defect prediction tasks; (2) class overlap is the primary factor contributing to instance hardness, which can be characterized through feature, structural, instance, and multiresolution overlap; (3) feature and structural overlap are primary factors contributing to data complexity in defect prediction from a dataset-level perspective, with the imbalance ratio exhibiting relevance to dataset hardness only when accounting for misclassification costs of different classes; (4) no universal data preprocessing method exists, and data preprocessing may not always reduce data complexity, i.e., data normalization can inadvertently worsen structural overlap issues, whereas feature selection can reduce structural overlap while increasing feature overlap. Lastly, we seek to incorporate data complexity information into the training process to mitigate the impact of class overlap, and present ways that instance hardness values, instance hardness measures, and dataset complexity measures can be used to enhance learning and improve the defect prediction performance.

In the future, we plan to extend this empirical study to cross-project defect prediction, as well as unsupervised and semi-supervised learning-based defect prediction. Furthermore, we intend to conduct further analysis on the joint-effect of class imbalance and class overlap on defect prediction tasks. At the same time, we intend to explore the integration of instance hardness measures and dataset complexity measures with active learning methods to enhance the learning process.

\begin{acks}
The authors extend their sincere appreciation to the editor and anonymous reviewers for their perceptive feedback and constructive recommendations, which have substantially elevated the quality of this paper. The first author is especially grateful to his girl, Qing Qin, whose trust and patience have ignited an unparalleled passion and dedication within him, inspiring him to persevere and carry out each experiment in this paper.

\end{acks}

\bibliographystyle{ACM-Reference-Format}
\bibliography{reference}


%

%

\end{document}